\documentclass[11pt]{article}
\pdfoutput=1 % if your are submitting a pdflatex (i.e. if you have
\usepackage{jheppub} % for details on the use of the package, please
\usepackage[T1]{fontenc} % if needed

\usepackage{bm}
\usepackage{bigdelim}
\usepackage[table,dvipsnames]{xcolor}
  \def\red{\textcolor{red}}
  \def\blue{\textcolor{blue}}
  \def\purple{\textcolor{purple}}
  
\usepackage{hhline}

\def\alphas{\alpha_{\rm s}}
\def\Nc{N_{\rm c}}
\def\Nf{N_{\rm f}}
\def\CA{C_{\rm A}}
\def\qhatA{\hat q_{\rm A}}
\def\Re{\operatorname{Re}}
\def\sgn{\operatorname{sgn}}
\def\Beta{\operatorname{B}}
\def\gammaE{\gamma_{\rm\scriptscriptstyle E}}
\def\eps{\epsilon}

\def\Ybar{\overline{Y}}

\def\B{{\bm B}}
\def\grad{{\bm\nabla}}

\def\ix{{\rm i}}
\def\fx{{\rm f}}
\def\xx{{\rm x}}
\def\xbx{{\bar{\rm x}}}
\def\yx{{\rm y}}
\def\ybx{{\bar{\rm y}}}

\def\Ax{\ybx}
\def\bx{\yx}

\def\seq{{\rm seq}}
\def\new{{\rm new}}

\def\yfrak{{\mathfrak y}}
\def\bbI{\mathbb{I}}

\def\sh{\operatorname{sh}}
\def\ch{\operatorname{ch}}

\def\coth{\operatorname{coth}}

\def\altx{{\mathfrak x}}
\def\di{p}
\def\dj{q}
\def\dk{r}
\def\dl{s}
\def\qmark{{\bm ?}}

\title{The LPM effect in sequential bremsstrahlung: analytic
  results for sub-leading (single) logarithms}

\author[a]{Peter Arnold,}
\author[b,c]{Tyler Gorda,}
\author[d]{Shahin Iqbal%
\footnote{
  During the bulk of this work, Shahin Iqbal was on leave from
  the National Centre for Physics, Quaid-i-Azam University Campus,
  Islamabad, Pakistan.
}}

\affiliation[a]{Department of Physics, University of Virginia,
  P.O.\ Box 400714, 
  Charlottesville, VA 22904, U.S.A.}
\affiliation[b]{Technische Universit{\"a}t Darmstadt, Department of
  Physics, D-64289 Darmstadt, Germany}
\affiliation[c]{Helmholtz Research Academy for FAIR,
  D-64289 Darmstadt, Germany}
\affiliation[d]{Institute of Particle Physics,
    Central China Normal University, Wuhan, 430079, China}

\emailAdd{parnold@virginia.edu}
\emailAdd{tyler.gorda@physik.tu-darmstadt.de}
\emailAdd{smi6nd@virginia.edu}

\abstract{
Consider the in-medium splitting $g \to gg$ of a very high-energy
gluon traversing a QCD medium,
accounting for the Landau-Pomeranchuk-Migdal (LPM) effect.
It has been known for some time that soft radiative corrections to that
splitting generate a double-log correction to the splitting rate,
whose effects can be absorbed into running of the medium parameter
$\hat q$
describing the rate of transverse momentum kicks to high-energy particles
due to small-angle scattering from the medium.
Less has been known about sub-leading, {\it single} logarithms
in this context.
In this paper, we find analytic formulas for those single logs
(with various caveats and clarifications).
}

\begin{document} 
\maketitle
\flushbottom

%%%%%%%%%%%%%%%%%%%%%%%%%%%%%%%%%%%%%%%%%%%%%%%%%%%%%%%%%%%%%%%%%%%%%%%%%%%%%%%

\section{Introduction}
\label{sec:intro}

\subsection {Background and motivation}
\label{sec:background}

When passing through matter, high energy particles lose energy by
showering, via the splitting processes of hard bremsstrahlung and pair
production.  At very high energy, the quantum mechanical duration of
each splitting process, known as the formation time, exceeds the mean
free time for collisions with the medium, leading to a significant
reduction in the splitting rate known as the Landau-Pomeranchuk-Migdal
(LPM) effect \cite{LP1,LP2,Migdal}.%
\footnote{
  The papers of Landau and Pomeranchuk \cite{LP1,LP2} are also available in
  English translation \cite{LPenglish}.
}
The generalization of the LPM effect from QED to QCD was originally
carried out by
Baier, Dokshitzer, Mueller, Peigne, and Schiff \cite{BDMPS1,BDMPS2,BDMPS3}
and by Zakharov \cite{Zakharov1,Zakharov2}
(BDMPS-Z).
A long-standing problem in field theory has
been to understand how to implement this effect in cases where
the formation times of two consecutive splittings overlap.
Several authors \cite{Blaizot,Iancu,Wu} have previously analyzed this issue
for QCD at leading-log order.  They found that (i) soft gluon bremsstrahlung
produces a double-log enhancement of overlap effects and that (ii) this
enhancement can be absorbed into a previously-discovered running
\cite{Wu0} of the medium parameter $\hat q$
(sometimes called the ``jet quenching parameter'')
describing transverse momentum
diffusion of a high-energy particle moving through the medium.
In a series of papers \cite{2brem,seq,dimreg,4point,QEDnf,qedNfstop,qcd},
we and collaborators have worked on a
program to evaluate the effects of
overlapping formation times without leading-log or soft
bremsstrahlung approximations.
We have verified \cite{qcd} that our
results contain the previously known double logarithm.
That double logarithm is accompanied by a
sub-leading single logarithm, which was
extracted numerically in ref.\ \cite{qcd}.
The purpose of the current paper is
to now extract an analytic result for the single logarithm.

The overarching goal of the program
\cite{2brem,seq,dimreg,4point,QEDnf,qedNfstop,qcd},
including the contribution of this paper, is to
ultimately provide a theoretical calculation that will
determine whether or not in-medium jets can be treated as
the evolution of a collection of individual high-energy
partons, or whether formation times overlap so drastically that
the number distribution of high-energy partons in the shower at
any moment is not even approximately a sensible concept.%
\footnote{
  For more discussion of the motivations and current state of this program,
  see sections 1 of refs.\ \cite{qedNfstop,qcd}.
}
This is a problem whose answer, for a first pass,
would be interesting even in the simplest, idealized situation.
So, though the framework developed in
refs.\ \cite{2brem,seq,dimreg,4point,QEDnf,qedNfstop,qcd} can in
principle be used to study overlapping formation times in a variety of
situations relevant to high-energy parton splitting in medium,
explicit calculations so far have studied the simplest
theoretical case, as we do here.
We study the
case of an infinite, static, homogeneous medium, which
in practice means that we assume that the medium properties are
constant over the length and time scales of the relevant formation time.
We also work in the $\hat q$ approximation (sometimes referred to as
the multiple scattering approximation).
We work in the large-$\Nc$ limit, where $\Nc$ is the number of quark colors.
We will also only consider quantities that have been integrated over the
transverse momenta of all the daughters of each splitting.
Throughout, we focus on gluon splitting, generated by $g \to gg$.
Overlap effects arise in double gluon splitting $g \to gg \to ggg$
and in corresponding virtual corrections to single gluon splitting.
As one example of a useful thought experiment for evaluating the
significance of overlap effects, ref.\ \cite{qedNfstop} proposed
computing their effect on the {\it shape} of the distribution
$\langle \eps(z) \rangle$ of where the energy of the shower is
deposited along the direction $z$ of the high-energy parton that
initiated the in-medium shower.%
\footnote{
  $\eps(z)$ refers to
  the energy deposited per unit length in the $z$ direction.
  Here ``deposited'' does not simply mean ``energy loss'' from a leading parton.
  For a quark-gluon plasma, it refers to where some subset of particles in
  the cascade have split in energy down to $E \sim T$, so
  that they thermalize with the plasma at that location $z$.
  In our notation,
  $\eps(z)$ represents the distribution of deposited energy for a single
  example of a random
  shower created by a single
  initial gluon of a given energy $E_0$ in our simplified
  case of an infinite, uniform medium.
  $\langle \eps(z) \rangle$ represents the statistical
  average over all such showers.
}
The ``shape'' of the distribution means the features you see if you
ignore the overall scale of the $z$ axis.
The motivation for studying the shape was to find characteristics of
shower development that are independent of the size of $\hat q$
(so that they do not depend on effects that can be absorbed into
$\hat q$) and which are infrared (IR) safe.
As a warm-up for a future QCD calculation, ref.\ \cite{qedNfstop}
explicitly computed aspects of the shape
for large-$\Nf$ QED in the $\hat q$ approximation.
However, ref.\ \cite{qedNfstop} also noted that there are potential
issues with this proposal for QCD --- technical issues that are
associated with the soft, sub-leading single logarithms that
are the subject of this paper.

Our motivation for finding analytic expressions for single logs are
several-fold.  Within the context of the $\hat q$ approximation,
the double log enhancements, and the sub-leading single logs,
will appear as double-log and single-log infrared (IR) divergences
in computations of energy loss and shower development.
(i) We hope that having analytic expressions will aid
in understanding and resolving potential issues in finding IR-safe
$\hat q$-independent measures of the importance of overlap corrections.
(ii) In order to go beyond leading (or
even sub-leading) log analysis of overlap effects in calculations
of shower development, one will need to
subtract out these IR divergences and handle them separately.  But it's
difficult and expensive to accurately extract the single logs
numerically \cite{qcd}, and so this subtraction will be much simpler
with analytic results in hand for the single logs.
(iii) The extracted double-log and single-log effects may then need
to be resummed, as was originally done for double logs in ref.\ \cite{Wu0}.
Having analytic expressions (as opposed to numerics)
for single logs at first order may
facilitate developing the resummation at next-to-leading-log order.
(iv) It's useful to understand whether single logs arise only from
the boundaries of integration regions that produce double logs, or whether
there is an additional, independent source of IR divergences.
(v) With analytic expressions, one may attempt to investigate whether there
is a simple, natural way to absorb single logs (and not just double
logs) into a redefinition \cite{Wu0,Blaizot,Iancu,Wu} of the medium
parameter $\hat q$.  In this paper, we will focus just on deriving
the analytic expressions for single logs.  We leave application to
IR-safe $\hat q$-independent characteristics of shower development (i),
implementation of
the subtractions (ii), and exploration of resummation (iii) to later work.
We will resolve the question (iv) of whether there are any
single IR logs that are independent from double logs.
The question (v) of whether the single logs can be naturally absorbed
into a redefinition of $\hat q$ is addressed in a companion paper \cite{logs2}.

% ----------------------------------------------------------------------------

\subsection {Infrared cut-offs}
\label{sec:IRcutoff}

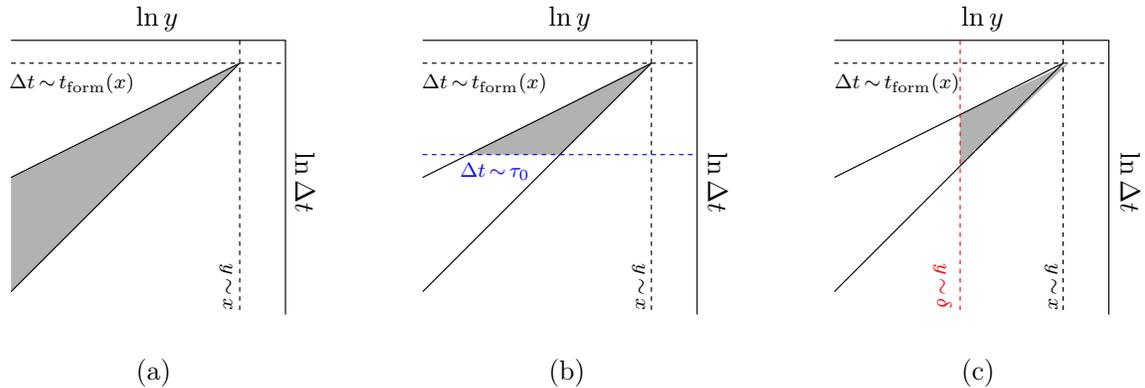
\begin {figure}[t]
\begin {center}
  \resizebox{0.99\linewidth}{!}{
  \begin{picture}(450,150)(0,0)
    \put(0,25){\includegraphics[scale=0.5]{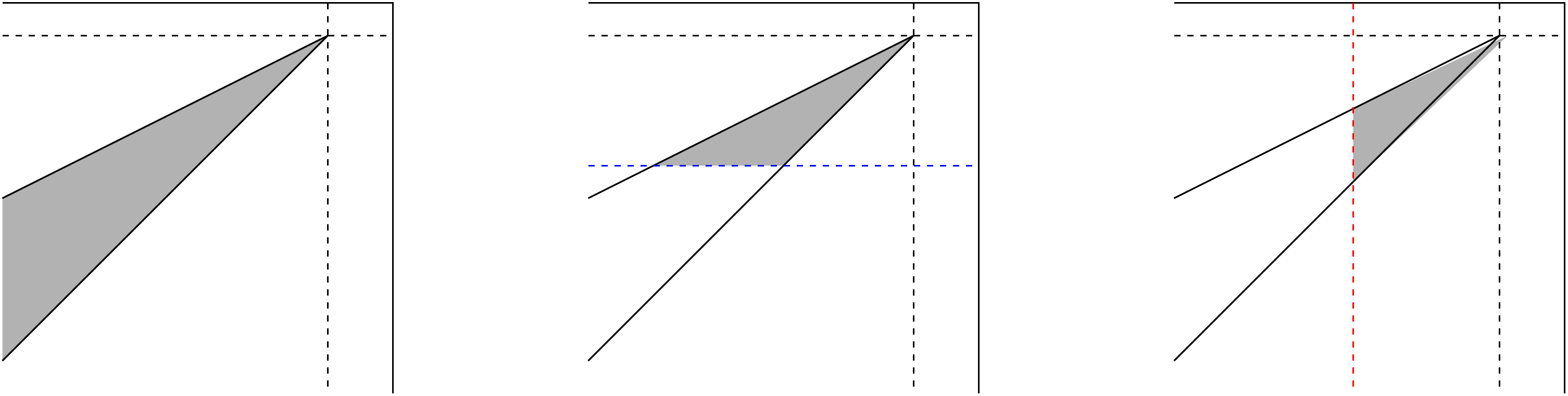}}
% fig a
    \put(0,115){$\scriptstyle{\Delta t\,\sim\,t_{\rm form}(x)}$}
    \put(83,46){\rotatebox{-90}{$\scriptstyle{y\,\sim\,x}$}}
    \put(50,140){$\ln y$}
    \put(113,90){\rotatebox{-90}{$\ln\Delta t$}}
    \put(50,0){(a)}
% fig b
    \put(163,115){$\scriptstyle{\Delta t\,\sim\,t_{\rm form}(x)}$}
    \put(246,46){\rotatebox{-90}{$\scriptstyle{y\,\sim\,x}$}}
    \put(213,140){$\ln y$}
    \put(276,90){\rotatebox{-90}{$\ln\Delta t$}}
    \put(213,0){(b)}
    \put(178,80){$\color{blue}\scriptstyle{\Delta t\,\sim\,\tau_0}$}
% fig c
    \put(326,115){$\scriptstyle{\Delta t\,\sim\,t_{\rm form}(x)}$}
    \put(409,46){\rotatebox{-90}{$\scriptstyle{y\,\sim\,x}$}}
    \put(376,140){$\ln y$}
    \put(439,90){\rotatebox{-90}{$\ln\Delta t$}}
    \put(376,0){(c)}
    \put(365,46){\rotatebox{-90}{$\color{red}\scriptstyle{y\,\sim\,\delta}$}}
%
%  \put(0,0){.}
%  \put(0,150){.}
%  \put(450,0){.}
%  \put(450,150){.}
  \end{picture}
  }
  \caption{
     \label {fig:region}
     The region of integration (\ref{eq:region})
     giving rise to a double log in the
     $\hat q$ approximation with (a) no cut-off, (b) the cut-off
     $\Delta t \sim \tau_0$ used in earlier literature, and (c)
     the IR regulator $y \sim \delta$ used in our calculations.
     See text for discussion.
     Above, $t_{\rm form}(x) \sim \sqrt{x E/\hat q}$ is the formation
     time associated with a single splitting $E \to x,(1{-}x)E$.
  }
\end {center}
\end {figure}

Our claim to have calculated IR logarithms has an important technical
caveat.  The caveat
was originally discussed in section 3.2 of ref.\ \cite{qcd},
but we review it here.  A schematic of the integration
region that gives rise
to the IR double log is shown in fig.\ \ref{fig:region}.
Here, $y$ is the energy fraction%
\footnote{
  Technically, calculations are carried out in Light Cone Perturbation
  Theory (LCPT) and $y$ is the longitudinal light-cone momentum
  fraction of the softest daughter compared to the particle that initiates
  the double splitting.  However, since the parents and daughters have
  high energy and are nearly collinear, we can for most purposes think of
  $y$ as the energy fraction.
}
of the softest daughter gluon
in a double splitting process $g \to ggg$
(with energies $E \to xE, yE, (1{-}x{-}y)E$) or else the
softest virtual gluon in a virtual correction to a single
splitting process $g \to gg$ (with energies $E \to xE, (1{-}x)E$).
$\Delta t$ is the time scale associated with the emission of
that softest gluon.  (Specifically, rates equal the product
of the amplitude and conjugate amplitude for a process, and
$\Delta t$ is the time separation
between emission of the $y$ gluon in the amplitude and emission of
the $y$ gluon in the conjugate amplitude.)
The shaded region in fig.\ \ref{fig:region}a shows the full region
that gives rise to the double-log IR divergence in a strict
$\hat q$ approximation (here specialized to the case of an infinite medium).
This region corresponds to
\begin {subequations}
\label {eq:region}
\begin {equation}
   \frac{yE}{\hat q L}
   \ll \Delta t
   \ll t_{\rm form}(y) \,,
\label {eq:tregion}
\end {equation}
where $t_{\rm form}(y) \sim \sqrt{yE/\hat q}$ is the formation time
associated with the soft emission and
\begin {equation}
  L \equiv t_{\rm form}(x)
\label {eq:L}
\end {equation}
is the formation length scale of the underlying single splitting process.%
\footnote{
  \label{foot:L}
  The original work \cite{Wu0} on double log corrections to $\hat q$
  directly studied transverse momentum diffusion of a high-energy
  particle traversing a medium of some length $L$ and studied
  the effect of a soft radiation on that transverse momentum.
  In contrast, the IR double logs in our application (like the IR
  double logs in refs.\ \cite{Blaizot,Iancu,Wu}) correspond to
  the effect of a soft radiation occurring on top of an underlying
  hard single splitting process $g \to gg$.  In that case, the
  relevant analog of the medium size $L$ is the relevant scale of
  the formation length for the underlying splitting process.
  For a medium that is thin enough, that scale also turns out to
  be $L$, but for the infinite-medium case that we treat in our
  work, the relevant formation time is parametrically of order
  $t_{\rm form}(x) \sim \sqrt{x E/\hat q}$,
  where $x E$ is the least-energetic daughter
  of the underlying single-splitting process $E \to xE, (1{-}x)E$.
}
%Physically,
%the first inequality corresponds to $k_\perp$ ordering of emissions;%
%\footnote{
%  That is, $k_{\perp,y} \ll k_{\perp,x}$.
%  A rough way to see the equivalence is to note that the high-energy
%  nearly-collinear emission of the $y$ gluon would, in vacuum, violate
%  energy conservation by $\Delta E \sim k_{\perp,y}^2/2\omega_y$, where
%  $\omega_y$ is the energy of the $y$ gluon.
%  By the uncertainty principle, the $y$ emission
%  must then complete in a time $\Delta t \sim 1/\Delta E$, and so
%  $k_{\perp,y}^2 \sim \omega_y/\Delta t$.
%  A similar argument for the underlying $x$ emission gives typical
%  $k_{\perp,x}^2 \sim \omega_x/L$, where $L$ is the $x$ formation time.
%  $k_\perp$ ordering is then equivalent to $\omega_y\,\Delta t \ll \omega_x L$.
%  In our notation, where $\omega_y \simeq yE$ and $\omega_x \simeq xE$,
%  that's $yL/x \ll \Delta t$.
%  One may then use $L = t_{\rm form}(x) \sim \sqrt{\hat q/xE}$ to
%  get the form (\ref{eq:tregion}).
%}
The second inequality just says that the time of the
$y$ emission must fit within the $y$ formation time $t_{\rm form}(y)$.
These inequalities (\ref{eq:tregion}) can
be equivalently expressed as a range on $y$:
\begin {equation}
   \frac{\hat q (\Delta t)^2}{E}
   \ll y
   \ll \frac{\hat q L\,\Delta t}{E}
   \,.
\label {eq:yregion}
\end {equation}
\end {subequations}

\begin {figure}[t]
\begin {center}
  \begin{picture}(260,225)(-80,-25)
    \put(60,75){\includegraphics[scale=0.5]{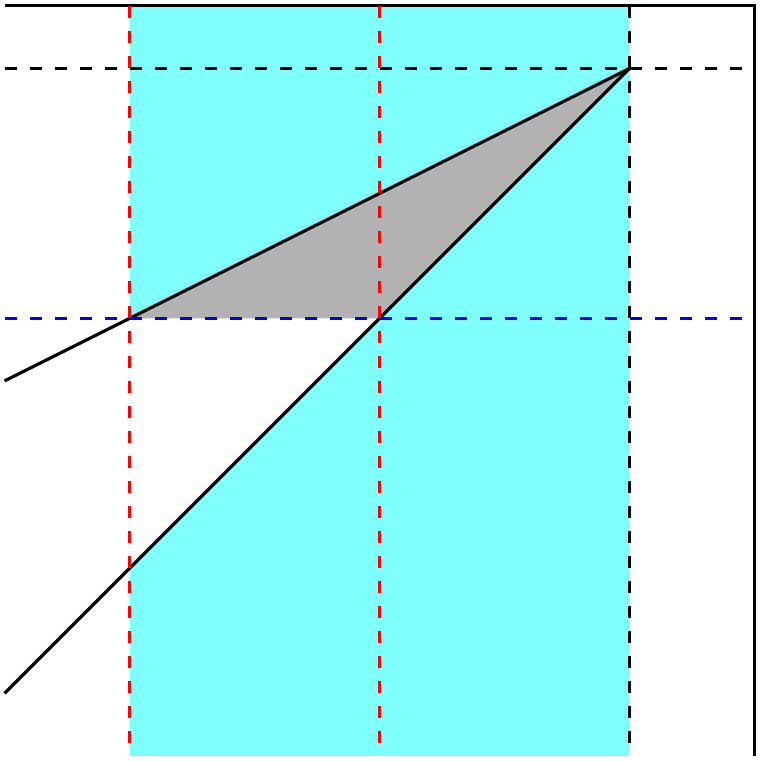}}
    \put(-77,173){$\scriptstyle{
                   \Delta t\,\sim\,t_{\rm form}(x)
                   \sim \sqrt{x E/\hat q}
                   \sim \sqrt{x E/g^4 T^3}
                 }$}
    \put(0,137){$\color{blue}{\scriptstyle{
       \Delta t\,\sim\,\tau_0 \sim 1/g^2T
     }}$}
    \put(75,75){\rotatebox{-90}{$\color{red}{
       \scriptstyle{yE\,\sim\,\hat q\tau_0^2\,\sim\,T}
     }$}}
    \put(111,75){\rotatebox{-90}{$\color{red}{
       \scriptstyle{
          yE\,\sim\,\hat q \tau_0^{} \, t_{\rm form}(x)\,\sim\,\sqrt{x E T}
       }
     }$}}
    \put(148,75){\rotatebox{-90}{$\scriptstyle{y\,\sim\,x}$}}
    \put(110,190){$\ln y$}
    \put(173,140){\rotatebox{-90}{$\ln\Delta t$}}
  %\put(-80,-30){.}
  %\put(-80,195){.}
  %\put(180,-25){.}
  %\put(180,195){.}
  \end{picture}
  \caption{
     \label {fig:region2}
     Parametric scales associated with various features of
     fig.\ \ref{fig:region}b. The light blue region,
     combined with the gray region, shows where
     the $\hat q$ approximation is useful in our (theoretically idealized)
     application.  In the lower light-blue region, $\hat q$-approximation
     propagators over the short time of the $y$ emission are
     approximately vacuum propagators.  We have only shown $y < x$ above
     because, in our discussion here, $y$ represents the softest gluon.
%%[[Say something about $yE \lesssim T$?]]
  }
\end {center}
\end {figure}
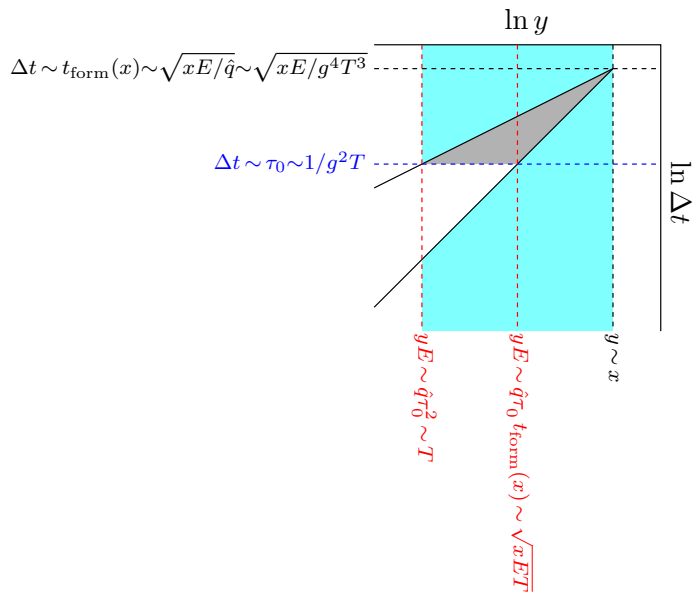

The $\hat q$ approximation is an approximation that assumes multiple
scattering from the medium.
A calculation in the original work of Liou, Mueller, and Wu
\cite{Wu0} on double-log contributions to
$\hat q$ may be interpreted as the observation that the
$\hat q$ approximation therefore breaks down for time scales $\lesssim$
the characteristic mean-free-time $\tau_0$ for the high-energy
partons to elastically scatter from the medium.  The actual
region that contributes to the double logarithms would then be
cut off as shown in fig.\ \ref{fig:region}b.%
\footnote{
  Our characterization of the double log region in
  fig.\ \ref{fig:region}b can be translated to the corresponding
  region $A{=}A_1{+}A_2$ of fig.\ 2 of ref.\ \cite{Wu0} as follows:
  their medium size $L$ plays the role of our formation time
  $L \equiv t_{\rm form}(x)$
  (see our footnote \ref{foot:L}), and
  their variables $\omega,t,l_0$ are our $yE,\Delta t,\tau_0$.
  The reason that one of the region boundaries in
  their fig.\ 2 is curved and ours are all straight is because
  our axes are logarithmic, as indicated in our figure.
}
(We more explicitly identify some of the scales in this figure,
for the case of a weakly-coupled thermalized quark-gluon plasma,
in fig.\ \ref{fig:region2}.  For a strongly-coupled quark-gluon
plasma, simply erase the explicit factors of $g$.)

For technical reasons, our previous work \cite{qcd} has implemented
a different type of IR regularization: an IR cut-off
$(p^+)^{\rm min} = (P^+)^{\rm initial} \delta$ on longitudinal light-cone momenta
(equivalent, given our high-energy approximations, to an IR
cut-off $E\delta$ on energy), where $\delta$ is a small number.
This cut-off is depicted in fig.\ \ref{fig:region}c.
One motivation for this cut-off is that it seemed technically the easiest
to implement.  Unlike, for example, a sharp cut-off on time, there is
also less
question about
consistency of the regulator when adding together different diagrams.
Another motivation is that our regulator allowed our calculations to correctly
reproduce the renormalization and running of the QCD coupling constant,
arising from UV divergences as $\Delta t \to 0$.  This is
possible because the $\hat q$ approximation is
valid somewhat more generally than just $\Delta t \gg \tau_0$ because
propagators in the $\hat q$ approximation correctly reproduce simple vacuum
propagators as $\Delta t \to 0$.  Because of this, the actual range of
usefulness of the $\hat q$ approximation in our application is depicted by
the combination of the gray and light blue regions in fig.\ \ref{fig:region2}.%
\footnote{
  This statement about the validity of the $\hat q$ approximation is
  a theoretical idealization, applicable to the theoretical limit of
  extremely high energies in the theorists' limit of a medium
  wide enough to completely contain the corresponding formation time.
}

Our strategy for now is to use the simple IR cut-off of
fig.\ \ref{fig:region}c, with the intention
of later finding IR-safe applications to energy loss.
To do better, and handle the full physics of fig.\ \ref{fig:region2},
one would need to go beyond the $\hat q$ approximation to handle the
transition between the gray-shaded region of fig.\ \ref{fig:region2}
and the unshaded region
beneath it.  As mentioned earlier, the underlying framework of our
formalism can in principle be applied without making the $\hat q$
approximation, but that would require a leap in calculational complexity.

\begin {figure}[t]
\begin {center}
  \includegraphics[scale=0.8]{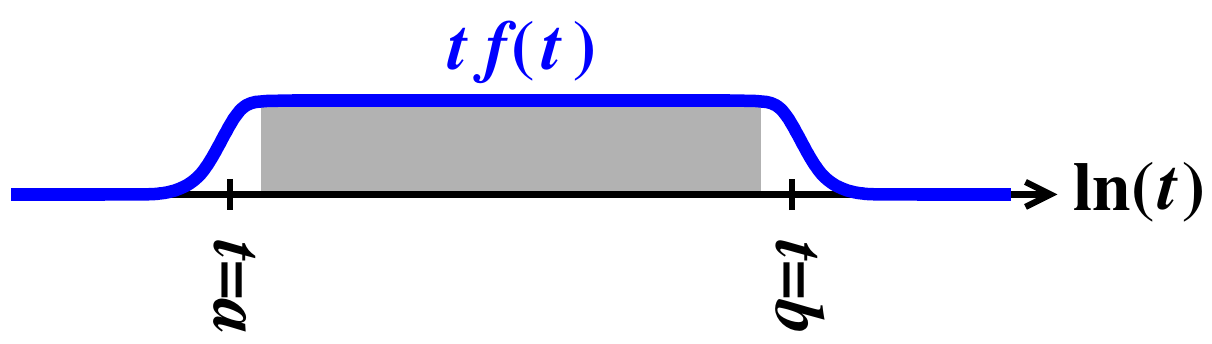}
  \caption{
     \label {fig:ftoy}
     A simple calculus example of an integrand $f(t)$
     that gives rise to a leading
     logarithm $\ln(b/a)$ from the shaded region, plus a sub-leading
     additive constant $c$ determined by the details of the behavior
     at the boundaries of that region.
     Note that this qualitative plot is of $t\,f(t)$ vs.\ $\ln t$,
     and so the logarithmic region corresponds to a flat plateau.
  }
\end {center}
\end {figure}

\begin {figure}[t]
\begin {center}
  \includegraphics[scale=0.5]{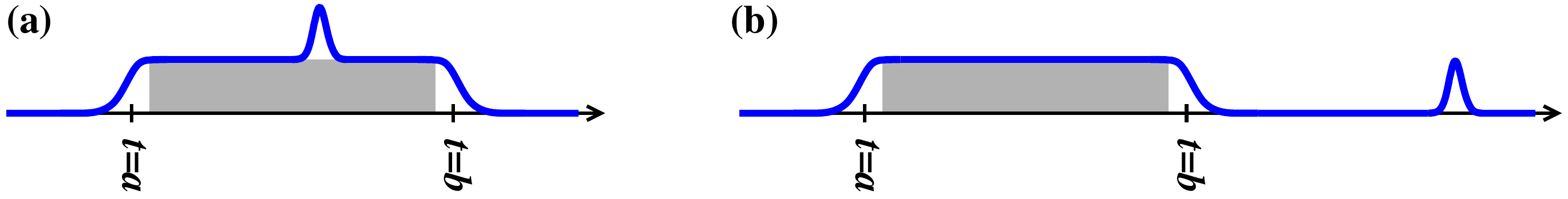}
  \caption{
     \label {fig:ftoy2}
     Like fig.\ \ref{fig:ftoy} but showing cases where there are
     additional contributions to the sub-leading additive constant
     $c$ that do not correspond to the boundaries of the logarithmic
     contribution.
  }
\end {center}
\end {figure}

In our case we have a two-dimensional integral (over $y$ and
$\Delta t$) that gives a double
log, and we are interested in determining a sub-leading single log.
To set the stage, let's instead
first review what would happen in
a simple calculus example of a {\it one}-dimensional integral
that gives a {\it single} log, and imagine that we were interested
in determining a sub-leading additive constant.
Consider a function $f(t)$ that is approximately proportional to
$1/t$ over a range $(a,b)$ that spans several orders of magnitude,
and suppose that $f$ falls off quickly outside of this region.
This is depicted in fig.\ \ref{fig:ftoy} by a qualitative plot of
$t \times f(t)$ vs.\ $\ln t$.
At leading log order,
$\int_0^\infty dt \> f(t) \approx \ln(b/a)$, coming from the
shaded region of fig.\ \ref{fig:ftoy}.
Beyond that approximation, we have
$\int_0^\infty dt \> f(t) \simeq \ln(b/a) + c$, where the additive
constant $c$ is determined by exactly
how $f(t)$ falls off at $t \sim a$ and at
$t \sim b$ --- that is, by how it falls off at the boundary of the
logarithmic region $(a,b)$.
Note that, though the boundaries of the
region that give rise to the logarithm will contribute to $c$,
there could be {\it other} contributions to $c$ that have nothing
to do with the logarithm.  For instance, a function
$f(t)$ might look like
figs.\ \ref{fig:ftoy2}a or b, where the extra hump would give an additional
contribution to $c$.

A similar situation applies to the case of leading double logarithms,
which arise from integrals proportional to
\begin {equation}
   \iint \frac{dy\>d(\Delta t)}{y \, \Delta t}
\label {eq:Lint0}
\end {equation}
over the shaded regions of fig.\ \ref{fig:region}.  Sub-leading single
logarithms then arise from the details of how the integrand falls off
at the {\it boundaries} of the shaded integration region.  (And
one must check whether there are any additional contributions to
sub-leading logs that are not associated with that boundary.)

We can now see how the choice of IR regulator impacts the interpretation
of results for single logs by comparing our IR regulator of
fig.\ \ref{fig:region}c with the situation of fig.\ \ref{fig:region}b.
For the boundaries that are in common between the two figures, the
contributions to single logarithms will be the same; for boundaries
that differ, the contributions to single logarithms will also differ.
So, though parts of the calculation are common to any analysis, the
full result for our single logarithm will depend on our particular
choice (fig.\ \ref{fig:region}c) of IR regulator.
(If an IR-regulated expression is not good enough for a particular
application,
and one needs a truly complete result for single logarithms,
then one would have to go beyond the $\hat q$ approximation in order
to compute the single log contribution from the $\Delta t\sim \tau_0$
boundary in fig.\ \ref{fig:region}b.)

% ----------------------------------------------------------------------------

\subsection {Reproducing the known double log result}

In ref.\ \cite{qcd}, we found numerically that, with our IR regularization,
the double logarithm arising from a single splitting overlapping
a second, softer splitting could be absorbed into the usual
BDMPS-Z single splitting rate by replacing
\begin {equation}
  \qhatA
  \longrightarrow
  \qhatA^{\,\rm eff}(\delta) = 
  \left[
     1 + \frac{\CA\alphas}{4\pi} \ln^2\delta 
  \right]
  \qhatA .
\label {eq:qhateffDBL}
\end {equation}
Ref.\ \cite{qcd} discusses how, after accounting for our different
choice of IR regularization, this is equivalent to
earlier (analytic) results from the literature
\cite{Wu0,Blaizot,Iancu,Wu} that
\begin {equation}
  \hat q^{\,\rm eff}(L) = 
  \left[
     1 + \frac{\CA\alphas}{2\pi} \ln^2\!\left( \frac{L}{\tau_0} \right) 
  \right]
  \hat q .
\label {eq:qhateffORIGINAL}
\end {equation}
We'll present here a somewhat simpler way to summarize the equivalence.
The leading-log results
(\ref{eq:qhateffDBL}) and (\ref{eq:qhateffORIGINAL}) have the
same form:
\begin {equation}
  \hat q^{\,\rm eff} = 
  \left[
     1 + \frac{\CA\alphas}{\pi} {\cal L}
  \right]
  \hat q ,
\label {eq:qhateffL}
\end {equation}
where ${\cal L}$ is the {\it area} of the shaded double-log region
${\cal R}$ plotted on log-log plots such as fig.\ \ref{fig:region}.
Specifically, ${\cal L}$ is given by the integral (\ref{eq:Lint0}),
which can be rewritten as%
\footnote{
  To be more specific, the area of the triangular shaded region in
  fig.\ \ref{fig:region}c can be found by calling the left-hand boundary
  $y{\sim}\delta$ the ``base'' of the triangle.  Using (\ref{eq:tregion}),
  this base has (logarithmic) length
  $\ln t_{\rm form}(\delta) - \ln(E\delta/\hat qL) \simeq
   -\tfrac12\ln\delta + \log\bigl(t_{\rm form}(x)/(E/\hat q)^{1/2}\bigr)$,
  which, at leading-log level for small $\delta$ but fixed $x$ is just
  $-\tfrac12\ln\delta$.  The corresponding ``height'' of the triangle
  perpendicular to the base is $\ln x - \ln\delta \approx -\ln\delta$,
  and so ${\cal L} \approx \tfrac14\ln^2\delta$ at leading log, so that
  (\ref{eq:qhateffL}) gives (\ref{eq:qhateffDBL}).
  Similarly, for the area of the triangular shaded region in
  fig.\ \ref{fig:region}b, take the base to be the horizontal boundary
  $\Delta t \sim \tau_0$, with length given by (\ref{eq:yregion}) as
  $\ln(\hat q L \tau_0/E) - \ln(\hat q \tau_0^2/E) = \ln(L/\tau_0)$.
  The height is $\ln L - \ln\tau_0 = \ln(L/\tau_0)$, and so
  ${\cal L} \approx \tfrac12\ln^2(L/\tau_0)$, so that (\ref{eq:qhateffL}) gives
  (\ref{eq:qhateffORIGINAL}).
} 
\begin {equation}
   {\cal L} 
   \equiv \iint_{\cal R} \frac{dy\>d(\Delta t)}{y \, \Delta t}
   = \iint_{\cal R} d(\ln y) \> d\bigl(\ln(\Delta t)\bigr)
   = \mbox{area on log-log plot} .
\label {eq:Lint}
\end {equation}
With this rewriting, the coefficient $\CA\alphas/\pi$ of the double
log ${\cal L}$ in (\ref{eq:qhateffL}) is universal, and the difference
in the double log due to our choice of IR regulator can be understood as
packaged into the area (\ref{eq:Lint}).

% ----------------------------------------------------------------------------

\subsection {Results for single log}

We find that the only IR single logarithms are those that are associated
with the boundaries of the double-log region.%
\footnote{
  Though there are no other IR single logarithms (where IR means
  associated with soft emission), there is an additional
  non-IR logarithm ---
  namely the UV logarithm associated with renormalization of
  the coupling constant \cite{qcd}.
}
On a related note, the IR single logarithms arise (on net) only from
the time-ordered diagrams that contributed to double logarithms
in the leading-log work of refs.\ \cite{Blaizot,Iancu,Wu}, which are
but a subset of the full set of diagrams \cite{qcd} needed for
a more general calculation that avoids large-logarithm approximations.
(This conclusion required finding and correcting a phase error in our
earlier work \cite{qcd}, which is explained in appendix \ref{app:error}.)

The simplest (though slightly misleading)
way to express our final result for IR single logs is to simply
generalize (\ref{eq:qhateffDBL}) to include single logarithms.
Our result is
\begin {equation}
  \qhatA \longrightarrow
  \qhatA^{\,\rm eff}(\delta,x) = 
  \left[
     1
     + \tfrac{\CA\alphas}{2\pi}
         \bigl( \tfrac12 \ln^2\delta + \bar s(x) \, \ln\delta \bigr)
  \right]
  \qhatA
\label {eq:qhateffSNGL}
\end {equation}
for our application, where%
\footnote{
  The notation $\bar s(x)$ for the single-log coefficient is inherited
  from ref.\ \cite{qcd}.  You may think of $s$ as standing for ``single'' log.
  For our present purpose, the bar is merely a vestige of
  earlier notational choices.
} 
\begin {equation}
  \bar s(x) =
      - \ln\bigl(16\,x(1{-}x)(1{-}x{+}x^2)\bigr)
      + 2 \, \frac{ \bigl[
                 x^2 \bigl( \ln x - \frac{\pi}{8} \bigr)
                 + (1{-}x)^2 \bigl( \ln(1{-}x) - \frac{\pi}{8} \bigr)
               \bigr] }
             { (1-x+x^2) }
   \,.
\label {eq:sbar}
\end {equation}
$x$ and $1{-}x$ are the energy fractions of the two daughters in the
underlying single splitting process.
The appearance of the factors
$1{-}x{+}x^2$ above is related to the fact that the formation time
for the underlying single-splitting process is%
\footnote{
  More precisely, (\ref{eq:tform0}) is
  $|\Omega_0|^{-1}$, where $\Omega_0$ is
  the complex harmonic oscillator frequency (\ref{eq:Om0}) associated with
  making the $\hat q$ approximation to the LPM effect
  in single $g{\to}gg$ splitting.
  The phrase ``formation time'' by itself is often used to refer
  parametrically to this time scale and does not, as far as we know,
  have a universally established convention for what factors of 2
  to include in (\ref{eq:tform0}).
}
\begin {equation}
   \left[
      \frac{\qhatA}{2 E}
      \left( -1 + \frac{1}{x} + \frac{1}{1{-}x} \right)
   \right]^{-1/2}
   = \sqrt{ \frac{2x(1{-}x)E}{(1{-}x{+}x^2) \qhatA} }
   \,.
\label {eq:tform0}
\end {equation}

We have extracted the single log coefficient (\ref{eq:sbar}) from
the $y{\to}0$ limit of the generic-$y$ formulas of ref.\ \cite{qcd}.
Those generic-$y$ formulas were derived in the large-$\Nc$ limit.
Since the result \cite{Wu0,Blaizot,Iancu,Wu} for double logs does not
depend on the large-$\Nc$ limit,
one might reasonably hope that the same is true for single logs.
%One check of this expectation should come soon from calculations
%of the $1/\Nc^2$ corrections for overlapping double gluon bremsstrahlung
%\cite{1overN}.

We gave a caveat above that expressing the final answer in the form
(\ref{eq:qhateffSNGL}) is slightly misleading.  That's because our
single log has complicated $x$ dependence, whose form depends specifically on
our application of $\hat q$ to overlap effects on jet quenching in the
infinite-medium limit.  Because of this, it does not clearly have as
``universal'' a form as the double log result (\ref{eq:qhateffL}).
The question of universality will be addressed in ref.\ \cite{logs2}.

\begin {figure}[t]
\begin {center}
  \includegraphics[scale=0.55]{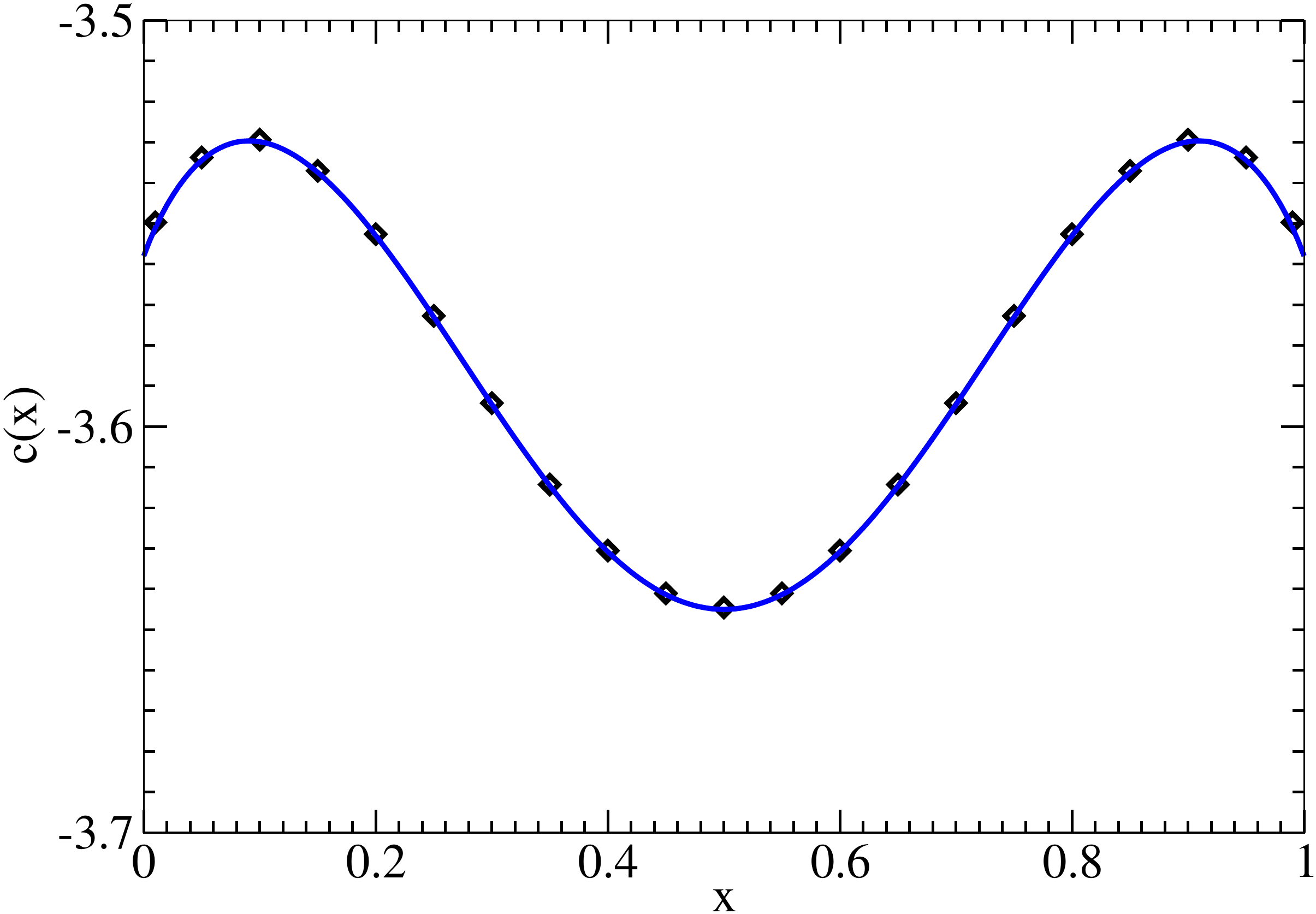}
  \caption{
    \label {fig:cnew}
    A test of numerical vs.\ analytic extraction of the single log
    coefficient $\bar s(x)$.  The plot is of $c(x) \equiv \bar s(x) -
    \ln\bigl( -1 + \frac{1}{x} + \frac{1}{1-x} \bigr)$ vs. $x$, where
    the subtraction of the logarithm (based on the ``educated guess''
    for $\bar s(x)$ in ref.\ \cite{qcd}) makes the fine details more
    visible by removing the logarithmic divergence of $\bar s(x)$ as
    $x \to 0$ or $1$.  The numerical extraction of $c(x)$ is shown by
    the diamonds and is the same
    as fig.\ 20 of ref.\ \cite{qcd} except corrected by a
    $4\pi$ downward shift, as explained in appendices \ref{app:error}
    and \ref{app:cshift} of this paper.
    (We have also very slightly improved the accuracy.)
    The solid curve is our
    analytic result, based on (\ref{eq:sbar}).  }
\end {center}
\end {figure}

Fig.\ \ref{fig:cnew} shows a test of our (corrected) numerical
extraction of the single log coefficient in ref.\ \cite{qcd} vs.\ our
new analytic result (\ref{eq:sbar}) above.  They are in full agreement.
%%[[Work on the accuracy of the numerical extraction
%%in order to make the agreement look dead on?]]

% --------------------------------------------------------------------------

\subsection {Outline}

In the next section, we review the diagrams that produce double
logarithmic IR behavior, which we call the ABC diagrams.  These are the
same diagrams which (on net) produce the IR single logarithms.
We also summarize the result for each ABC diagram's individual contribution
to the single log coefficient $\bar s(x)$.
In section \ref{sec:nonABC},
we discuss how {\it other} (non-ABC) diagrams, which
contribute to overlapping gluon splitting in more general situations,
do not contribute to the IR logarithms that arise when one of the
splittings is soft.

There are two qualitatively different types of ABC diagrams, named A3
and A1, to which other ABC diagrams can be related (with some caveats).
In section \ref{sec:A3}, we give an overview of the extraction of the
small-$y$ behavior of the A3 diagram
from the complicated, general-$y$ formulas of
refs.\ \cite{2brem,qcd}.  In section \ref{sec:ij}, we discuss the
almost-symmetry that allows us to relate the result for that diagram
to a sub-class of other ABC diagrams.  In sections \ref{sec:A1}\
and \ref{sec:ii}, we repeat that procedure for the A1 diagram
and relate it to the remaining ABC diagrams.
A large variety of details are left for appendices.

Our conclusion, such as it is, is given in section \ref{sec:conclusion}.

% ===========================================================================

\section{Diagrams}

\subsection{Relevant diagrams and total rate}

\begin {figure}[t]
\begin {center}
  \includegraphics[scale=0.6]{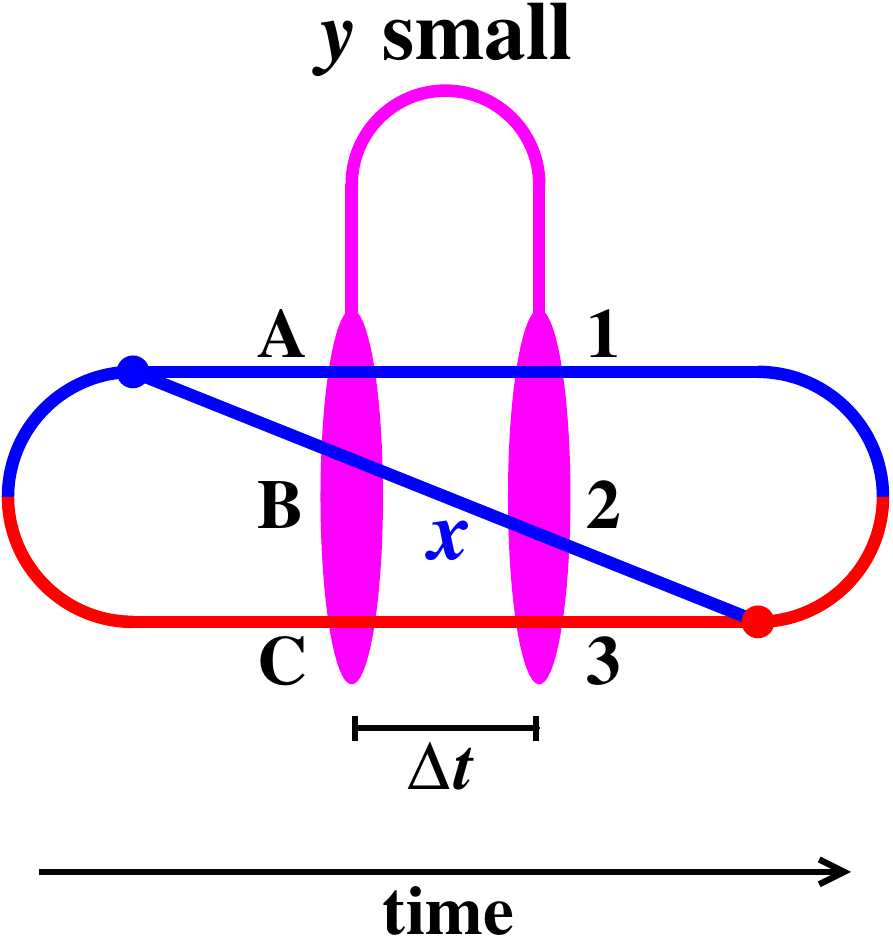}
  \caption{
     \label{fig:ABCdiags}
     The nine time-ordered
     rate diagrams which (together with their complex conjugates)
     produce the double logarithm.  In our analysis, all of the lines are
     gluons.
     The labeling A, B, C, 1, 2, and 3 is our naming convention for where
     the relatively-soft $y$ gluon may connect the three harder gluon lines.
     The magenta color of the $y$ gluon is used to indicate that the
     line could be either blue or red depending on how it is
     connected (see fig.\ \ref{fig:ABCexamples}).
  }
\end {center}
\end {figure}

\begin {figure}[t]
\begin {center}
  \includegraphics[scale=0.5]{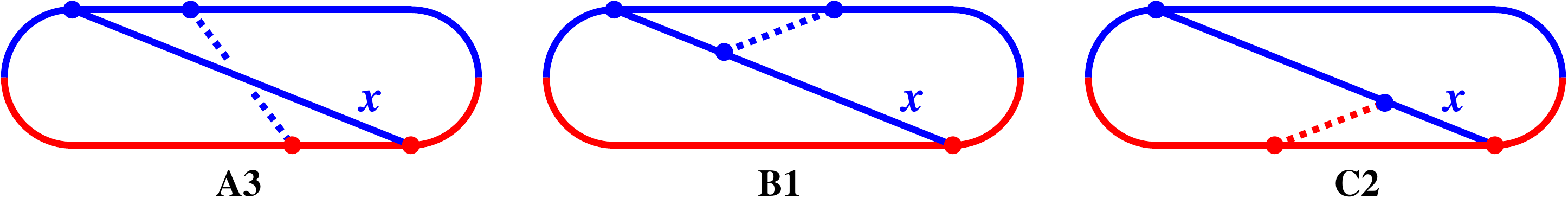}
  \caption{
     \label{fig:ABCexamples}
     Three examples of fig.\ \ref{fig:ABCdiags}.  The soft $y$ gluon
     is denoted by the dashed line.
     A3 and C2 contribute to the rate for double splitting $g\to ggg$,
     whereas B1 is an example of a virtual (i.e.\ loop) correction to single
     splitting $g\to gg$.
     In the nomenclature of
     ref.\ \cite{qcd}, A3 is called $xy\bar y\bar x$,
     B1 is $yxy\bar x$ with renaming
     $y \leftrightarrow z \equiv 1{-}x{-}y$ of the
     loop momentum fraction variable, and C2 is $z\bar y y\bar z$
     (the $x \leftrightarrow z$ permutation of $x\bar y y\bar x$). 
  }
\end {center}
\end {figure}

The nine time-ordered diagrams which (on net) produce the double
logarithm \cite{Blaizot, Iancu, Wu} are collectively depicted by
fig.\ \ref{fig:ABCdiags}.
Three particular examples are shown in
fig.\ \ref{fig:ABCexamples} and are drawn in the style of
refs.\ \cite{2brem,qcd}.  The blue part represents a contribution to
the amplitude for the process (either double splitting $g{\to}ggg$
or a virtual correction to single splitting $g{\to}gg$); the red part
represents a contribution to the conjugate amplitude; and the entire
diagram represents a contribution to the rate.
Only high-energy parton lines are shown: each line
implicitly interacts multiple times with the medium, and there is
implicit averaging of the rate over the randomness of the medium.
Following ref.\ \cite{Blaizot}, we name the diagrams A1, A2, ..., C3.
We'll refer to them collectively as the ``ABC diagrams.''
The double log arises from the case where the $y$ gluon is soft
compared to the other daughters of the double splitting.  But the $y$
gluon should be understood to still be high energy, with energy
large compared to the energy scales of the QCD medium (e.g.\
energy large compared to the temperature $T$ of a quark-gluon plasma).
The ABC diagrams are the diagrams where
emission of the soft
$y$ gluon is completely contained within the time interval spanned
by the underlying single-splitting process that creates the $x$ gluon.

Generically, differential rates from individual ABC diagrams
have small-$y$ expansions of the form \cite{2brem,qcd}
\begin {equation}
   \frac{d\Gamma}{dx\,dy} =
   \frac{\# \ln y + \#}{y^{3/2}}
   +
   \frac{\# \ln y + \#}{y}
   +
   O(y^{-1/2}) ,
\label {eq:smallybehavior}
\end {equation}
where ``$\#$'' represents various $x$-dependent coefficients specific to
the diagram.
These rates have already been integrated over the ``$\Delta t$''
of fig.\ \ref{fig:region}a.
The $O(y^{-3/2})$ terms above generate IR power-law divergences
when integrated over $y$.
The $y^{-1}\ln y$ and $y^{-1}$ terms respectively
generate double and single logs, and the $O(y^{-1/2})$ terms are
IR finite.  When one adds all of the ABC diagrams together,
the power-law divergences cancel \cite{qcd},
leaving just the terms that generate
logs.

Specifically,
in this paper, we will find that the rate corresponding to the sum
of all ABC diagrams has the small-$y$ expansion
\begin {equation}
  \left[ \frac{d\Gamma}{dx\,dy} \right]_{\rm ABC}
  \simeq
  - \frac{\CA\alphas}{4\pi y}
  \left[ \frac{d\Gamma}{dx} \right]_{\rm LO}
    \bigl(
      \ln y
      + \bar s(x)
    \bigr)
    ,
\label {eq:total2}
\end {equation}
where $\bar s(x)$ is our result (\ref{eq:sbar}) for the single-log
coefficient, and where%
\footnote{
  (\ref{eq:LO}) is written here in the form used in
  appendix A.1 of ref.\ \cite{qcd}.
}
\begin {equation}
  \left[ \frac{d\Gamma}{dx} \right]_{\rm LO}
  =
  \frac{\alphas}{\pi} \, P(x) \,
  \Re(i\Omega_0)
\label {eq:LO}
\end {equation}
is the leading-order BDMPS-Z result for single splitting $g{\to}gg$
in the $\hat q$ approximation.  Above,
\begin {equation}
  \Omega_0 \equiv
  \sqrt{
    \frac{-i \qhatA}{2E}
    \Bigl( -1 + \frac{1}{x} + \frac{1}{1{-}x} \Bigr)
  }
  =
  \sqrt{
    \frac{-i \qhatA (1{-}x{+}x^2)}{2x(1{-}x)E}
  }
\label {eq:Om0}
\end {equation}
gives the complex harmonic oscillator frequency
associated with the single splitting process, and
$P(x)$ is the DGLAP $g{\to}gg$ splitting function%
\footnote{
  For technical clarification concerning (\ref{eq:Pgg}), see
  comments after eq.\ (A.5) in ref.\ \cite{qcd}.
  In particular, $P(x)$ should be defined as the absolute value
  of the right-hand side in cases where one intends
  to ``front-end'' transform the diagram in a way that $x$ or $1{-}x$
  might become negative.
}
\begin {equation}
  P(x)
  = \CA \frac{1 + x^4 + (1-x)^4}{x(1-x)}
  = \CA \frac{2(1{-}x{+}x^2)^2}{x(1-x)}
  \,.
\label {eq:Pgg}
\end {equation}

To first order in the $\alphas$ associated with high-energy splitting,
the IR log corrections (\ref{eq:total2}) can be absorbed into the
leading-order single splitting rate (\ref{eq:LO}) by the redefinition
\begin {equation}
  \qhatA \longrightarrow
  \left[
     1
     - \frac{\CA\alphas}{2\pi} \int_\delta dy \>
         \frac{ \ln y + \bar s(x) }{y}
  \right]
  \qhatA ,
\end {equation}
since $\Omega_0 \propto \sqrt{\qhatA}$.
As discussed earlier, we have chosen to regulate the IR with a sharp
lower cut-off $\delta$ on $y$.  The $\delta$ dependence of the above
integral gives our previously quoted result (\ref{eq:qhateffSNGL})
for $\qhatA^{\,\rm eff}$.

% ---------------------------------------------------------------------------

\subsection{Results for individual ABC diagrams}
\label {sec:ABCsummary}

It will be useful to break down our results diagram by diagram,
which is how we attacked finding the small-$y$ expansion.

The results have an appealing symmetry that allows one to
cover all nine diagrams with just two equations.
We refer to each specific ABC diagram with a pair of integers
$\di,\dj$ where $\di=1,2,3$ represents A,B,C and $\dj=1,2,3$ represents
$1,2,3$.
So, for example, the B3 diagram corresponds to $\di,\dj=2,3$.
We consider here the labels
A1, A2, ..., C3
to be short-hand for the corresponding diagram
of fig.\ \ref{fig:ABCdiags} {\it plus} its complex conjugate.
We also use the notation%
\footnote{
  For readers wondering at our unusual choice of font for
  $(\altx_1,\altx_2,\altx_3)$: We've used this to avoid confusion
  with the 4-particle values of $(x_1,x_2,x_3,x_4)$ used later
  (and also because, even for purely 3-particle situations, the
  particular ordering
  of $(\altx_1,\altx_2,\altx_3)$ is different from the convention
  $(x_1,x_2,x_3)$ used in past work, such as section 4.2 of ref.\ \cite{2brem}).
}
\begin {equation}
   (\altx_1,\altx_2,\altx_3) \equiv (1{-}x,x,-1)
\label {eq:altx}
\end {equation}
for the longitudinal momentum fractions, in the $y{\to}0$ limit,
of the three hard lines corresponding to A,B,C or 1,2,3
in fig.\ \ref{fig:ABCdiags}.

We will need the complex harmonic oscillator frequencies $\Omega_0$
and $\Omega_y$ associated,
in the $y{\to}0$ limit, with (i) the underlying $x$ emission process and
(ii) the soft $y$ emission process.
$\Omega_0$ was given already in (\ref{eq:Om0}), but here it will
be useful to note that it can be written symmetrically in terms
of the momentum fractions (\ref{eq:altx}) as
\begin {equation}
  \Omega_0 \equiv
  \sqrt{
    \frac{-i \qhatA}{2E}
      \Bigl( \frac{1}{\altx_1} + \frac{1}{\altx_2} + \frac{1}{\altx_3} \Bigr)
  }
  =
  \sqrt{
    \frac{-i \qhatA (\altx_1^2+\altx_2^2+\altx_3^2)}{4|\altx_1\altx_2\altx_3| E}
  }
  .
\label {eq:altOm0}
\end {equation}
The other frequency is
\begin {equation}
  \Omega_y \equiv \sqrt{ \frac{-i \qhatA}{2y E} } .
\label {eq:Omy}
\end {equation}
It can be useful to note that
\begin {equation}
  \altx_1^2+\altx_2^2+\altx_3^2 = 2(1-x+x^2) .
\end {equation}

With this notation, our results for
the differential rates associated with the
ABC diagrams split into two cases.
For $\di{\not=}\dj$ diagrams,
\begin {subequations}
\label {eq:eachABC}
\begin {multline}
  \left[ \frac{d\Gamma}{dx\,dy} \right]_{\di\dj}
  \simeq
  \frac{\CA\alphas^2\,P(x)}{2\pi^2 y}
  \Re
  \biggl[
    \red{i\Omega_y}
    \Bigl\{
       - \ln\bigl( \tfrac{\Omega_y}{2\pi\Omega_0} \bigr) - \gammaE
       - \tfrac{i\pi}{2} \delta_{\di3}
    \Bigr\}
\\
    + \blue{i\Omega_0}
    \biggl\{
      2 \Bigl[
        \tfrac{1}{\eps} + \ln\bigl(\tfrac{\pi\mu^2}{\Omega_0 E}\bigr)
      \Bigr]
      - \bigl(1{+}\tfrac{\altx_\dk^2}{2(\altx_1^2+\altx_2^2+\altx_3^2)}\bigr)
        \left[
          \ln\bigl(
             \tfrac{|\altx_1 \altx_2 \altx_3| y\Omega_y}{\Omega_0}
          \bigr)
          + \tfrac{i\pi}{2} \delta_{\di3}
        \right]
\\
      + \tfrac{\altx_\dk^2}{2(\altx_1^2+\altx_2^2+\altx_3^2)}
          \bigl[ 1-\ln2+2\ln(2|\altx_\di \altx_\dj|) \bigr]
    \biggr\}
  \biggr] ,
\label {eq:iNEj}
\end {multline}
where $\altx_\dk$ above represents the longitudinal momentum fraction in
(\ref{eq:altx}) that is not $\altx_\di$ or $\altx_\dj$,%
\footnote{
  If one prefers $\altx_\dk$ defined by an equation:
  $|\altx_\dk| = |\epsilon_{\di\dj\dl} \altx_\dl|$.
}
and $\delta_{ij}$ is a Kronecker delta.

For the diagrams A1, B2, and C3, which involve
gluon self-energy loops,
\begin {multline}
  \left[ \frac{d\Gamma}{dx\,dy} \right]_{\di\di}
  \simeq
  -2 \times
  \frac{\CA\alphas^2\,P(x)}{2\pi^2 y}
  \Re
  \biggl[
    \red{i\Omega_y}
    \Bigl\{
       - \ln\bigl( \tfrac{\Omega_y}{2\pi\Omega_0} \bigr) - \gammaE
       - \tfrac{i\pi}{2} \delta_{\di3}
    \Bigr\}
\\
    + \blue{i\Omega_0}
    \biggl\{
      2 \Bigl[
        \tfrac{1}{\eps} + \ln\bigl(\tfrac{\pi\mu^2}{\Omega_0 E}\bigr)
      \Bigr]
      -
      \left[
          \ln\bigl(
             \tfrac{|\altx_1 \altx_2 \altx_3| y\Omega_y}{\Omega_0}
          \bigr)
          + \tfrac{i\pi}{2} \delta_{\di3}
      \right]
\\
      + \tfrac{\altx_\dj^2+\altx_\dk^2}{4(\altx_1^2+\altx_2^2+\altx_3^2)}
          (1 - \ln2)
    \biggr\}
  \biggr]
\label {eq:ii}
\end {multline}
\end {subequations}
(no sum on $\di$), where in this formula $\altx_\dj$ and $\altx_\dk$
represent the two longitudinal
momenta in (\ref{eq:altx}) that are not $\altx_\di$.%
\footnote{
  If one prefers the right-hand side of (\ref{eq:ii}) to be written
  solely in terms of the index $\di$:
  $\altx_\dj^2 + \altx_\dk^2 = (\altx_1^2 + \altx_2^2 + \altx_3^2) - \altx_\di^2$.
}

Above, we've used the color red to indicate the leading terms in the
small-$y$ expansion.
These are the $O(y^{-3/2})$ terms of (\ref{eq:smallybehavior}), which generate
IR power-law divergences.  The color blue indicates the sub-leading
$O(y^{-1})$ terms, which generate the IR double and single logarithms.

We use dimensional regularization, and the
IR log-divergent terms in (\ref{eq:eachABC})
contain a UV-divergent piece $1/\eps$.
The appearance of such mixed UV-IR divergences in individual diagrams
is a well-known annoyance of Light-Cone Perturbation Theory (LCPT),
which was used to calculate our original generic-$y$ results
in ref.\ \cite{qcd}.  These mixed divergences must cancel when one
sums all the ABC diagrams.  And they do indeed cancel, as does the dependence
of the IR logs on the UV-renormalization scale $\mu$.

Note that the permutation symmetry between results for
different $\di{\not=}\dj$ diagrams in (\ref{eq:iNEj}) is slightly spoiled
by the $\frac{i\pi}{2} \delta_{\di 3}$ terms,
and similarly for permutation symmetry of
the diagrams of (\ref{eq:ii}).
These additional terms appear only in the
C1, C2, and C3 diagrams.  They originate from the fact that
the $y$ gluon propagator in time-ordered ABC diagrams (such as
the explicit examples in fig.\ \ref{fig:ABCexamples}) are colored
red (conjugate amplitude) for the three C diagrams, as opposed
to blue (amplitude) for the A and B diagrams.
As a result, the complex phases that appear in the calculation are
different, leading to $i\pi$ terms in the formula.
(For more discussion of $i\pi$ terms, see ref.\ \cite{logs2}.)

The sum of the small-$y$ rates (\ref{eq:eachABC}) is
\begin {multline}
  \left[ \frac{d\Gamma}{dx\,dy} \right]_{\rm ABC}
  \simeq
  \frac{\CA\alphas^2\,P(x)}{2\pi^2 y}
  \Re
  \biggl[
    \blue{i\Omega_0}
    \biggl\{
      -
      \left[
        \ln\bigl( \tfrac{|\altx_i \altx_j \altx_k| y\Omega_y}{\Omega_0} \bigr)
        + \tfrac{i\pi(\altx_1^2+\altx_2^2)}{4(\altx_1^2+\altx_2^2+\altx_3^2)}
      \right]
\\
      + \frac{ 2[ \altx_1^2 \ln(2|\altx_2 \altx_3|)
                  + \altx_2^2 \ln(2|\altx_3 \altx_1|)
                  + \altx_3^2 \ln(2|\altx_1 \altx_2|) ] }
             { (\altx_1^2+\altx_2^2+\altx_3^2) }
    \biggr\}
  \biggr] ,
\end {multline}
which is
\begin {multline}
  \left[ \frac{d\Gamma}{dx\,dy} \right]_{\rm ABC}
  \simeq
  \frac{\CA\alphas^2\,P(x)}{2\pi^2 y}
  \Re
  \biggl[
    \blue{i\Omega_0}
    \biggl\{
      - \ln\bigl( \tfrac{x(1{-}x) y\Omega_y}{4\Omega_0} \bigr)
\\
      + \frac{ \bigl[
                 (1{-}x)^2 \bigl( \ln x - \frac{i\pi}{8} \bigr) 
                 + x^2 \bigl( \ln(1{-}x) - \frac{i\pi}{8} \bigr)
                 + \ln\big(x(1{-}x)\bigr)
               \bigr] }
             { (1-x+x^2) }
    \biggr\}
  \biggr] .
\label {eq:total1}
\end {multline}
Note that the IR power-law divergences have canceled.%
\footnote{
  See the discussion in appendix E.3 of ref.\ \cite{qcd}.
}
Eq.\ (\ref{eq:total1}) can be massaged into our final result
(\ref{eq:sbar})
using (i) the formulas for $\Omega_y$ and $\Omega_0$ and (ii) the
fact that $\Omega_0$ has phase $e^{-i\pi/4}$ implies
$\Re(i\Omega_0 \times i) = -\Re(i\Omega_0)$.

% ===========================================================================

\section{non-ABC Diagrams}
\label {sec:nonABC}

\subsection{Overview}

On net, our single logs will come from the same diagrams
as fig.\ \ref{fig:ABCdiags},
from the boundaries of the double-log integration regions.
However, there are many other diagrams that {\it individually}
contribute to single and even to double logs.  The IR-log contributions
from those other diagrams, however, cancel in groups, as summarized
in table \ref{tab:cancel} (to be explained shortly).%
\footnote{
  Table \ref{tab:cancel} in this paper, describing the cancellation of
  IR logs of non-ABC diagrams, has a form that is somewhat related to
  table 1 of ref.\ \cite{qcd}, which summarized the cancellation of
  power-law IR divergences.  Readers interested in both can find a
  description of how they are related in appendix \ref{app:TableComparison}.
}
The fact that various
individual diagrams might have canceling IR-log contributions
is not surprising because individual diagrams also have canceling
IR {\it power-law} divergences in the $\hat q$ approximation.
These power-law divergences were first noted in ref.\ \cite{2brem},
and their explicit cancellation among all diagrams was verified in
ref.\ \cite{qcd}.%
\footnote{
   See in particular appendix E of ref.\ \cite{qcd}.
}

\begin{table}[t]
\vspace{5mm}
\begin {center}
\resizebox{\textwidth}{!}{%
\begin{tabular}{rcccccc}
\hline\hline
\multicolumn{1}{l}{\textbf{Real}}
& $~~~(x,y)~~~$ & $~~~(x,z)~~~$ & $~~~(z,y)~~~$ &
  $~~~(z,x)~~~$ & $~~~(y,x)~~~$ & $~~~(y,z)~~~$ \\
\hhline{~------}
 $2\Re(yx\bar x\bar y)$
    & \multicolumn{2}{|c}{\cellcolor{Cyan!60}$\alpha$}
    & \multicolumn{2}{|c}{\cellcolor{Cyan!40}$\alpha'$}
    & \multicolumn{1}{|c}{A3}
    & \multicolumn{1}{|c|}{B3} \\
  \cline{6-7}
  $2\Re(y\bar x x\bar y)$
    & \multicolumn{1}{|c}{\cellcolor{Cyan!60}}
    & \multicolumn{1}{c}{\cellcolor{Cyan!60}\tiny$*$}
    & \multicolumn{1}{|c}{\cellcolor{Cyan!40}}
    & \multicolumn{1}{c}{\cellcolor{Cyan!40}\tiny$*'$}
    & \multicolumn{1}{|c}{C1}
    & \multicolumn{1}{|c|}{C2} \\
  \hhline{~|~~|~~|--}
  $2\Re(x\bar y\bar x y)$
    & \multicolumn{1}{|c}{\cellcolor{Cyan!60}}
    & \multicolumn{1}{c}{\cellcolor{Cyan!60}\tiny$*$}
    & \multicolumn{1}{|c}{\cellcolor{Cyan!40}}
    & \multicolumn{1}{c}{\cellcolor{Cyan!40}\tiny$*'$}
    & \multicolumn{1}{|c}{\cellcolor{CarnationPink!60}$\beta$}
    & \multicolumn{1}{|c|}{\cellcolor{CarnationPink!40}$\beta'$} \\
  ${\cal A}_\seq${\small$(x,y)$}
    & \multicolumn{2}{|c}{\cellcolor{Cyan!60}}
    & \multicolumn{2}{|c}{\cellcolor{Cyan!40}}
    & \multicolumn{1}{|c}{\cellcolor{CarnationPink!60}}
    & \multicolumn{1}{|c|}{\cellcolor{CarnationPink!40}} \\
  \cline{2-7}
\noalign{\vskip 0.3em}
\hline
\multicolumn{1}{l}{\textbf{Virtual Class I}}
&         &             & \multicolumn{2}{c}{[$x{\to}1{-}x$ cousins]} &&
\\
& $(x,y)$ & $(x,z)^\bullet$ & $(1{-}x,y)$
                            & $\!\!\!(1{-}x,x{-}y)^\bullet\!\!\!$ && \\
\hhline{~----~~}
  $2\Re(yx\bar x y)$
    & \multicolumn{2}{|c}{\cellcolor{Cyan!60}$\alpha$}
    & \multicolumn{2}{|c|}{\cellcolor{Cyan!40}$\alpha'$}
    &&\\
  $2\Re(y\bar x xy)$
    & \multicolumn{1}{|c}{\cellcolor{Cyan!60}}
    & \multicolumn{1}{c}{\cellcolor{Cyan!60}\tiny$\dagger$}
    & \multicolumn{1}{|c}{\cellcolor{Cyan!40}}
    & \multicolumn{1}{c|}{\cellcolor{Cyan!40}\tiny$\dagger'$}
    &&\\
  $2\Re(\bar x yxy)$
    & \multicolumn{1}{|c}{\cellcolor{Cyan!60}}
    & \multicolumn{1}{c}{\cellcolor{Cyan!60}\tiny$\dagger$}
    & \multicolumn{1}{|c}{\cellcolor{Cyan!40}}
    & \multicolumn{1}{c|}{\cellcolor{Cyan!40}\tiny$\dagger'$}
    &&\\
  bkEnd(${\cal A}_\seq${\small$(x,y)$})
    & \multicolumn{2}{|c}{\cellcolor{Cyan!60}}
    & \multicolumn{2}{|c|}{\cellcolor{Cyan!40}}
    &&\\
  \hhline{~|--|--|~~}
  $2\Re(yxy\bar x)$
    & \multicolumn{1}{|c}{\cellcolor{CarnationPink!60}$\beta$}
    & \multicolumn{1}{|c}{B1}
    & \multicolumn{1}{|c}{\cellcolor{CarnationPink!40}$\beta'$}
    & \multicolumn{1}{|c|}{A2}
    &&\\
  \hhline{~|--|--|~~}
  $2\Re(xyy\bar x)$
    & \multicolumn{2}{|c}{$\longleftarrow ~{\rm A1}~ \longrightarrow$}
    & \multicolumn{2}{|c|}{$\longleftarrow ~{\rm B2}~ \longrightarrow$}
    &&\\
  \cline{2-5}
\noalign{\vskip 0.3em}
\hline
\multicolumn{1}{l}{\textbf{Virtual Class II}}
  & $(x,y)$ & $(1{-}x,y)^\bullet$ &&&& \\
\hhline{~--~~~~}
  frEnd(${\cal A}_\seq${\small$(y,x)$})
    & \multicolumn{1}{|c}{\cellcolor{CarnationPink!60}$\beta$}
    & \multicolumn{1}{|c|}{\cellcolor{CarnationPink!40}$\beta'$}
    &&&&\\
  \hhline{~--~~~~}
  $2\Re(x\bar y\bar y\bar x)$
    & \multicolumn{2}{|c|}{$\longleftarrow ~{\rm C3}~ \longrightarrow$}
    &&&&\\
  \cline{2-3}
\noalign{\vskip 0.3em}
\hline\hline
\end{tabular}
}
\end {center}
\caption{
  \label{tab:cancel}
  The shaded regions (labeled $\alpha$, $\alpha'$, $\beta$, and $\beta'$)
  depict groups of cancellations among IR single and double logs
  from real ($g \to ggg$) and virtual (next-to-leading-order $g \to gg$)
  processes as $y\to0$.  The unshaded entries do not cancel and are labeled
  according to fig.\ \ref{fig:ABCdiags}.  The fist column gives the names
  of diagrams, in the convention of ref.\ \cite{qcd}.  For the real diagrams,
  each row covers
  six distinct diagrams given by permutations of the daughters
  $(x,y,z)$ of the $g{\to}ggg$ process, where the column headers denote
  what $(x,y)$ is permuted into relative to the diagram name in the first
  column, and where $z \equiv 1{-}x{-}y$.
  A column header $(\xi,\chi)$ means to
  replace $(x,y) \to (\xi,\chi)$ in the diagram listed in the first column.
  Column headers with bullet superscripts
  ($\bullet$) do not represent different diagrams but instead represent
  a different labeling of the internal lines of the un-bulleted header
  to the left, corresponding to a different soft limit ($y{\to}0$) of that
  diagram.
  For each Class I virtual diagram, another distinct diagram
  can be generated by substituting $x\to 1{-}x$ (see ref.\ \cite{qcd}),
  as also indicated by the column headers $(\xi,\chi)$.
  ${\cal A}_{\rm seq}$ represents only one
  of the two large-$\Nc$ color routings of ``sequential'' diagrams
  (see ref.\ \cite{seq} for details).
  bkEnd and frEnd refer to back-end and front-end transformations,
  as described in refs.\ \cite{QEDnf,qcd}.
  The notations $*$, $*'$, $\dagger$ and $\dagger'$ on certain pairs
  of table entries are explained in the text.
  The labels A1, A2, ..., C3 above are shorthand for the corresponding
  diagrams of fig.\ \ref{fig:ABCdiags} {\it plus} their complex conjugates.
  The meaning of the double-sized boxes for A1, B2, and C3 is explained
  in appendix \ref{app:TableComparison}.
}
\end{table}

The simplest demonstration of the cancellation of single IR logs for
diagrams {\it other} than the ABC diagrams of fig.\ \ref{fig:ABCdiags}
is the comparison in fig.\ \ref{fig:cnew} of the total numerical result
(extracted from the sum of {\it all}\/ diagrams) to our analytic
result (\ref{eq:sbar}), which we will derive by only considering
the ABC diagrams.  However, we may also discuss diagrammatically
why the IR logs of the non-ABC diagrams cancel.

Throughout this paper, the longitudinal momentum fractions
of the three daughters of a double splitting
process $g{\to}ggg$ will be referred to as $(x,y,z)$, where
\begin {equation}
   z \equiv 1-x-y .
\end {equation}

% ----------------------------------------------------------------------------

\subsection{Cancellation of IR logs for non-ABC diagrams}
\label {sec:cancel}

First, let's summarize the form of the $y{\to}0$ expansion of
individual diagrams (in the $\hat q$ approximation).
Generically, differential rates from individual diagrams
have small-$y$ expansions of the same form (\ref{eq:smallybehavior})
as the ABC diagrams:
\begin {equation}
   \frac{d\Gamma}{dx\,dy} =
   \frac{\# \ln y + \#}{y^{3/2}}
   +
   \frac{\# \ln y + \#}{y}
   +
   O(y^{-1/2}) .
\end {equation}
Some individual diagrams additionally have more singular
$O(y^{-5/2})$ behavior, but this will not be relevant:
those diagrams appear in combinations where the $O(y^{-5/2})$
terms cancel in a way that will not affect the arguments below.
(Specifically, the $y^{-5/2}$ terms cancel for the pair of entries in
table \ref{tab:cancel} marked ``$*$'', and similarly for the
pairs marked $*'$, $\dagger$, and $\dagger'$.)

\begin {figure}[t]
\begin {center}
  \includegraphics[scale=0.5]{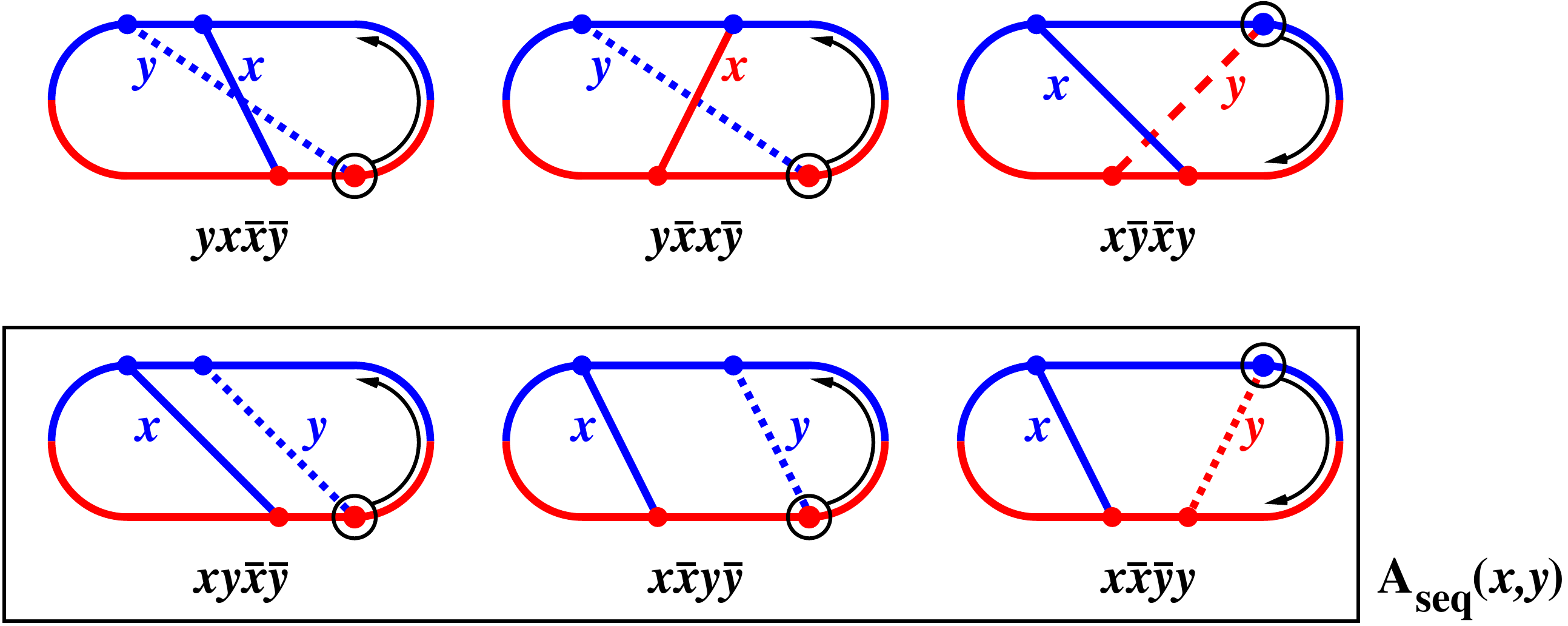}
  \caption{
     \label{fig:alpha1}
     The $g{\to}ggg$ diagrams corresponding to the $(x,y)$ column
     of the upper darker-blue $(\alpha)$ rectangle of table \ref{tab:cancel}.
     The black circles and arrows depict the action of a back-end
     transformation on these diagrams.
  }
\end {center}
\end {figure}

\begin {figure}[t]
\begin {center}
  \includegraphics[scale=0.50]{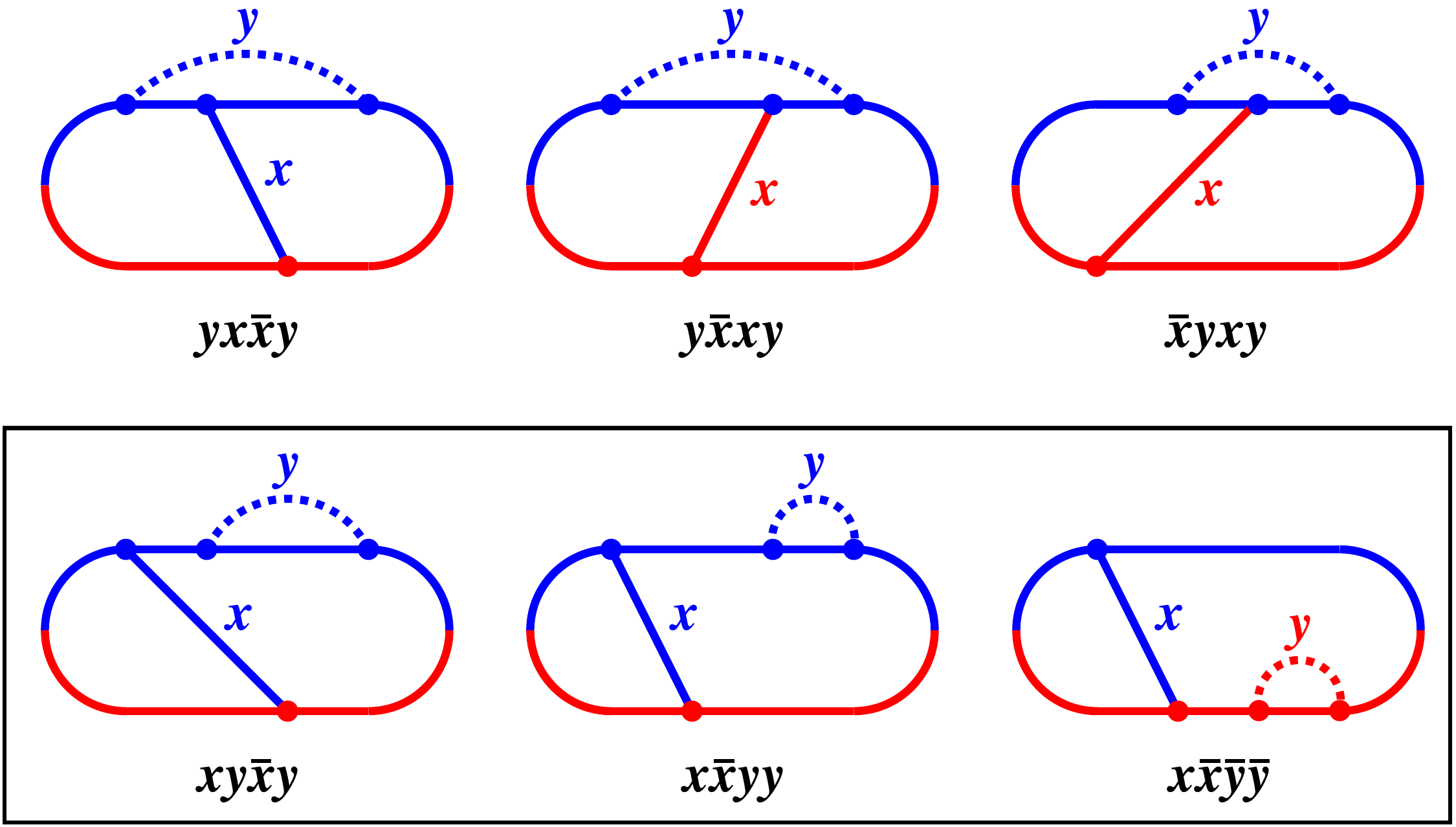}
  \caption{
     \label{fig:alpha2}
     The virtual $g{\to}gg$ diagrams corresponding to the $(x,y)$ column
     of the lower darker-blue $(\alpha)$ rectangle of table \ref{tab:cancel}.
  }
\end {center}
\end {figure}

We now argue that the two darker-blue rectangles in table \ref{tab:cancel}
(the two rectangles labeled $\alpha$) give canceling contributions to
IR logs.
The real ($g {\to} ggg$) double splitting diagrams corresponding
to the $(x,y)$ column of the top rectangle in
table \ref{tab:cancel} are shown in
fig.\ \ref{fig:alpha1}.  The ${\cal A}_\seq(x,y)$ entry of the table
corresponds to the sum of the entire bottom row of the figure,
with the caveat \cite{seq}%
\footnote{
  Specifically, see sections 2.2.1 and 3.1 of ref.\ \cite{seq}.
}
that ${\cal A}_\seq(x,y)$ includes only one of the
two large-$\Nc$ color routings of those bottom diagrams.
[The other color routing is represented by the $(x,z)$ column entry for
${\cal A}_\seq$.]  The virtual diagrams for $g{\to}gg$
corresponding to the lower darker-blue ($\alpha$) rectangle
in table \ref{tab:cancel}
are shown in fig.\ \ref{fig:alpha2}.
These diagrams are ``back-end'' transformations of the diagrams in
fig.\ \ref{fig:alpha1}, which means that the drawings of the diagrams
are related by sliding the latest-time (right-most) vertex around
the back of the diagram from the amplitude to the conjugate amplitude,
or vice versa.%
\footnote{
  See in particular section 4.1 of ref.\ \cite{QEDnf} and section 2.2
  of ref.\ \cite{qcd}.
  Since we take $2\Re(\cdots)$ of these diagrams at the end of the day,
  it doesn't matter that the back-end transformation
  of $x\bar y\bar x y$ is the complex conjugate of
  $\bar x y x y$ and is not $\bar x y x y$ itself.
}
As discussed in refs.\ \cite{QEDnf,qcd}, the differential
rates for such a pair of diagrams have identical magnitudes but opposite
signs.  This does not
mean that their effects exactly cancel, however,
because rates for $g{\to}ggg$ processes and rates for virtual
corrections to $g{\to}gg$ processes appear differently in applications.
However, in applications to shower development and energy loss
(see examples in ref.\ \cite{qcd}%
\footnote{
  Specifically, see ref.\ \cite{qcd} sections 1.2 and 3.1 and appendix F.
}%
), the differential $g{\to}ggg$ rates and
$g{\to}gg$ rates are integrated against functions that become the same
in the soft limit $y{\to}0$, reflecting the fact that
if one of the daughters of $g{\to}ggg$ becomes arbitrarily soft, then
effects of $g{\to}ggg$ become physically indistinguishable from
those of $g{\to}gg$.
Generically, then, the effects of two back-end related diagrams should
cancel at leading order in $y{\to}0$, and the corrections to that
cancellation should be suppressed by an additional factor of $y$
(coming from the Taylor expansion in $y$ of whatever functions that
$g{\to}ggg$ and virtual $g{\to}gg$ processes are multiplied by
in the application of interest).
Since the sum of diagrams in fig.\ \ref{fig:alpha1} and the sum of
diagrams in fig.\ \ref{fig:alpha2} each behave like
(\ref{eq:smallybehavior}) and so are $O(y^{-3/2})$, that means
that their combined effect in applications will be order
$y \times y^{-3/2} = y^{-1/2}$, which is not singular enough to
contribute to IR logarithms.

\begin {figure}[t]
\begin {center}
  \includegraphics[scale=0.50]{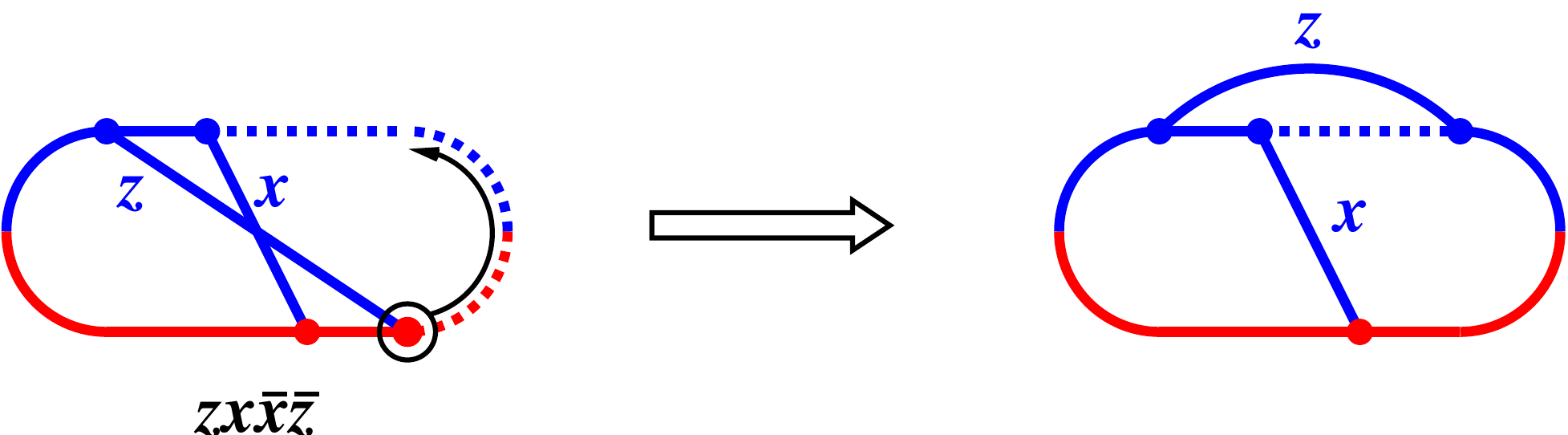}
  \caption{
     \label{fig:alphaxz}
     The back-end transformation of the $y{\leftrightarrow}z$ permutation of
     $yx\bar x\bar y$.  The dotted lines indicate the soft gluon ($y{\to}0$).
     The virtual diagram on the right is the {\it same} diagram as the
     $yx\bar xy$ diagram of fig.\ \ref{fig:alpha2}
     (since $y$ is just a loop variable and not a fixed, external momentum
     fraction),
     but a different virtual gluon line is soft in the diagram here.
  }
\end {center}
\end {figure}

A similar argument applies to the sum of diagrams in the
$(x,z)$ column of the same two darker-blue rectangles ($\alpha$)
in table \ref{tab:cancel}.  However, the $(x,z)$ column for
the {\it virtual}\/ entries does not represent a different diagram than
the $(x,y)$ entries but instead represents different IR limits
of the same diagrams.  An example is shown by the diagram
on the right-hand
side of fig.\ \ref{fig:alphaxz}.  This figure also shows the
distinct $g{\to}ggg$ diagram that is the back-end partner.

So, in total,
the darker-blue ($\alpha$) rectangles do not contribute to IR logs.

\begin {figure}[t]
\begin {center}
  \includegraphics[scale=0.50]{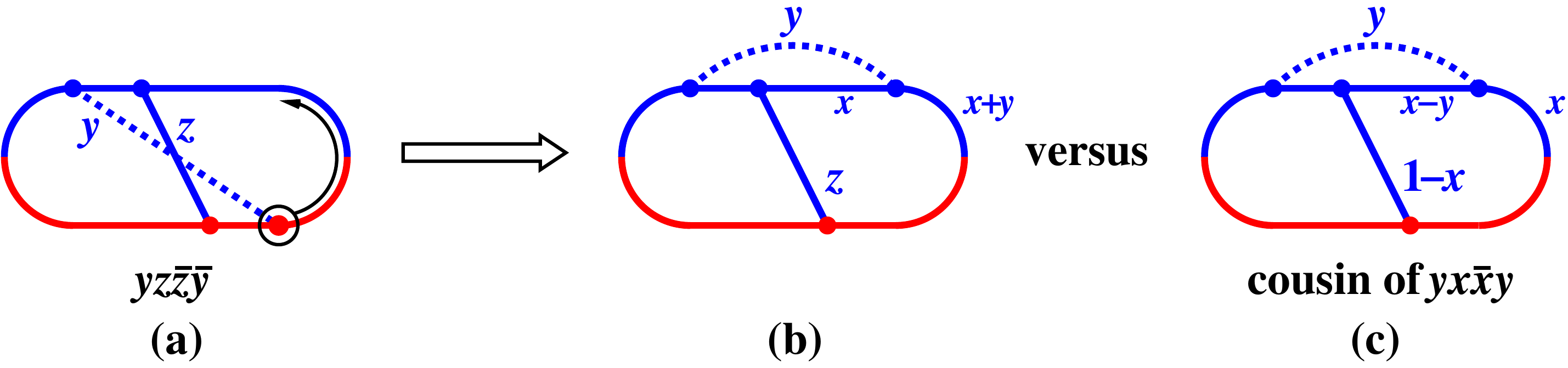}
  \caption{
     \label{fig:alphaprime}
     (a--b) The back-end transformation of the $x \leftrightarrow z$
     permutation of 
     $yx\bar x\bar y$
     compared to
     (c) the $x \to 1{-}x$ cousin of the Class I virtual diagram
     $yx\bar x y$.
  }
\end {center}
\end {figure}

The cancellation between the lighter-blue ($\alpha'$) rectangles is
somewhat similar, but an additional argument is needed.
As an example, consider the IR cancellation of (a) the
$(z,y)$ column entry of the $2\Re(y x \bar x \bar y)$
row of table \ref{tab:cancel}
with (c) the $(1{-}x,y)$ column entry of the
$2\Re(yx\bar x y)$ row.
Fig.\ \ref{fig:alphaprime} shows, correspondingly, the diagrams (a)
$y z\bar z\bar y$, (b) its back-end transform, and (c)
$yx\bar x y$ with $x \to 1{-}x$.
If we were adding (a) and its back-end transformation (b) for a given $y$,
they would cancel just like in our previous discussion.  But in the
case of $g{\to}gg$ processes, the variables we use to specify the momentum
fractions of the two daughters should be fixed ($x$ and $1{-}x$) and should
not change when we later integrate over loop variables (in this case $y$).
So we want to add (a) with (c), not (a) with (b).
Unlike (a) and (b), the differential rates $d\Gamma/dx\,dy$ associated
with (a) and (c) are not exactly the negative of each other.
However, this is inessential because the momentum fractions of lines in
(b) and (c) match up in the $y{\to}0$ limit: the differences are
suppressed by relative factors of $y$.  As argued earlier in a different
context, a correction of relative order $y$ to
(\ref{eq:smallybehavior}) will not affect IR logarithms.
So the lighter-blue ($\alpha'$) entries will have the same cancellation
of IR logarithms as the darker-blue ($\alpha$) entries.

\begin {figure}[t]
\begin {center}
  \includegraphics[scale=0.44]{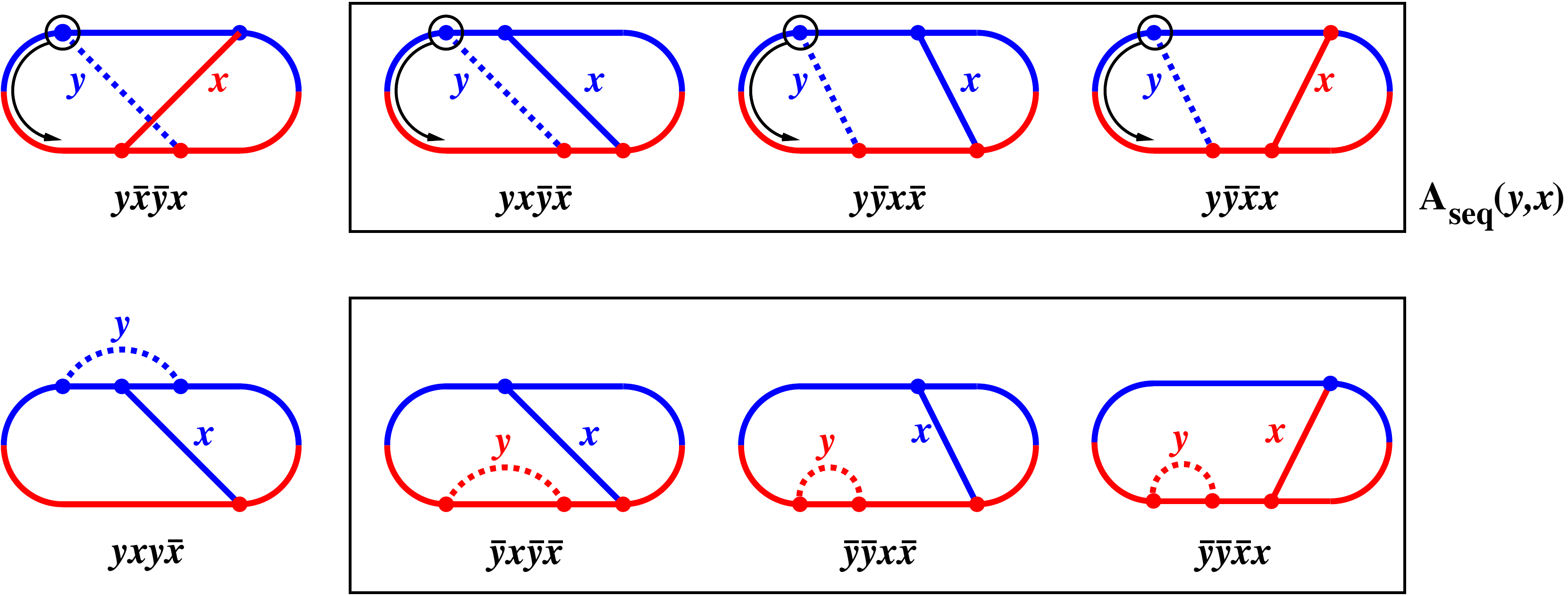}
  \caption{
     \label{fig:beta}
     Diagrams corresponding to the darker-pink rectangles $(\beta)$
     in table \ref{tab:cancel}.  The $g{\to}ggg$ diagrams (top row)
     are related by front-end transformation to the virtual $g{\to}gg$
     diagrams of the bottom row.
  }
\end {center}
\end {figure}

The diagrams corresponding to the darker-pink rectangles ($\beta$) of
table \ref{tab:cancel} are shown in fig.\ \ref{fig:beta}.  Here the
real and virtual diagrams are related by a {\it front}-end transformation
\cite{QEDnf,qcd},
which graphically corresponds to sliding the earliest-time (left-most)
vertex around the front of the diagram.  Like the back-end transformation,
the front-end transformation introduces an overall minus sign.
However, it {\it also} involves a relabeling of momentum fractions
as%
\footnote{
  See in particular section 4.2 of ref.\ \cite{QEDnf} and section 2.2 of
  ref.\ \cite{qcd} (but note that our fig.\ \ref{fig:beta} here
  already implements the step
  $x\leftrightarrow y$ discussed for figs.\ 9 and 10 of ref.\ \cite{qcd}).
}
\begin {equation}
  (x,y,E) \longrightarrow
  \Bigl( \frac{x}{1{-}y} \,,\, \frac{{-}y}{1{-}y} \,,\, (1{-}y)E \Bigr) ,
\label {eq:frontendy}
\end {equation}
in the case of fig.\ \ref{fig:beta}.  Now note that in the
$y{\to}0$ limit of interest to extracting IR logs, this transformation
simplifies to
\begin {equation}
  (x,y,E) \longrightarrow (x,{-}y,E)
\label {eq:frontendsmally}
\end {equation}
up to corrections that, for each, are suppressed by an additional factor
of $y$.  So, {\it except}\/ for $y\to-y$, the front-end related
diagrams in fig.\ \ref{fig:beta} should cancel each other, similar
to what happened for the back-end transformations in the $\alpha'$
group discussed previously.

Discussing the effect of $y\to-y$ on the cancellation of
front-end related diagrams is subtle.  To make front-end
transformations like (\ref{eq:frontendy}) successfully relate diagrams,
one must be very careful \cite{QEDnf,qcd} to figure out, in
formulas for diagrams, which instances of longitudinal momentum
fractions like $y$ should be written as $|y|$ and which as simply $y$.
We will not attempt to justify or re-derive it here.  The $y$ in the
denominator of the IR log term
\begin {equation}
   \frac{\# \ln y + \#}{y}
\label {eq:smallybehavior2}
\end {equation}
from (\ref{eq:smallybehavior}) turns out to be a $|y|$.%
\footnote{
   The $1/|y|$ factor comes from the small-$y$ behavior of the
   $g{\to}gg$ DGLAP splitting function $P(y)$ associated with
   the $y$ emission.  To implement front-end transformations,
   DGLAP splitting functions should involve the absolute values
   of the particle momentum fractions, as is implemented in
   ref.\ \cite{qcd} eqs.\ (A.5) and (A.23).  See footnote 35
   or ref.\ \cite{qcd}.
}
That leaves the possibility that $y\to-y$ could take $\ln y$
to $\ln(-y)$ in (\ref{eq:smallybehavior2}).
Since a front-end transformation also negates the diagram,
that means that the sum of a pair of front-end related diagrams could
have a non-zero result proportional to
\begin {equation}
   \frac{\ln y}{y} - \frac{\ln(-y)}{y} = \pm \frac{i\pi}{y} \,.
\label {eq:leftover}
\end {equation}
The details are complicated.  In general, we find both $\ln y$ and
$\ln|y|$ terms in our small-$y$ expansions of diagrams.
As an example, we carry out this analysis for ${\cal A}_\seq(y,x)$
in appendix \ref{app:Aseq}, and show how the IR single log of
${\cal A}_\seq(y,x)$ does not
cancel that of its front-end transformation but instead leaves behind a
left-over $i\pi$ term like (\ref{eq:leftover}).
However, we have not taken the time to do the same analytically
for the other front-end related pair of diagrams also marked $\beta$
(darker pink) in table \ref{tab:cancel}: $2\Re(y\bar x\bar yx)+2\Re(yxy\bar x)$.
Instead, we have just extracted the small-$y$
behavior of that sum numerically and find that the IR log exactly
cancels that of the ${\cal A}_\seq(y,x)$ pair.
That is, altogether,
the IR logs from the four diagrams in the dark-pink rectangles ($\beta$)
cancel each other.
Regrettably, we do not know a simpler way to argue that the left-over terms
(\ref{eq:leftover}) must cancel among these diagrams.

Finally, the IR logs from the lighter-pink rectangles ($\beta'$)
cancel each other similarly.

% ----------------------------------------------------------------------------

\subsection{Diagrams with 4-gluon or instantaneous vertices}

\begin {figure}[t]
\begin {center}
  \includegraphics[scale=0.5]{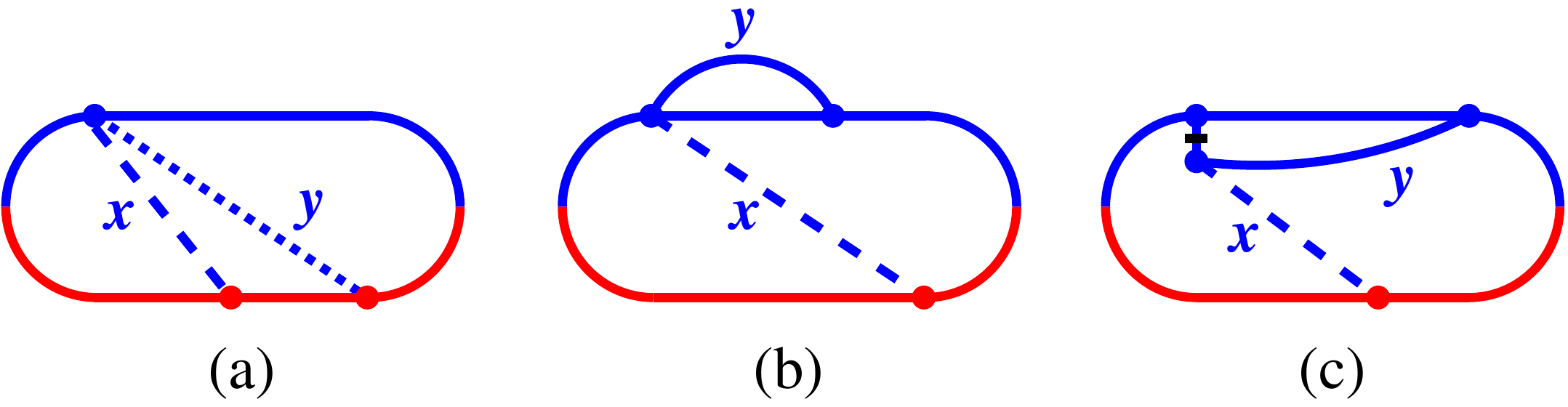}
  \caption{
     \label{fig:4Iexamples}
     Examples of diagrams that involve 4-gluon or LCPT instantaneous
     vertices, which have been left out of our discussion.
  }
\end {center}
\end {figure}

In our discussion of non-ABC diagrams above, we have left out
diagrams that contain 4-gluon vertices \cite{4point},
examples of which are shown in figs.\ \ref{fig:4Iexamples}a and b.
As mentioned earlier, our analysis of diagrams makes
use of Light-Cone Perturbation Theory.  In LCPT, there are non-local,
effective 4-gluon interactions that are instantaneous in light-cone
time.  We have left out diagrams that contain these interactions as well,
an example of which is shown in fig.\ \ref{fig:4Iexamples}c.
(See ref.\ \cite{QEDnf} for examples of the analysis of such diagrams,
for generic $y$, in the case of large-$\Nf$ QED.)
There is no reason to suspect that
diagrams involving 4-gluon or LCPT instantaneous vertices
would contribute to (net) logarithms, and we find that
they do not \cite{qcdI}.

% ===========================================================================

\section {The A3 diagram}
\label {sec:A3}

\subsection {Scales and form}

We start with a calculation of the IR logs associated with the A3
diagram (including adding its complex conjugate).  This corresponds to the
$y{\to}0$ limit of $2\Re(xy\bar y\bar x)$.  Throughout this paper, we
do not assume that $x$ is small, and we will
treat $x$ and $1{-}x$ as order $y^0$.  We will also not bother, in
parametric formulas, to show the constant, dimensionful factor
$\sqrt{E/\hat q}$ associated with formation times.  So, the
time interval (\ref{eq:tregion}) that gives rise to double logs
will be summarized as simply
\begin {equation}
   y \ll \Delta t \ll \sqrt{y} \,.
\label {eq:tregion2}
\end {equation}

For given values of $x$ and $y$, ref.\ \cite{qcd} gives formulas
for the various diagrams in terms of one-dimensional integrals
of the form
\begin {equation}
  {\rm diagram}(x,y)
  = \int_0^\infty d(\Delta t) \> f(x,y,\Delta t) ,
\label {eq:form0}
\end {equation}
where $f$ is a quite complicated algebraic function specific
to each diagram.  (We'll give more details as needed.)
There is a complication, however.
The integral above has a UV divergence associated with $\Delta t \to 0$.
The regulated divergences give rise to what are called
``pole terms'' in refs.\ \cite{2brem,seq,dimreg,qcd}.
To handle the UV divergence separately, we split (\ref{eq:form0}) into
\begin {equation}
  {\rm diagram}(x,y)
  =
  \lim_{a\to 0}
    \left[
      \int_0^a d(\Delta t) \> f(x,y,\Delta t)
      + \int_a^\infty d(\Delta t) \> f(x,y,\Delta t)
    \right]
\label {eq:form1}
\end {equation}
and regulate the first integral with dimensional regularization.

We should mention that in the treatment of divergent integrals
for generic $y$ in
ref.\ \cite{qcd}, we needed to evaluate the second
integral numerically.  That motivated further reorganization
of the second integral to improve numerical accuracy by
removing its sensitivity to the tiny
cut-off $a$.  That numerics-motivated reorganization is unnecessary here
because our goal is to derive {\it analytic} results for the small-$y$
expansion.%
\footnote{
  The further reorganization is explained in section 4.3.2 of
  ref.\ \cite{QEDnf} and appendix D.1 of ref.\ \cite{qcd}.
  Since that is unnecessary here, we will not include
  the corresponding ``${\cal D}_2$'' subtractions in our formulas for
  diagrams.  Similarly, what we call our ``pole pieces'' in this paper
  do not have those ${\cal D}_2$ subtractions added back in.
  We should
  clarify that, even for numerics, no such ${\cal D}_2$ subtractions are needed
  for some {\it combinations} of diagrams such as eq.\ (A.12) of
  ref.\ \cite{qcd}, which combines the A3 diagram with other
  $g{\to}ggg$ crossed diagrams.  For that combination, the
  un-subtracted integral is insensitive to arbitrarily small $a$
  \cite{2brem,dimreg}.
}

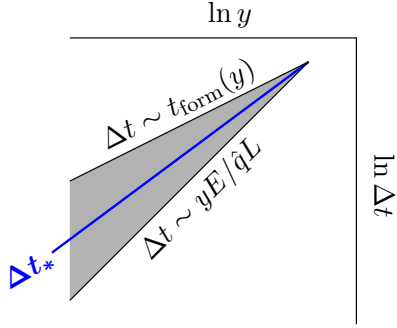
\begin {figure}[t]
\begin {center}
  \begin{picture}(152,125)(0,0)
    \put(22,5){\includegraphics[scale=0.5]{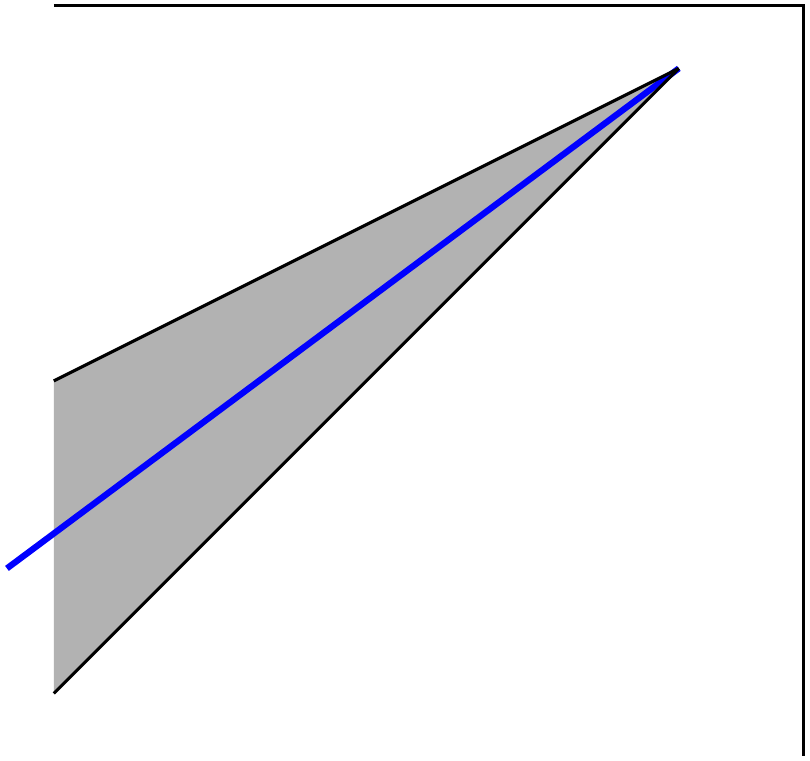}}
    \put(0,17){\rotatebox{30}{
      $\color{blue}{\bm{\Delta t_*}}$
    }}
    \put(37,70){\rotatebox{27}{
      $\Delta t \sim t_{\rm form}(y)$
    }}
    \put(52,27){\rotatebox{45}{
      $\Delta t \sim yE/\hat q L$
    }}
    \put(82,120){$\ln y$}
    \put(145,70){\rotatebox{-90}{$\ln\Delta t$}}
  %\put(0,0){.}
  %\put(0,125){.}
  %\put(152,0){.}
  %\put(152,125){.}
  \end{picture}
  \caption{
     \label {fig:dtstar}
     A choice of intermediate scale $\Delta t_*(y)$
     splitting the (shaded) double log region of fig.\
     \ref{fig:region}.  Remember that we
     (parametrically) define
     $L = t_{\rm form}(x)$ in our application.
  }
\end {center}
\end {figure}

To calculate IR single logs, imagine choosing a dividing
time $(\Delta t)_*$
somewhere in the parametric range (\ref{eq:tregion2}),%
\footnote{
  A similar strategy was used in ref.\ \cite{Wu0}.
  (But their analysis,
  which was in the context of transverse momentum broadening,
  did not need to confront the UV divergences that we have for
  individual ABC diagrams.  So they did not need any dividing scale analogous
  to our $a$.)
}
and
further divide (\ref{eq:form1}) into pieces
\begin {multline}
  {\rm diagram}(x,y)
  =
  \lim_{a\to 0}
    \left[
      \int_0^a d(\Delta t) \> f(x,y,\Delta t)
      + \int_a^{(\Delta t)_*} d(\Delta t) \> f(x,y,\Delta t)
    \right]
\\
    + \int_{(\Delta t)_*}^\infty d(\Delta t) \>f(x,y,\Delta t) .
\label {eq:form2}
\end {multline}
For small $y$, we then make a $\Delta t \ll \sqrt{y}$ approximation
to the $\Delta t$ integral over the interval $[a,(\Delta t)_*]$
in (\ref{eq:form2}) and a
$\Delta t \gg y$ approximation to the integral over
$[(\Delta t)_*,\infty]$.
That is, we split the calculation of double and single logs by
dividing the double log region into two, as divided pictorially
by the blue line in fig.\ \ref{fig:dtstar}.

Later, our small-$y$ expansions of these integrals will turn out
to be more compact if we formally restrict the choice of the dividing
time $(\Delta t)_*$ to the narrower range
\begin {equation}
   y^{2/3} \ll (\Delta t)_* \ll y^{1/2} \,.
\label {eq:t*region}
\end {equation}
So $(\Delta t)_* \sim y^{7/12}$, for example.
Note that there are no unusual fractional powers of $y$ in
our final results (\ref{eq:eachABC}) for small-$y$ expansions;
the choice (\ref{eq:t*region}) is just a convenience for managing
expansions of intermediate results before the eventual cancellation
of $(\Delta t)_*$ dependence in the combination (\ref{eq:form2}).

The ``interesting'' part of the $[a,(\Delta t)_*]$
and $[(\Delta t)_*,\infty]$ integrals in (\ref{eq:form2})
will be the behavior of the integrands at $\Delta t \sim y$ and
$\Delta t \sim \sqrt{y}$, respectively.  So, we will loosely refer
to these two integrals as the ``$\Delta t \sim y$'' and
``$\Delta t \sim \sqrt{y}\,$'' contributions, even though
it is actually the entire range (\ref{eq:tregion2}) that generates
the double log.

Our strategy in this paper will be to
take as our starting point the general formulas \cite{qcd} for diagrams,
which do not assume small $y$, and from them analytically extract the $y{\to}0$
expansion (\ref{eq:smallybehavior}).
The analysis will be similar to the analysis of ref.\ \cite{qcd} appendix
E.4 of IR power-law divergences,
which corresponds to the $O(y^{-3/2})$ terms in the
small-$y$ expansion (\ref{eq:smallybehavior}).
But now we must push the analysis to higher order in the expansion
to find the IR log (order $y^{-1}$) terms as well.

It will aid our discussion to
write out explicitly
the highest-level structure of the integrands
$f$ in terms of the same notation used in ref.\ \cite{qcd}.
Specifically, (\ref{eq:form0}) for the A3 diagram (including its
complex conjugate) is
\begin {equation}
  \left[ \frac{d\Gamma}{dx\,dy} \right]_{\rm A3}
  =
      \int_0^\infty d(\Delta t) \>
      2\Re C(-1,y,z,x,\alpha,\beta,\gamma,\Delta t) ,
\label {eq:A3intC}
\end {equation}
with
\begin {equation}
   C = D - \lim_{\hat q\to 0} D .
\label {eq:Cdef}
\end {equation}
Eq.\ (\ref{eq:Cdef})
represents subtracting out the result the diagram would have in
vacuum, which is merely a trick for simplifying calculations.%
\footnote{
  \label{foot:vacuum}
  See the discussion in section 5.4 in ref.\ \cite{2brem}.
  Note that our calculations are for splitting of on-shell
  high-energy partons in the medium.  In consequence, the
  total rate for splitting in the vacuum must be zero.
}
The symbols $(\alpha,\beta,\gamma)$ above are functions of $x$ and $y$
and represent various combinations of helicity-dependent DGLAP splitting
functions; their detailed formulas are not important for now.
The other arguments $(-1,y,z,x)$ of $C$ are the longitudinal
momentum fractions of the four high-energy gluon lines in the A3 diagram
of fig.\ \ref{fig:ABCexamples} during the time $\Delta t$ spanned
by the $y$ emission.

We will not write out the detailed generic-$y$
formulas for $D$, $\alpha$, $\beta$, and $\gamma$ here.
Readers may find them in appendix A of ref.\ \cite{qcd}, and we will
use them in various appendices.  However, here in the main text,
it will be useful to know that $\gamma$ dominates over $\alpha$ and
$\beta$ in the small-$y$ limit, where
\begin {equation}
  \gamma = \frac{2\,P(x)}{\CA x^2(1{-}x)^3 y^3} \, \bigl[ 1 + O(y) \bigr] .
\label {eq:gammaapprox}
\end {equation}
[Some of our intermediate calculations will be sensitive to the
relative $O(y)$ corrections in (\ref{eq:gammaapprox}), but ultimately
we will not need an explicit formula for those corrections.]

% ---------------------------------------------------------------------------

\subsection {$\Delta t \sim y$ contribution to A3 diagram}
\label {sec:A3y}

Unfortunately, an annoying complication arises in the small-$y$
expansion.  Though the A3 diagram overall has the small-$y$ expansion
(\ref{eq:smallybehavior}), which starts with $O(y^{-3/2})$,
the $\Delta t\sim y$ integration in (\ref{eq:form2}) produces
a spurious $y^{-2}$ divergence which is {\it canceled}\/ by a similar
$y^{-2}$ divergence of the pole term in (\ref{eq:form2}).
Because of this, in order to extract the $y^{-1}$ behavior of IR logs,
we must expand the $\Delta t \sim y$ integral, and so its integrand,
to
higher relative order in $y$ than hoped.
This complication was previously encountered in the calculation of
IR power-law divergences in ref.\ \cite{qcd} appendix E.4.

We originally planned to use a symbolic algebra program
to Taylor expand the integrand $f(x,y,\Delta t)$
in powers of $y \sim \Delta t$.  That is
awkward
because of the very complicated formula for $f = 2\Re C$ and issues of branch
cuts, and because automated expansions tend to produce long, complicated
results that are hard to organize and simplify.
In the end, we decided to do expansions by hand.

\begin {figure}[t]
\begin {center}
  \includegraphics[scale=0.80]{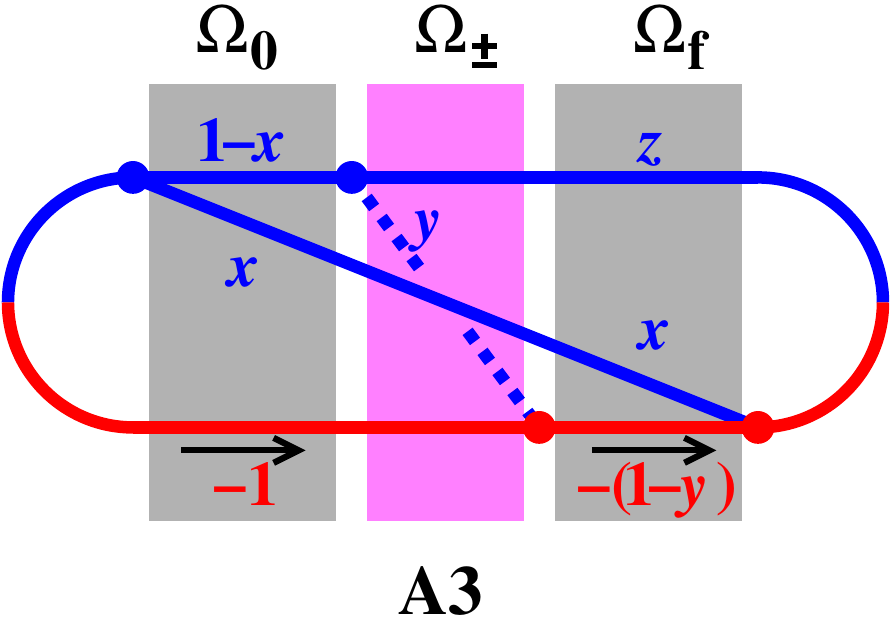}
  \caption{
     \label{fig:A3}
     The A3 diagram with shading denoting
     the initial 3-particle evolution (left),
     the 4-particle evolution (middle), and the final 3-particle
     evolution (right).  Our names $\Omega$
     for the corresponding complex harmonic
     oscillator frequencies are marked atop each region.
  }
\end {center}
\end {figure}

The steps leading to the small-$y$ expansion of the
$\Delta t$-integrand of the A3 diagram
for $\Delta t \sim y$ are summarized in appendix \ref{app:A3y}.
In the specific case of the A3 diagram, we obtain
\begin {multline}
  D
  \simeq
  - \frac{\CA^2\alphas^2}{8\pi^2(\Delta t)^2}
  (xyz)^2(1{-}x)(1{-}y)
  \gamma
\\ \times
  \biggl\{
     ( \red{1} + \xi - 2 s\xi\tau^2 )
       \ln\Bigl( 
         \frac{x y}{(1{-}x)(1{-}y)}
         (1 + \tau)
       \Bigr)
     + \red{ \frac{1}{1+\tau} }
\\
     + \xi
       \left[
          - \frac{2\tau}{1+\tau}
          - \frac{(1+s)\tau^3}{(1+\tau)^2}
       \right]
  \biggr\} ,
\label {eq:DA3}
\end {multline}
where we have found it convenient to define the variables
\begin {equation}
   s \equiv \frac{x^2}{12(1{-}x{+}x^2)} \,,
   \qquad
   \xi \equiv \frac{xy}{2z} \,,
   \qquad
   \tau \equiv \frac{i(\Omega_0{+}\Omega_\fx)\Delta t}{2\xi}
\label {eq:sxitau} .
\end {equation}
$\Omega_0$ and $\Omega_\fx$ are the complex harmonic oscillator
frequencies associated with the initial and final 3-particle evolution,
as indicated by the gray areas in fig.\ \ref{fig:A3}.
They are given by (\ref{eq:Om0}) and its generalization to other
momentum fractions:
\begin {equation}
   \Omega_0 = \Omega_{-1,x,1{-}x} ,
   \quad
   \Omega_\fx = \Omega_{-(1-y),x,z} ,
\end {equation}
where
\begin {equation}
   \Omega_{\zeta_1,\zeta_2,\zeta_3}
   = \sqrt{ 
     -\frac{i \hat q_{\rm A}}{2E}
     \left( \frac{1}{\zeta_1} + \frac{1}{\zeta_2} + \frac{1}{\zeta_3} \right)
   } .
\label {eq:Om3}
\end {equation}

Note that $\xi$ in (\ref{eq:sxitau})
is proportional to $y$ and so is small in the
small-$y$ limit, but
$s$ is order $y^0$. The rescaled time variable
$\tau$ is order $y^0$ for the case
$\Delta t \sim y$ we are currently working on.
The leading terms in (\ref{eq:DA3}) are written in red
(or with a red-colored factor) and will
give rise, after integration over $\Delta t$, to the spurious
$y^{-2}$ divergence mentioned earlier.
The other terms (without red) will give rise to $O(y^{-1})$ terms
after integration over $\Delta t$ and so will contribute to IR
logarithms.

We should explain a technical point at this juncture:
Why have we not fully expanded the small-$y$ dependence of
(\ref{eq:DA3})?  Note, for instance, the overall factor
$(1{-}y)$.  If we wanted to, for example, we could have expanded
\begin {equation}
  (1{-}y) \times \red{ \frac{1}{1+\tau} }
  \simeq
  \red{ \frac{1}{1+\tau} } - \frac{y}{1+\tau} \,,
\label{eq:dumbexpansion}
\end {equation}
in (\ref{eq:sxitau}) in order to more clearly separate leading and
sub-leading terms.  It turns out
that leaving (\ref{eq:dumbexpansion})
unexpanded not only makes the expression (\ref{eq:DA3}) more
compact but will also later simplify handling sub-leading
terms in the cancellation of the spurious
leading divergence with the pole terms in (\ref{eq:form2}).
For similar reasons, it is advantageous to avoid expanding the
overall factor of $\gamma=\gamma(x,y)$, as well as the
factors of $1{-}y$ and $1{+}\tau$ in the argument of the logarithm
in (\ref{eq:DA3}) [noting that the definition
(\ref{eq:sxitau}) of $\tau$ depends on $y$ through the definitions
of $\xi$ and $\Omega_\fx$ (and $z \equiv 1{-}x{-}y$)].

The next step is to extract the vacuum limit of (\ref{eq:DA3}) so that
we can make the vacuum subtraction (\ref{eq:Cdef}).
Since harmonic oscillator
frequencies $\Omega$ are proportional to $\sqrt{\hat q}$,
the vacuum limit for fixed $\Delta t$ is the $\tau \to 0$ limit
by virtue of the definitions (\ref{eq:sxitau}).
For A3, the vacuum-subtracted version $C$ of (\ref{eq:DA3}) is then
\begin {multline}
  C
  \simeq
  - \frac{\CA^2\alphas^2}{8\pi^2}
  (xyz)^2(1{-}x)(1{-}y)
  \gamma
\\ \times
  \frac{1}{(\Delta t)^2}
  \biggl\{
     ( \red{1} + \xi )
         \left[ \ln(1+\tau) - \frac{\tau}{1+\tau} \right]
     - 2 s\xi\tau^2 \ln\bigl(2\xi(1+\tau)\bigr)
\\
     + \xi
       \left[
          - \frac{\tau}{1+\tau} 
          - \frac{(1+s)\tau^3}{(1+\tau)^2}
       \right]
  \biggr\} ,
\end {multline}
where we've reorganized terms a bit.  We've also used
\begin {equation}
   \frac{x y}{(1{-}x)(1{-}y)}
   = 2\xi \, [1 + O(y)]
\end {equation}
to simplify the argument of the (sub-leading)
$s\xi\tau^2 \ln\bigl(\cdots\bigr)$ term.

Next, we need to integrate $\Delta t$ over $[a,(\Delta t)_*]$,
as in the relevant term of (\ref{eq:form2}).
Changing integration variables from $\Delta t$ to $\tau$ gives
\begin {multline}
 \int_a^{(\Delta t)_*} d(\Delta t)\>2 \Re C
 \simeq
\\
 - \frac{\CA^2\alphas^2}{8\pi^2} 
 (xyz)^2(1{-}x)(1{-}y)
 \gamma
 \Re\Biggl[
  i(\Omega_0{+}\Omega_\fx)
  \int_{\tau_a}^{\tau_*} d\tau \>
  \biggl\{
     ( \red{\tfrac{1}{\xi}} + 1 )
         \left[ \frac{\ln(1+\tau)}{\tau^2} - \frac{1}{\tau(1+\tau)} \right]
\\
     - 2 s \ln\bigl(2\xi(1+\tau)\bigr)
     - \frac{1}{\tau(1+\tau)} 
     - \frac{(1+s)\tau}{(1+\tau)^2}
  \biggr\}
 \Biggr] .
\label {eq:int0CA3}
\end {multline}
From (\ref{eq:t*region}) and (\ref{eq:sxitau}),
the rescaled dividing time $\tau_*$ is large $(\tau_* \gg 1)$
and in the range
\begin {equation}
   y^{-1/3} \ll \tau_* \ll y^{-1/2} .
\label {eq:tau*region}
\end {equation}
The rescaled UV cut-off $\tau_a$ should be treated as arbitrarily small.

The indefinite version $\int d\tau \> \{ \cdots \}$ of the integral
gives
\begin {equation}
  {\mathfrak c}(\tau)
  \equiv
     - ( \red{\tfrac{1}{\xi}} + 1 )
         \frac{\ln(1+\tau)}{\tau}
     - 2 s \tau \ln\bigl(2\xi(1+\tau)\bigr)
     + 2 s \tau
     - \ln\tau
     - 3s\ln(1+\tau)
     - \frac{1+s}{1+\tau}
  \,.
\label {eq:ctau}
\end {equation}
Then (\ref{eq:int0CA3}) becomes
\begin {multline}
 \int_a^{(\Delta t)_*} d(\Delta t)\>2\Re C
 \simeq
 - \frac{\CA^2\alphas^2}{8\pi^2}
 (xyz)^2(1{-}x)(1{-}y)
 \gamma
\\ \times
 \Re\Bigl[
  i(\Omega_0{+}\Omega_\fx)
  \bigl\{
    \red{\tfrac{1}{\xi}}
    + 2
    + s
    + \ln\bigl( \tfrac{i\Omega_0 a}{\xi} \bigr)
    + {\mathfrak c}(\tau_*)
  \bigr\}
 \Bigr] ,
\label {eq:intC1A3}
\end {multline}
where, expanding for large $\tau_*$ with (\ref{eq:tau*region}),
\begin {equation}
  {\mathfrak c}(\tau_*) \simeq
  - \frac{1}{\xi} \left(
        \frac{\ln\tau_*}{\tau_*} + \frac{1}{\tau_*^2}
    \right)
  - 2 s \tau_* \ln(2\xi\tau_*)
  + 2 s \tau_*
  - (1+3s) \ln\tau_*
  - 2s
\label {eq:cbigstar}
\end {equation}
up to corrections parametrically smaller than $1$.
Those corrections are too small to affect IR logarithms because
the overall factors multiplying ${\mathfrak c}(\tau_*)$
are of order $y^2\gamma = O(y^{-1})$, so that any
further suppression corresponds to IR-convergent additions to
the small-$y$ expansion (\ref{eq:smallybehavior}).

The red $1/\xi$ term in (\ref{eq:intC1A3}) gives
a contribution of order $y^2\gamma/\xi = O(y^{-2})$ to
(\ref{eq:intC1A3}).
This is the spurious $y^{-2}$ divergence mentioned earlier.

% ---------------------------------------------------------------------------

\subsection {UV piece of the A3 diagram}

The dimensionally-regulated
UV contribution to A3 in (\ref{eq:form2}) can be assembled from
formulas in ref.\ \cite{qcd}.
In appendix \ref{app:A3pole}, we display the generic-$y$ formula for this
contribution and then expand in $y$.  The result, through
next-to-leading order in $y$, is
\begin {align}
  \left[ \frac{d\Gamma}{dx\,dy} \right]_{\rm A3}^{(\Delta t<a)}
  \equiv {}&
  2\Re \left[ \frac{d\Gamma}{dx\,dy} \right]_{xy\bar y\bar x}^{(\Delta t<a)}
\nonumber\\
   \simeq{}&
   \frac{\CA^2 \alphas^2}{8\pi^2}
   (x y z)^2 (1{-}x) (1{-}y) \gamma
   \Re\biggl[
     (i\Omega_0 + i\Omega_\fx)
\nonumber\\ & ~ \times
     \left\{
        \red{\frac{1}{\xi}}
        + \frac{2}{\eps}
        + \ln\Bigl(\frac{\mu^4 a}{i\Omega_0 E^2}\Bigr)
        + 1
        + \ln(2\pi^2)
        - \ln(e^{-i\pi} x y z)
     \right\}
   \biggr]
  .
\label {eq:A3poley}
\end {align}
We again find it convenient to leave some elements unexpanded,
such as $\gamma$ and $\Omega_\fx$.

If we add (\ref{eq:A3poley}) to (\ref{eq:intC1A3}), the spurious
$y^{-2}$ divergences (the red terms) cancel.
With them out of the way,
one may then use the {\it leading} approximation (\ref{eq:gammaapprox})
to $\gamma$.  The result for the total
$\Delta t < (\Delta t)_*$ contribution to A3 is then
\begin {multline}
 \left[ \frac{d\Gamma}{dx\,dy} \right]_{\rm A3}^{(\Delta t<a)}
 +
 \int_a^{(\Delta t)_*} d(\Delta t)\>2\Re C
 \simeq
\\
 \frac{\CA\alphas^2\,P(x)}{2\pi^2 y}
 \Re\Biggl[
  i\Omega_0
  \biggl\{
    2 \Bigl[ \frac{1}{\eps} + \ln\Bigl(\frac{\pi\mu^2}{\Omega_0 E}\Bigr) \Bigl]
    - 1 + s - 2\ln(1{-}x) - \kappa(\tau_*)
  \biggr\}
 \Biggr] ,
\label {eq:intC1A3+pole}
\end {multline}
where we find it convenient to isolate the $\tau_*$-dependent terms
of (\ref{eq:cbigstar}) as
\begin {equation}
  \kappa(\tau_*) \equiv
  - \frac{1}{\xi} \left(
        \frac{\ln\tau_*}{\tau_*} + \frac{1}{\tau_*^2}
    \right)
  - 2 s \tau_* \ln(2\xi\tau_*)
  + 2 s \tau_*
  - (1+3s) \ln\tau_* .
\label {eq:kappa}
\end {equation}
Note that the dependence on $a$ has canceled in (\ref{eq:intC1A3+pole}),
as it must.

% ---------------------------------------------------------------------------

\subsection {$\Delta t \sim \sqrt{y}$ contribution to A3 diagram}
\label {sec:A3sqrty}

Turn now to the last term in (\ref{eq:form2}).
We repeat the procedure of section \ref{sec:A3y}, but now with
time scale $\Delta t \sim \sqrt{y}$ instead of $\Delta t \sim y$.
Like before, we need a corresponding small-$y$ approximation of
the integrand $2\Re C$.  We again start with the function
$D$ of (\ref{eq:Cdef}).
Details of the expansion
are summarized in appendix \ref{app:A3sqrty}, with result
\begin {equation}
  D
  \simeq
  - \frac{\CA\alphas^2\,P(x)}{4\pi^2 y} \,
  \biggl[
      (\Omega_y\csc_y)^2 \ln S
    + \frac{\xi \Omega_y^3 \Delta t}{2 S} \csc_y (\csc_y+\cot_y)^2
  \biggr] ,
\label {eq:DDA3}
\end {equation}
where
\begin {equation}
  S
  \simeq
  \red{2 i \Omega_0 \, \Delta t}
    + (2+3s)(\Omega_0\,\Delta t)^2
    + \xi \Omega_y (\csc_y + \cot_y) \Delta t
\label {eq:S}
\end {equation}
and we introduce the short-hand notation
\begin {equation}
   \operatorname{trig}_y
   \equiv
   \operatorname{trig}(\Omega_y\,\Delta t) .
\end {equation}
The red term in (\ref{eq:S}) indicates the leading-order behavior,
which in this case will generate a $y^{-3/2}$ power-law divergence
like in (\ref{eq:smallybehavior}), not a $y^{-2}$ divergence like
in section \ref{sec:A3y}.%
\footnote{
  Because of this, one may dispense with the exact definition
  of $\xi$ from (\ref{eq:sxitau}) and just take
  $\xi \simeq x y/2(1{-}x)$ here.
}

A caution is needed: it may be tempting to immediately
expand the $\ln S$ of (\ref{eq:DDA3}) using (\ref{eq:S}).
The caution is that we need to subtract out the vacuum result for
$D$ in (\ref{eq:Cdef}), but the red term in (\ref{eq:S})
{\it vanishes} in vacuum (where $\hat q$ and so $\Omega_0$ are zero).
That means that the small-$y$ expansion of $S$ looks different
in vacuum:
\begin {equation}
   \lim_{{\hat q}\to0} S \simeq 2\xi .
\end {equation}
The vacuum subtracted version of $D$ is then
\begin {equation}
  C
  \simeq
  - \frac{\CA\alphas^2\,P(x)}{4\pi^2 y}
  \biggl\{
      (\Omega_y\csc_y)^2 \ln S
    - \frac{i \xi \Omega_y^3}{4 \Omega_0} \csc_y (\csc_y+\cot_y)^2
    - \frac{\ln(2\xi)}{(\Delta t)^2}
    - \frac{1}{(\Delta t)^2}
  \biggr\} .
\end {equation}
Only now is it safe to expand the logarithm as
\begin {equation}
  \ln S \simeq
  \red{\ln(2 i \Omega_0 \, \Delta t)}
    - i(1+\tfrac32 s) \Omega_0\,\Delta t
    -  \frac{i\xi}{2\Omega_0} \Omega_y (\csc_y + \cot_y)
  .
\label {eq:lnSexpand}
\end {equation}
To integrate over $\Delta t$, it is convenient to switch integration
variables to
\begin {equation}
  \tilde\tau \equiv i \Omega_y \, \Delta t .
\label {eq:tildetau}
\end {equation}
Using
\begin {equation}
   \xi\Omega_y^2 = 6s\Omega_0^2 + O(y)
\label{eq:ssimp}
\end {equation}
[taken from (\ref{eq:Omy}) and (\ref{eq:sxitau})]
to simplify parts of the expression, we find
\begin {multline}
  \int_{(\Delta t)_*}^\infty d(\Delta t) \> 2\Re C
  \simeq
\\
  - \frac{\CA\alphas^2\,P(x)}{2\pi^2 y}
  \Re\int_{\tilde\tau_*}^\infty d\tilde\tau \>
  \biggl\{
     \red{i\Omega_y}
     \biggl[
       \frac{1}{\sh^2\tilde\tau}
       \ln\Bigl( \frac{2\Omega_0\tilde\tau}{\Omega_y} \Bigr)
       - \frac{[\ln(2\xi)+1]}{\tilde\tau^2}
     \biggr]
\\
     - i\Omega_0
     \biggl[
       \frac{(1{+}\tfrac32 s)\tilde\tau}{\sh^2\tilde\tau}
       -
       \frac{\tfrac32 s(1{+}\ch\tilde\tau)(3{+}\ch\tilde\tau)}{\sh^3\tilde\tau}
     \biggr]
  \biggr\} .
\label {eq:CCint0}
\end {multline}
The ``interesting'' region $\Delta t \sim \sqrt y$ of integration corresponds
to $\tilde\tau \sim 1$, and the lower cut-off $\tilde\tau_*$ on the integration
is parametrically small ($\tilde\tau_* \ll 1$) by virtue of
(\ref{eq:Omy}) and (\ref{eq:t*region}).
The integrals of the terms in (\ref{eq:CCint0}) are
then much simpler because we only need their small-$\tau_*$ expansion.

To integrate, we find it convenient to first re-organize the
terms multiplying red $i\Omega_y$ in (\ref{eq:CCint0}) as
\begin {equation}
   \frac{1}{\sh^2\tilde\tau}
     \ln\Bigl( \frac{2\Omega_0\tilde\tau}{\Omega_y} \Bigr)
   - \frac{\ln(2\xi)}{\tilde\tau^2}
   =
   \frac{1}{\tilde\tau^2}
      \ln\Bigl( \frac{\Omega_0\tilde\tau}{\xi\Omega_y} \Bigr)
   + \Bigl(\frac{1}{\sh^2\tilde\tau} - \frac{1}{\tilde\tau^2}\Bigr)
         \ln\Bigl( \frac{2\Omega_0\tilde\tau}{\Omega_y} \Bigr) .
\label {eq:rewrite}
\end {equation}
Almost all of the $\tilde\tau$ integrals have results in terms of
elementary functions, and one may then expand the result in powers
of $\tilde\tau_* \ll 1$.  The one integral that is slightly more complicated
is%
\footnote{
  To derive (\ref{eq:Z2intTrick}), first write
  $\int_{\tilde\tau_*}^\infty = \int_0^\infty - \int_0^{\tilde\tau_*}$.
  Then do the $\int_0^\infty$ integral exactly.  It is given by eq.\ (E.18)
  of ref.\ \cite{qcd}, which is derived in the corresponding paragraph of
  appendix B of ref.\ \cite{qcd}.
  For the $\int_0^{\tilde\tau_*}$ integral,
  expand the integrand in powers of small
  $\tilde\tau \le \tilde\tau_* \ll 1$ and
  integrate that expansion term by term to the desired order.
}
\begin {equation}
  \int_{\tilde\tau_*}^\infty d\tilde\tau \>
  \Bigl( \frac{1}{\sh^2\tilde\tau} - \frac{1}{\tilde\tau^2} \Bigr)
  \ln(c\tilde\tau)
  =
  -\ln(c\pi) + \gammaE
  + \tfrac13 \tilde\tau_* \bigl( \ln(c\tilde\tau_*) - 1 \bigr)
  + O( \tilde\tau_*^3 \ln\tau_* ) .
\label {eq:Z2intTrick}
\end {equation}
The integration in (\ref{eq:CCint0}) yields
\begin {align}
  \int_{(\Delta t)_*}^\infty d&(\Delta t) \> 2\Re C(\Delta t)
  \simeq
\nonumber\\ &
  \frac{\CA\alphas^2\,P(x)}{2\pi^2 y}
  \Re
  \biggl[
    \red{i\Omega_y}
    \left\{
       \ln\Bigl( \frac{2\pi\Omega_0}{\Omega_y} \Bigr) - \gammaE
    \right\}
\nonumber\\ & \qquad\qquad\qquad
    + i\Omega_0
    \left\{
      1
      - \ln2
      + 2 s
      - (1{+}3 s) 
          \ln\Bigl( \frac{\xi\Omega_y}{\Omega_0} \Bigr)
      + \kappa(\tau_*)
    \right\}
  \biggr] ,
\label {eq:CCint}
\end {align}
where $\kappa(\tau_*)$ is again (\ref{eq:kappa}).
Though it was convenient to set up the original integral
(\ref{eq:CCint}) in terms of the small parameter $\tilde\tau_*$,
it's nonetheless convenient to have written the
final result (\ref{eq:CCint}) in
terms of the $\tau_*$ instead, converting between
them using their definitions (\ref{eq:sxitau}) and (\ref{eq:tildetau}).
This way, we will be able to easily see the cancellation of
the splitting-time dependence $\kappa(\tau_*)$ when summing all
contributions to A3.

% ---------------------------------------------------------------------------

\subsection {A3 result}

The total result (\ref{eq:form2}) for the small-$y$ expansion of A3
is the sum of (\ref{eq:intC1A3+pole}) and (\ref{eq:CCint}):
\begin {align}
  \left[ \frac{d\Gamma}{dx\,dy} \right]_{\rm A3}
  \simeq
  \frac{\CA\alphas^2\,P(x)}{2\pi^2 y}
  \Re &
  \biggl[
    \red{i\Omega_y}
    \Bigl\{
       \ln\left( \tfrac{2\pi\Omega_0}{\Omega_y} \right) - \gammaE
    \Bigr\}
\nonumber\\ &
    + i\Omega_0
    \biggl\{
      2 \Bigl[
        \tfrac{1}{\eps} + \ln\left(\tfrac{\pi\mu^2}{\Omega_0 E}\right)
      \Bigr]
      - (1{+}3s) \ln\left( \tfrac{\xi\Omega_y}{\Omega_0} \right)
\nonumber\\ & \qquad\qquad
      - 2\ln(1{-}x)
      - \ln2
      + 3 s
    \biggr\}
  \biggr] .
\end {align}
Using the definitions (\ref{eq:sxitau}) and some algebra,
this result demonstrates the A3 case ($\di,\dj=1,3$)
of the more general formula
(\ref{eq:iNEj}) presented earlier.

% ============================================================================

\section{Other \boldmath$\di{\not=}\dj$ ABC diagrams}
\label{sec:ij}

\begin {figure}[t]
\begin {center}
  \includegraphics[scale=0.60]{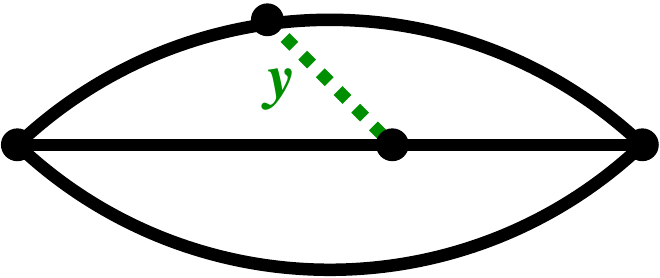}
  \caption{
     \label{fig:iNEj}
     If colors (red vs.\ blue) are not shown, and if the momentum fractions
     associated with the underlying $g{\to gg}$ process (black lines above)
     are not shown, then all of the ABC diagrams can be drawn in the
     same form, shown above.
  }
\end {center}
\end {figure}

Return now to the depiction in fig.\ \ref{fig:ABCdiags} of the
ABC diagrams.  Ignoring the color differences between red and blue
lines, and the label ``$x$'', all the diagrams look the same in
the case $\di{\not=}\dj$\,: they can all be drawn in the form of
fig.\ \ref{fig:iNEj}.
So one might guess that the results for these diagrams are all
related by permutations of the momentum fractions (\ref{eq:altx})
of the three lines of the underlying, hard $g{\to}gg$ splitting
process.  The one fly in the ointment is that, as mentioned
in section \ref{sec:ABCsummary}, there {\it are} differences
between blue (amplitude) and red (conjugate amplitude) lines,
arising from complex phases.  As in the discussion of
section \ref{sec:cancel},
these phases can give rise to ``$i \pi$'' terms in our interference
diagrams which,
when multiplied by other phases, give rise to ``$\pi$'' terms when
one takes $2\Re(\cdots)$ to add in the interference diagram's
complex conjugate.  Starting from our result for A3, written in
the form of the $\di,\dj=1,3$ case of (\ref{eq:iNEj}), one might
{\it guess} that the $\di{\not=}\dj$ ABC diagrams as a group give
\begin {multline}
  \left[ \frac{d\Gamma}{dx\,dy} \right]_{\di\dj}
  \simeq
  \frac{\CA\alphas^2\,P(x)}{2\pi^2 y}
  \Re
  \biggl[
    \red{i\Omega_y}
    \Bigl\{
       - \ln\bigl( \tfrac{\Omega_y}{2\pi\Omega_0} \bigr) - \gammaE
       + \pi \times \purple{\qmark_{\di\dj}}
    \Bigr\}
\\
    + \blue{i\Omega_0}
    \biggl\{
      2 \Bigl[
        \tfrac{1}{\eps} + \ln\bigl(\tfrac{\pi\mu^2}{\Omega_0 E}\bigr)
      \Bigr]
      - \bigl(1{+}\tfrac{\altx_\dk^2}{2(\altx_1^2+\altx_2^2+\altx_3^2)}\bigr)
        \left[
          \ln\bigl(
             \tfrac{|\altx_\di \altx_\dj \altx_\dk| y\Omega_y}{\Omega_0}
          \bigr)
          + \pi \times \purple{\qmark_{\di\dj}}
        \right]
\\
      + \tfrac{\altx_\dk^2}{2(\altx_1^2+\altx_2^2+\altx_3^2)}
          \bigl[ 1-\ln2+2\ln(2|\altx_\di \altx_\dj|) \bigr]
    \biggr\}
  \biggr]
\label {eq:iNEjguess}
\end {multline}
for some $(\di,\dj)$-dependent
values of the question marks ``$\qmark_{\di\dj}$'' above.
In our guess, we have only included such $\pi$ terms
in the case of logarithms whose arguments depend on $y$, since it
is the color of the $y$ line that can change between different
ABC diagrams.

The goal of this paper is to simply find the formula for the single
logs, in the most efficient way we can manage.
Rather than derive (\ref{eq:iNEjguess}) and carefully analyze every
branch cut and complex phase, the most efficient method for us was
to simply take (\ref{eq:iNEjguess}) as an ansatz and then use numerical
extraction of the $y{\to}0$ limits of generic-$y$ formulas
\cite{qcd} to determine the question marks,
which we expect to be simple fractions.
Doing that produces and verifies our final results (\ref{eq:iNEj}).%
\footnote{
  The $\pi$ terms associated with the power-law divergences
  [the red $i\Omega_y$ terms in (\ref{eq:iNEjguess})] were
  previously extracted in appendix E of ref. \cite{qcd}.
  Some (A3,B3,C1,C2) were extracted analytically, but the derivation was
  only shown for the case A3.  Other $\pi$ terms (A2,B1) were determined
  by numerics.  (See in particular table 1 of ref.\ \cite{qcd}, where
  bolded entries indicate an analytic derivation.)
  With the formulas for the power-law divergences then in hand, one
  may subtract all of (\ref{eq:iNEjguess})
  except for the $i\Omega_0 \pi \times\qmark_{\di\dj}$ from the generic-$y$
  numerical result for each diagram, and then
  numerically determine that remaining $i\Omega_0 \pi \times\qmark_{\di\dj}$
  term for small $y$.  (We went down to
  $y \sim 10^{-10}$ for these numerical extractions, which requires using
  very-high precision arithmetic.)
}

Most of the $\di{\not=}\dj$ ABC diagrams are related to each other
(even for $y$ not small)
through combinations of front- and back-end transformations and/or
changing $x \to 1{-}x$.  These relations are shown in
table \ref{tab:otherABC}.
They provide another way to understand
the symmetry of (\ref{eq:iNEjguess}) and could in principle be
used to analytically derive most of the $i\pi$ terms if one were
willing to carefully repeat the extraction of the small-$y$ limit
of A3 in a way consistent with front-end transformation,
keeping track of all the relevant phases and branch cuts.

 \begin{table}[t]
%\vspace{5mm}
\begin {center}
\begin{tabular}{l}
\hline\hline
\noalign{\vskip 0.3em}
  ${\rm B3} = {\rm A3}|_{x\to1-x}$ \\
  ${\rm C1} = {\rm bkEnd}[ {\rm frEnd}({\rm A3}) ]^*$ \\
  ${\rm C2} = {\rm C1}|_{x\to1-x}$ \\
  ${\rm A2} = {\rm B1}|_{x\to1-x}$ \\
%  ${\rm B1}$ ~~ \mbox{[see text]} \\
\noalign{\vskip 0.3em}
\hline\hline
\end{tabular}
\end {center}
\caption{
  \label{tab:otherABC}
  A sequence of relations to relate other $\di{\not=}\dj$ ABC diagrams to
  A3 and B1.
}
\end{table}

\begin {figure}[t]
\begin {center}
  \includegraphics[scale=0.40]{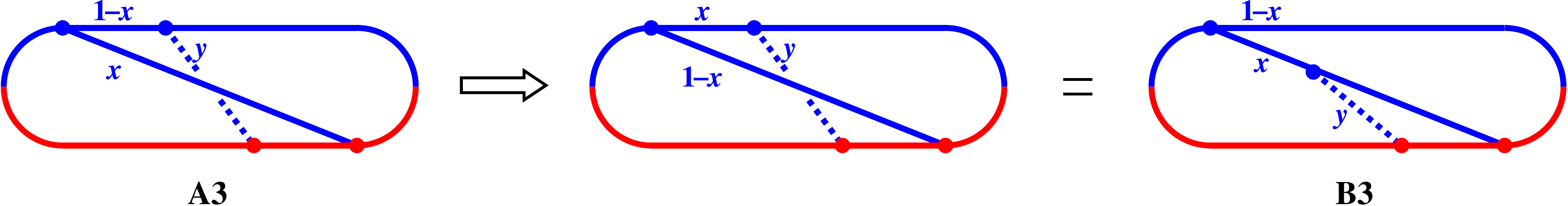}
  \caption{
     \label{fig:B3}
     Obtaining the B3 diagram from the A3 diagram by $x \to 1{-}x$.
  }
\end {center}
\end {figure}

The simplest relations are the $x \to 1{-}x$ relations in the table.
Consider B3 as an example.
If one changes $x \to 1{-}x$ in the A3 diagram, the result is
just a different way of drawing the B3 diagram, as shown in
fig.\ \ref{fig:B3}.
This permutes $(\altx_1,\altx_2,\altx_3) \equiv (1{-}x,x,-1)$
to $(\altx_2,\altx_1,\altx_3)$,
which exactly matches (\ref{eq:iNEj}) for the cases of A3 and B3.

\begin {figure}[t]
\begin {center}
  \includegraphics[scale=0.40]{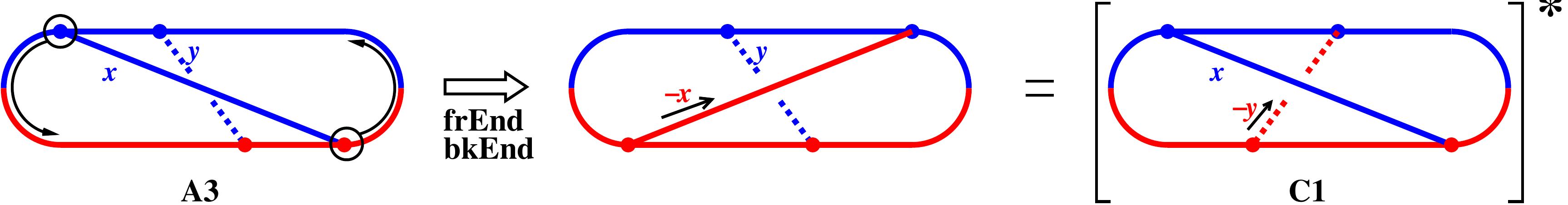}
  \caption{
     \label{fig:C1}
     Obtaining the C1 diagram from the A3 diagram by
     front- and back-end transformation and complex conjugation.
  }
\end {center}
\end {figure}

A more subtle example is the C1 diagram.  This can be obtained
from A3 by doing both a front-end and back-end transformation
and also complex conjugation (which exchanges blue and
red), as shown in fig.\ \ref{fig:C1}.
[The complex conjugation isn't important since the
results we quote in (\ref{eq:iNEj}) take $2\Re(\cdots)$ of the
diagrams shown, in order to include the complex conjugate diagrams.]
Here we are front-end transforming the $x$ emission vertex, unlike
the front-end transformation of the $y$ vertex discussed in the
context of (\ref{eq:frontendy}).  The corresponding transformation
is \cite{qcd}
\begin {equation}
  (x,y,E) \longrightarrow
  \Bigl( \frac{-x}{1{-}x} \,,\, \frac{y}{1{-}x} \,,\, (1{-}x)E \Bigr) .
\label {eq:frontendx}
\end {equation}
This takes longitudinal momenta
\begin {equation}
  \bigl((1{-}x)E, xE, -E\bigr)
  \to \bigl(E, -xE, -(1{-}x)E \bigr)
\end {equation}
and so, up to signs, corresponds to a permutation of the momenta of
the underlying, hard $g{\to}gg$ process.  This is how, in this way of
looking at things, the symmetry (except for $i\pi$ terms) of
(\ref{eq:iNEj}) comes about.

One could relate {\it all}\/ of the ABC diagrams using such transformations
if it were possible to add to table \ref{tab:otherABC} a rule
for B1 in terms of any other $\di{\ne}\dj$ ABC diagram.
Just looking at the momentum fractions of lines in the relevant
diagrams, it's possible that something like
${\rm B1} = \operatorname{bkEnd}[\operatorname{frEnd}({\rm B3})]|_{y{\to}-y}^*$
might work.  But we have not made any study of the validity
of this type of
transformation (which includes the extra step $y\to-y$); so
we leave it aside.

% ===========================================================================
% ===========================================================================

\section {The A1 diagram}
\label {sec:A1}

\subsection {Scales and form}

Analyzing the $y{\to}0$ expansion of the A1 diagram, which involves
a gluon self-energy loop in the amplitude,  has both similarities
and differences with the previous analysis of the A3 diagram.

One difference is that there is nothing special about the scale
$\Delta t \sim y$.  Instead, we return to the decomposition
\begin {equation}
  {\rm diagram}(x,y)
  =
  \lim_{a\to 0}
    \left[
      \int_0^a d(\Delta t) \> f(x,y,\Delta t)
      + \int_a^\infty d(\Delta t) \> f(x,y,\Delta t)
    \right]
\label {eq:form1B}
\end {equation}
of (\ref{eq:form1}), where the $\Delta t$ integral is split into
just a regularized UV contribution ($\Delta t < a$) and everything
else.  The only interesting physics scale for this diagram is
$\Delta t \sim t_{\rm form}(y)$, which in our parlance is $t \sim y^{1/2}$.

\begin {figure}[t]
\begin {center}
  \includegraphics[scale=0.60]{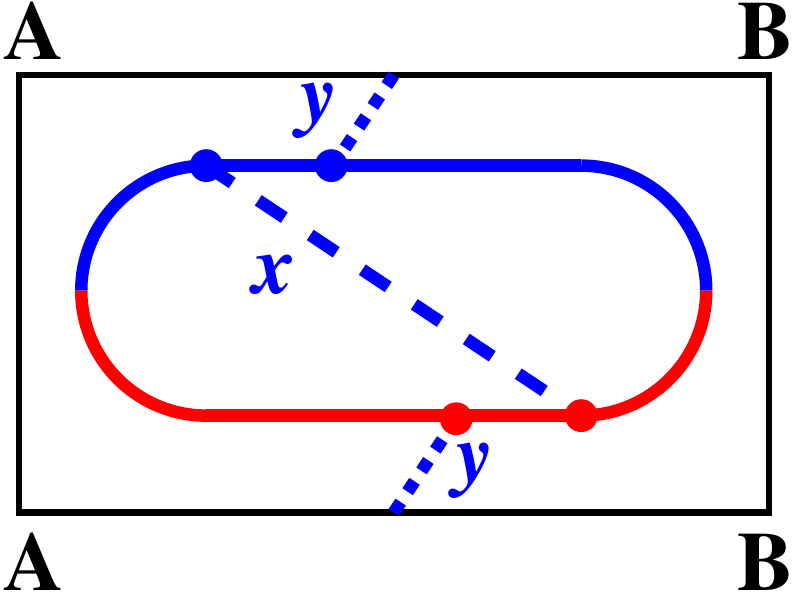}
  \caption{
     \label{fig:A3cylinder}
     The A3 diagram drawn on a cylinder,
     as in section 4.3 of ref.\ \cite{2brem}.
  }
\end {center}
\end {figure}

The other difference is that the generic-$y$ formula for the A1 diagram
\cite{qcd}, calculated in the large-$\Nc$ limit, has contributions from
{\it two} different large-$\Nc$ color routings, depicted in
fig.\ \ref{fig:A1}.%
\footnote{
  For a brief discussion of
  specifics concerning the A1 diagram $xyy\bar x$, see
  appendix D.4 of ref.\ \cite{qcd}.
}
This is in contrast to the A3 diagram (fig.\ \ref{fig:A3cylinder}),
which has only one possible large-$\Nc$ color routing.

\begin {figure}[t]
\begin {center}
  \includegraphics[scale=0.60]{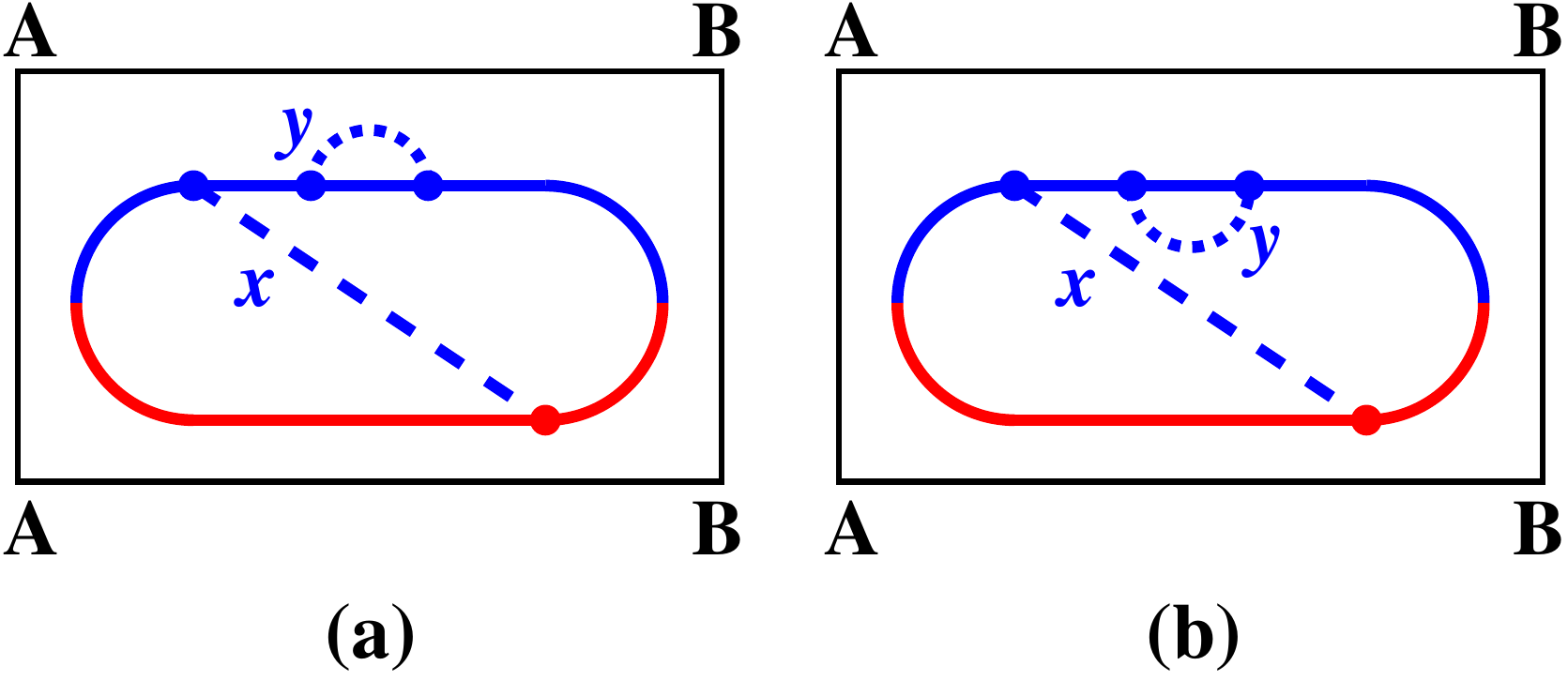}
  \caption{
     \label{fig:A1}
     The two large-$\Nc$ color routings of the A1 diagram, analogous
     to the treatment of the $xy\bar x\bar y$ diagram in section 2.2.1
     of ref.\ \cite{seq}.
     Here we imagine drawing our time-ordered large-$\Nc$ diagrams on a
     cylinder, as in ref.\ \cite{seq}: The top edge AB of the
     each rectangle should be identified with the bottom edge AB.
     In the large-$\Nc$ limit,
     medium interactions (not drawn) are correlated between each pair of
     high-energy particle lines that are neighbors as one
     circles the cylinder.
  }
\end {center}
\end {figure}

The sum of the two color routings is written in the generic-$y$
formulation of ref.\ \cite{qcd} as
\begin {equation}
  2\Re \left[ \frac{d\Gamma}{dx\, dy} \right]_{x y y\bar x}
  =
  \tfrac12 \bigl[ A_\new(x,y) + A_\new(x,z) \bigr] .
\label {eq:dGfundQCD}
\end {equation}
where the explicit factor of $\tfrac12$ represents the symmetry factor
for the amplitude loop of the A1 diagram (the blue
gluon self-energy loop in
fig.\ \ref{fig:A1}). Eq.\ (\ref{eq:dGfundQCD}) is the appropriate
normalization if one intends to integrate over {\it all}\/ values
of the loop momentum fraction $y$, which for the A1 diagram would
be $\int_0^{1-x} dy$.  However, in our discussion, we have used
the letter $y$ to always represent the {\it softest} gluon.
So, we should instead think of the loop integral as running
only over {\it half} the values, as $\int_0^{(1-x)/2} dy$ [with only
$y{\to}0$ now corresponding to a virtual gluon becoming soft], and so
we compensate by normalizing our ``differential rate'' in this paper as
\begin {equation}
  \left[ \frac{d\Gamma}{dx\, dy} \right]_{\rm A1}
  \equiv
  2\Re \left[ \frac{d\Gamma}{dx\, dy} \right]_{x y y\bar x}
  \equiv
  A_\new(x,y) + A_\new(x,z)
\label {eq:dGfund}
\end {equation}
instead of (\ref{eq:dGfundQCD}).

In what follows, we will first focus on color routing (a) of
fig.\ \ref{fig:A1} and the calculation of the small-$y$ limit
of $A_\new(x,y)$.
We'll later write the answer in a general form that will allow
us to obtain the small-$y$ limit of the other
color routing $A_\new(x,z) = A_\new(x,1{-}x{-}y)$
by a permutation symmetry.

Following the treatment in \cite{QEDnf,qcd},
we
will {\it not}\/ use the trick of subtracting out the vacuum piece
for the A1 diagram.%
\footnote{
  Similar to the discussion in footnote $\ref{foot:vacuum}$, the total
  vacuum piece for single splitting $g{\to}gg$ must vanish, in the
  context of this calculation.  That doesn't necessarily mean that
  the vacuum piece need
  vanish for each, individual {\it time-ordered}\/ diagram,
  such as A1, that contributes to $g{\to}gg$.  But it does vanish,
  even for individual diagrams, in dimensional regularization, since
  there is then no scale in vacuum that can
  make up the dimensions of the answer.
  So there is no harm in subtracting vacuum pieces for some diagrams
  and not for others, as convenient, provided one also calculates the
  ``pole terms'' (from the integral over $\Delta t < a$) correspondingly.
}
The analog of the A3 diagram's (\ref{eq:A3intC}) then has the form
\begin {equation}
  A_\new(x,y)
  =
   \int_0^\infty d(\Delta t) \>
   2\Re
   \widetilde D_{\rm new}(-1,y,z,x,\bar\alpha,\bar\beta,\bar\gamma,\Delta t) ,
\end {equation}
which should be split up, following (\ref{eq:form1B}), into pieces
\begin {equation}
  A_\new(x,y)
  =
  A_\new^{(\Delta t < a)}(x,y)
  +
   \int_a^\infty d(\Delta t) \>
   2\Re
   \widetilde D_{\rm new}(-1,y,z,x,\bar\alpha,\bar\beta,\bar\gamma,\Delta t)
  .
\label {eq:Anewsplit}
\end {equation}
The formulas for the relevant combinations
$(\bar\alpha,\bar\beta,\bar\gamma)$ of helicity-dependent DGLAP
splitting functions
are given in appendix A of ref.\ \cite{qcd}.
In the small-$y$ limit, $\bar\alpha$ dominates over $\bar\beta$ and
$\bar\gamma$ with
\begin {equation}
  \bar\alpha = \frac{2\,P(x)}{\CA x^2(1{-}x)^4 y^3} \, \bigl[ 1 + O(y) \bigr],
\label {eq:baralphaapprox}
\end {equation}
analogous to the case of (\ref{eq:gammaapprox}) for the A3 diagram.

There is one modification of our $\widetilde D_\new$ here compared to the
$D_\new$ of ref.\ \cite{qcd}.
Unlike refs.\ \cite{QEDnf,qcd}, we will be doing the integral
in the last term of
(\ref{eq:form1B}) analytically rather than numerically, and so we
will not have to make additional subtractions to that integral to make
it better behaved for numerical integration.  This means leaving out
what were called ``${\cal D}_2$'' subtractions that were implemented in
ref.\ \cite{QEDnf,qcd}.%
\footnote{
  The ${\cal D}_2$ subtractions
  are explained in section 4.3.2 of
  ref.\ \cite{QEDnf} and appendix D of ref.\ \cite{qcd}.
  Our omission of them corresponds to leaving out the last term
  in eq.\ (A.64) of ref.\ \cite{qcd} and modifying pole terms accordingly.
  In particular, our $A_\new^{(\Delta t < a)}$ in (\ref{eq:Anewsplit}) above
  is a modified version of the $A_\new^{\rm pole}$ in eqs.\ (A.62) and
  (A.66) of ref.\ \cite{qcd}, so that $A_\new(x,y)$ will remain the same.
}
Leaving out those additional subtractions will slightly
simplify the small-$y$ analysis that we want to do here.

% ---------------------------------------------------------------------------

\subsection {$\Delta t \sim \sqrt{y}$ contribution to A1 routing (a)}
\label {sec:A1y}

We first need an expansion in $\Delta t \sim \sqrt{y}$
for the $\Delta t$-integrand,
similar to (\ref{eq:DDA3}).
Details are given in appendix \ref{app:A1sqrty}, with result,
now for the A1 diagram,
\begin {equation}
  \widetilde D_\new
  \simeq
  \frac{\CA\alphas^2\,P(x)}{4\pi^2 y} \,
  (\Omega_y\csc_y)^2
  \biggl[
    \ln S_\new
    - \frac{\xi \Omega_y}{4 i \Omega_0} \frac{(\csc_y-\cot_y)^2}{\csc_y}
  \biggr] ,
\label {eq:DDA1}
\end {equation}
where
\begin {equation}
  S_\new
  \simeq
  \red{2 i \Omega_0 \, \Delta t}
    + (2+3s)(\Omega_0\,\Delta t)^2
    + \xi \Omega_y (-\csc_y + \cot_y) \Delta t .
\label {eq:Snew}
\end {equation}
The only difference between $S_\new$ above and the $S$ of (\ref{eq:S})
is the minus sign on $\csc_y$ in the last term.
$s$ is the same as in (\ref{eq:sxitau}).
$\xi$ is also the same, but
its leading-order expression $\xi \simeq xy/2(1{-}x)$ will
be adequate in the current context.

Expanding $\ln S_\new$ in $y$, switching integration variables
to the $\tilde\tau$ of (\ref{eq:tildetau}), and again using
(\ref{eq:ssimp}), the
$[a,\infty]$ integral of
(\ref{eq:form1B}) is
\begin {multline}
  \int_a^\infty d(\Delta t) \> 2\Re \widetilde D_\new(\Delta t)
  \simeq
\\
  \frac{\CA\alphas^2\,P(x)}{2\pi^2 y}
  \Re\int_{\tilde\tau_a}^\infty d\tilde\tau \>
  \biggl\{
     \red{i\Omega_y}
     \biggl[
       \frac{1}{\sh^2\tilde\tau}
       \ln\Bigl( \frac{2\Omega_0\tilde\tau}{\Omega_y} \Bigr)
     \biggr]
%\\
     - i\Omega_0
     \biggl[
       \frac{(1{+}\tfrac32 s)\tilde\tau}{\sh^2\tilde\tau}
       + \frac{\tfrac32 s(1-\ch\tilde\tau)(3-\ch\tilde\tau)}{\sh^3\tilde\tau}
     \biggr]
  \biggr\} ,
\end {multline}
where $\tilde\tau_a = i \Omega_y a$ is arbitrarily small.
Performing the integration as in section \ref{sec:A3sqrty}, again using
(\ref{eq:rewrite}) and (\ref{eq:Z2intTrick}),
\begin {align}
  \int_{a}^\infty d(\Delta t) \> 2 \Re \widetilde D_\new
  \simeq
  \frac{\CA\alphas^2\,P(x)}{2\pi^2 y}
  \Re
  &
  \biggl[
    \frac{\ln(2i\Omega_0 a) + 1}{a}
    + {\red{ i\Omega_y }}
    \left\{
       - \ln\Bigl( \frac{2\pi\Omega_0}{\Omega_y} \Bigr) + \gammaE
    \right\}
\nonumber\\ &
    + i\Omega_0
    \left\{
      \ln(i\Omega_0 a)
      + \ln\Bigl( \frac{2\Omega_y}{\Omega_0} \Bigr)
      - 1
      + 3s(\ln2 - 1)
    \right\}
  \biggr] ,
\label {eq:AnewFinite}
\end {align}
ignoring corrections that vanish as $a\to0$.

% ---------------------------------------------------------------------------

\subsection {UV piece and total A1 routing (a)}

The dimensionally-regulated contribution corresponding to the
$A_\new^{(\Delta t<a)}$ term in (\ref{eq:Anewsplit}) can be assembled
from formulas in refs.\ \cite{QEDnf,qcd}.
In appendix \ref{app:A1pole}, we discuss the generic-$y$ formula and
then expand in $y$.  The result is
\begin {multline}
  A_\new^{(\Delta t < a)} \simeq
    \frac{\CA \alphas^2 \, P(x)}{2\pi^2 y} \,
    \Re \biggl[
       - \frac{\ln(2 i\Omega_0 a) + 1}{a}
       + i \Omega_0
       \biggl\{
         - 2\left(
               \frac{1}{\eps}
               + \ln\Bigl(\frac{\pi\mu^2}{E\Omega_0} \Bigr)
            \right)
         + \ln\bigl(x y (1{-}x) \bigr)
\\
         - \ln(i\Omega_0 a) - \ln 2 + 1
       \biggr\}
    \biggr] .
\label {eq:ApolenewNoD2LO}
\end {multline}
Adding this to (\ref{eq:AnewFinite}), the dependence on the UV
separation scale $a$ cancels, as it must.  We are left with
\begin {multline}
  A_\new(x,y)
  \simeq
  \frac{\CA\alphas^2\,P(x)}{2\pi^2 y}
  \Re
  \biggl[
    {\red{ i\Omega_y }}
    \left\{
       \ln\Bigl( \frac{\Omega_y}{2\pi\Omega_0} \Bigr)
       + \gammaE
    \right\}
\\
    + i\Omega_0
    \left\{
     - 2
       \left(
          \frac{1}{\eps}
          + \ln\Bigl(\frac{\pi\mu^2}{E\Omega_0} \Bigr)
       \right)
     + \ln\bigl(x y (1{-}x) \bigr)
     + \ln\Bigl( \frac{\Omega_y}{\Omega_0} \Bigr)
     + 3s (\ln2 - 1)
    \right\}
  \biggr] .
\label {eq:AnewResult}
\end {multline}

% ---------------------------------------------------------------------------

\subsection {The other color routing}

Using the longitudinal
momentum fractions $(\altx_1,\altx_2,\altx_3) \equiv (1{-}x,x,-1)$
of the underlying hard single-splitting process, we can algebraically
manipulate (\ref{eq:AnewResult}) into a form similar to
(\ref{eq:ii}):
\begin {multline}
  A_\new(x,y)
  \simeq
  -
  \frac{\CA\alphas^2\,P(x)}{2\pi^2 y}
  \Re
  \biggl[
    \red{i\Omega_y}
    \Bigl\{
       - \ln\bigl( \tfrac{\Omega_y}{2\pi\Omega_0} \bigr) - \gammaE
    \Bigr\}
\\
    + \blue{i\Omega_0}
    \biggl\{
      2 \Bigl[
        \tfrac{1}{\eps} + \ln\bigl(\tfrac{\pi\mu^2}{\Omega_0 E}\bigr)
      \Bigr]
      - \ln\bigl(
             \tfrac{|\altx_1 \altx_2 \altx_3| y\Omega_y}{\Omega_0}
          \bigr)
      + \tfrac{\altx_2^2}{2(\altx_1^2+\altx_2^2+\altx_3^2)}
          (1 - \ln2)
    \biggr\}
  \biggr] .
\label {eq:Anew11a}
\end {multline}
Now note that in the color routing for this result, fig.\ \ref{fig:A1}a,
the soft $y$ gluon's neighbors going around the cylinder are
$\altx_1=1{-}x$ (the line that the $y$ gluon leaves and reconnects to)
and $\altx_3=-1$.  And note that the $(\altx_1,\altx_2,\altx_3)$
appear symmetrically in (\ref{eq:Anew11a}) {\it except}\/ for the
$\altx_2^2$ factor in the last term.

Compare that to the other color routing, fig.\ \ref{fig:A1}b,
which corresponds to the $A_{\rm new}(x,z)$ in (\ref{eq:dGfund}).
Now the soft $y$ gluon's neighbors going around the cylinder are
$\altx_1=1{-}x$ (the line that the $y$ gluon leaves and reconnects to)
and $\altx_2=x$.  So we may expect that the result is the same
as (\ref{eq:Anew11a}) except with that $\altx_2^2$ factor replaced
by $\altx_3^2$:
\begin {multline}
  A_\new(x,z)
  \simeq
  -
  \frac{\CA\alphas^2\,P(x)}{2\pi^2 y}
  \Re
  \biggl[
    \red{i\Omega_y}
    \Bigl\{
       - \ln\bigl( \tfrac{\Omega_y}{2\pi\Omega_0} \bigr) - \gammaE
    \Bigr\}
\\
    + \blue{i\Omega_0}
    \biggl\{
      2 \Bigl[
        \tfrac{1}{\eps} + \ln\bigl(\tfrac{\pi\mu^2}{\Omega_0 E}\bigr)
      \Bigr]
      - \ln\bigl(
             \tfrac{|\altx_1 \altx_2 \altx_3| y\Omega_y}{\Omega_0}
          \bigr)
      + \tfrac{\altx_3^2}{2(\altx_1^2+\altx_2^2+\altx_3^2)}
          (1 - \ln2)
    \biggr\}
  \biggr] .
\label {eq:Anew11b}
\end {multline}
We have checked (both analytically and numerically) that this is correct.

Adding together the two color routings (\ref{eq:Anew11a}) and
(\ref{eq:Anew11b}) as in (\ref{eq:dGfund}) gives our final result
for the small-$y$ limit of the A1 diagram, which is the $\di{=}1$ case
of (\ref{eq:ii}).

% ===========================================================================

\section {Other \boldmath$\di{=}\dj$ ABC diagrams}
\label {sec:ii}

The B2 diagram is simply the A1 diagram with $x \to 1{-}x$, and so
corresponds to the $\di{=}2$ case of (\ref{eq:ii}).  The C3 diagram can
be related to the A1 diagram by front- and back-end transformation:
\begin {equation}
  {\rm C3} = {\rm bkEnd}[ {\rm frEnd}({\rm A1}) ]^* ,
\end {equation}
and the front-end transformations may introduce $i\pi$ terms, as previously
discussed in section \ref{sec:ij}.  Again, rather than carefully keep track
of all the branch cuts necessary to implement front-end transformations
in our small-$y$ expansions, we have determined the $i\pi$ terms by
generalizing the A1 result to
\begin {multline}
  \left[ \frac{d\Gamma}{dx\,dy} \right]_{\di\di}
  \simeq
  -2 \times
  \frac{\CA\alphas^2\,P(x)}{2\pi^2 y}
  \Re
  \biggl[
    \red{i\Omega_y}
    \Bigl\{
       - \ln\bigl( \tfrac{\Omega_y}{2\pi\Omega_0} \bigr) - \gammaE
       + \pi \times \purple{\qmark_{\di\di}}
    \Bigr\}
\\
    + \blue{i\Omega_0}
    \biggl\{
      2 \Bigl[
        \tfrac{1}{\eps} + \ln\bigl(\tfrac{\pi\mu^2}{\Omega_0 E}\bigr)
      \Bigr]
      -
      \left[
          \ln\bigl(
             \tfrac{|\altx_1 \altx_2 \altx_3| y\Omega_y}{\Omega_0}
          \bigr)
          + \pi \times \purple{\qmark_{\di\di}}
      \right]
\\
      + \tfrac{\altx_\dj^2+\altx_\dk^2}{4(\altx_1^2+\altx_2^2+\altx_3^2)}
          (1 - \ln2)
    \biggr\}
  \biggr]
\label {eq:iiguess}
\end {multline}
and then using numerical extraction of the $y{\to}0$ limit from
the generic-$y$ formulas of ref.\ \cite{qcd} to determine the question
marks.  Our final result is (\ref{eq:ii}).

% ===========================================================================

\section {Conclusion}
\label {sec:conclusion}

Our results were already summarized earlier in sections \ref{sec:intro}
and \ref{sec:ABCsummary}.  Our primary result is the expression (\ref{eq:sbar})
for the sub-leading, single IR logarithm arising from soft radiative
corrections to an underlying, hard, gluon splitting process $g{\to}gg$.
We remind readers that this result was calculated for an idealized situation,
with many caveats described in sections \ref{sec:background} and
\ref{sec:IRcutoff}, not least of
which was the assumption that the size of the medium is large compared
to relevant formation times.

%%%%%%%%%%%%%%%%%%%%%%%%%%%%%%%%%%%%%%%%%%%%%%%%%%%%%%%%%%%%%%%%%%%%%%%%%%%%%%

\acknowledgments

This work was supported, in part, by the U.S. Department
of Energy under Grants No.~DE-SC0007984 (Arnold and Gorda)
and DE-SC0007974 (Arnold);
by the Deutsche Forschungsgemeinschaft
(DFG, German Research Foundation) -- Project-Id 279384907 -- SFB 1245
(Gorda);
and by the National Natural
Science Foundation of China under
Grant Nos.\ 11935007, 11221504 and 11890714 (Iqbal).

%%\paragraph{Note added.} This is also a good position for notes added
%%after the paper has been written.

%%%%%%%%%%%%%%%%%%%%%%%%%%%%%%%%%%%%%%%%%%%%%%%%%%%%%%%%%%%%%%%%%%%%%%%%%%%%%%
\appendix
%%%%%%%%%%%%%%%%%%%%%%%%%%%%%%%%%%%%%%%%%%%%%%%%%%%%%%%%%%%%%%%%%%%%%%%%%%%%%%

\section{Correction to treatment of front-end virtual sequential
  diagrams in ref.\ \cite{qcd}}
\label {app:error}

In this appendix, we fix an error concerning $i\pi$ terms in the
result for the pole piece ${\cal A}_\seq^{\rm pole}(x,y)$
of sequential diagrams in
the original published version of eq.\ (A.37) of ref.\ \cite{qcd}.
The problem only manifests when one front-end transforms the
$g{\to}ggg$ sequential diagrams to get results for front-end virtual
sequential diagrams, as in eq.\ (A.60) of ref.\ \cite{qcd}.
The problem has to do with getting signs and phases correct, which
we never presented an explicit derivation of in ref.\ \cite{qcd}.
Here, we go through the derivation.

The sequential diagrams are given by the bottom row of
fig.\ \ref{fig:alpha1} (plus their complex conjugates), and
${\cal A}_\seq(x,y)$ represents one particular color-routing of those
diagrams, as described in ref.\ \cite{seq}.
The pole piece of those diagrams was computed in ref.\ \cite{dimreg},
where the possibility of front-end transformations was not considered.

In the process of generalizing ref.\ \cite{dimreg}
to (now correctly) handle signs and phases for front-end transformations,
we will {\it also} incorporate some inconsequential corrections \cite{QEDnf}
to the formulas
in the original published version of ref.\ \cite{dimreg}, with
explanation of those particular corrections left to footnotes.
The footnoted corrections are
inconsequential because they mildly affect intermediate equations but not the
final result for ${\cal A}_\seq^{\rm pole}(x,y)$.

The total pole piece ${\cal A}_\seq^{\rm pole}(x,y)$ is finite, but intermediate
steps are carried out, following ref.\ \cite{dimreg}, with the aid of
dimensional regularization with $d\,{\equiv}\,d_\perp{=}\,2{-}\eps$
transverse spatial dimensions.

% --------------------------------------------------------------------------

\subsection{$xy\bar x\bar y$ diagram}

A formula for one of the color routings (called $xy\bar x\bar y_2$)
of the $xy\bar x\bar y$ diagram shown in fig.\ \ref{fig:alpha1}
was given in eq.\ (5.14) of ref.\ \cite{dimreg}
as%
\footnote{
\label{foot:update}
  Eq.\ (\ref{eq:xyxyresultDR3}) has one of our ultimately-inconsequential
  corrections:
  the inclusion of the overall factor $(\mu/E)^{2\eps}$, where $\mu$ is the
  renormalization scale, as discussed in
  ref.\ \cite{QEDnf}.  See in particular the discussion surrounding
  eq.\ (F.31) of ref.\ \cite{QEDnf}.
}
\begin {align}
   \left[\frac{d\Gamma}{dx\,dy}\right]_{xy\bar x\bar y_2} &\simeq
   \left( \frac{\mu}{E} \right)^{\!2\eps}
   \frac{\CA^2 \alphas^2 M_\ix M_\fx^\seq}{2^{\frac{d}{2}+2}d\pi^d E^d} \,
     \frac{\Gamma^2( \tfrac{d+2}{4} )}
          {\Gamma(\frac{d}{2}) \sin(\frac{\pi d}{4})} \,
     \bigl(i\hat x_1 \hat x_2 \hat x_3 \hat x_4\Omega_\ix \sgn M_\ix\bigr)^{d/2}
\nonumber\\ & \hspace{10em} \times
     (d\bar\alpha + \bar\beta + \bar\gamma)
   \int \frac{d(\Delta t)}{(\Delta t)^{d/2}} \>
%\nonumber\\ &\quad
   \quad+\quad \{ \ix \leftrightarrow \fx^\seq \} ,
\label {eq:xyxyresultDR3}
\end {align}
where, for example, $M_\ix \equiv M_{E,x} \equiv x(1{-}x)E$
and $M_\fx^\seq=yz(1{-}x)E$.
Both of these $M$'s are positive for the $xy\bar x\bar y$ diagram, and
so the $\sgn M$ factors in (\ref{eq:xyxyresultDR3}) were replaced by
${+}1$ in the rest of the discussion of $xy\bar x\bar y$ in ref.\ \cite{dimreg}.
However, front-end transformations can negate $M_\ix$ and so, in the
application to virtual diagrams, we need to keep those factors of
$\sgn M$.%
\footnote{
  We must also use appropriate absolute value signs in the
  formulas for the DGLAP splitting functions $P(x)$ and
  DGLAP combinations $(\bar\alpha,\bar\beta,\bar\gamma)$,
  as in ref.\ \cite{qcd} eqs.\ (A.5) and (A.46).
}
Following through the subsequent development, and using%
\footnote{
  Eq.\ (\ref{eq:abcPP2DR}) is the corrected
  version of ref.\ \cite{dimreg} (5.17).
  See ref.\ \cite{qcd} (C.11) and the related discussion of
  ref.\ \cite{QEDnf} (F.32).  This completes our starting points for the
  ``inconsequential corrections'' which do not affect the original
  result of ref.\ \cite{dimreg} for ${\cal A}_\seq^{\rm pole}$.
}
\begin {equation}
   \bar\alpha + \tfrac1{d} \bar\beta + \tfrac1{d} \bar\gamma
   =
   \frac{P(x)}{\CA x^2(1{-}x)^2} \,\,
   \frac{P\bigl(\frac{y}{1{-}x}\bigr)}
        {\CA (1{-}x) y^2 (1{-}x{-}y)^2}
   \, ,
\label {eq:abcPP2DR}
\end {equation}
eq.\ (5.18) of
ref.\ \cite{dimreg} then generalizes to
\begin {align}
  2 \Re \biggl[\frac{d\Gamma}{dx\,dy}\biggr]^{(\Delta t<a)}_{xy\bar x\bar y}
  \simeq{}&
   \frac{\alphas^2 \mu^{2\eps} \, P(x) \, P(\yfrak)}{4\pi^2(1-x)}
   \, \frac{d}{2} \,
   \Beta(\tfrac12{+}\tfrac{d}{4},-\tfrac{d}{4})
   \left( \int_0^a \frac{d(\Delta t)}{(\Delta t)^{d/2}} \right)
\nonumber\\ & \quad \times
   \Re\Biggl[
     \left(
       \frac{\strut M_{(1-x)E,\yfrak}}{2\pi i} \,
       \frac{\bigl(|M|\Omega\bigr)_{E,x}}{2\pi}
     \right)^{\!\!\frac{d}{2}-1}
     i (\Omega\sgn M)_{E,x}
\nonumber\\ & \hspace{4em}
     +
     \left(
       \frac{\strut M_{E,x}}{2\pi i} \,
       \frac{\bigl(|M|\Omega\bigr)_{(1-x)E,\yfrak}}{2\pi}
     \right)^{\!\!\frac{d}{2}-1}
     i (\Omega\sgn M)_{(1-x)E,\yfrak}
     \,
   \Biggr]
   ,
\label {eq:xyxyDR}
\end {align}
where $\yfrak \equiv y/(1{-}x)$.
For front-end transformations,
the important distinction here relative to ref.\ \cite{dimreg} is
the appearance of $\sgn M$ factors above in
$\Omega\sgn M$ and $|M|\Omega = M\Omega\sgn M$.

% --------------------------------------------------------------------------

\subsection{$x\bar xy\bar y + x\bar x\bar y y$ diagrams}

Eq.\ (5.1) of ref.\ \cite{dimreg} analyzed the pole piece of
the last two diagrams of fig.\ \ref{fig:alpha1} (plus their complex
conjugates) and found the corresponding rate
\begin {align}
   \left[\Delta \frac{d\Gamma}{dx\,dy}\right]_{
        \begin{subarray}{} x\bar x y\bar y + x\bar x \bar y y \\
                        + \bar xx \bar yy + \bar x xy\bar y \end{subarray}
   }
   &= - \frac{1}{1-x}
   \int_0^\infty \! d(\Delta t_x)
   \int_0^\infty \! d(\Delta t_y) \>
   \tfrac12(\Delta t_x+\Delta t_y)
\nonumber\\& \qquad \times
     \Re \left[\frac{d\Gamma}{dx\,d(\Delta t_x)}\right]_{E,x} \,
     \Re \left[\frac{d\Gamma}{d\yfrak\,d(\Delta t_y)}\right]_{(1-x)E,\yfrak} ,
\label {eq:DxxyyetcDR}
\end {align}
where%
\footnote{
   Eq.\ (\ref{eq:GammaFoo}) is also updated along the same lines
   as footnote \ref{foot:update}.
}
\begin {align}
   \left[\frac{d\Gamma}{dx\,d(\Delta t)}\right]_{E,x} &\equiv
   \left( \frac{\mu}{E} \right)^{\eps}
   \frac{\alphas P(x)}{x^2(1-x)^2E^d}
   \grad_{\B^\xbx} \cdot \grad_{\B^\xx}
   \langle \B^\xbx,\Delta t | \B^\xx,0 \rangle_{E,x}
   \Bigr|_{\B^\xbx = \B^\xx = 0}
\nonumber\\
  &=
  - \left( \frac{\mu}{E} \right)^{\eps}
  \frac{\alphas P(x)}{x^2(1-x)^2E^d}
   \left( \frac{M\Omega\csc(\Omega \, \Delta t)}{2\pi i} \right)^{\!d/2}
   i d M\Omega\csc(\Omega \, \Delta t)
\label {eq:GammaFoo}
\end {align}
and the last equality comes from ref.\ \cite{dimreg}
eqs.\ (3.4--3.5).
The formulas for $M$'s and $\Omega$'s are such that the phase of
$\Omega$ is $e^{-i\pi/4}$ when $M > 0$ and $e^{+i\pi/4}$ when $M < 0$.
Given this, the results for one type of integral we need is
\begin {align}
   \frac{d\Gamma}{dx}
   &=
   \Re \int_0^\infty d(\Delta t) \>
   \left[\frac{d\Gamma}{dx\,d(\Delta t)}\right]_{E,x}
\nonumber\\
   &=
   \Re\left[
     - \left( \frac{\mu}{E} \right)^{\eps}
     \frac{\alphas P(x)}{x^2(1-x)^2E^d}
     \frac{i d M}{2} \left( \frac{|M|\Omega}{2\pi} \right)^{\!d/2}
     \Beta(\tfrac12{+}\tfrac{d}{4},-\tfrac{d}{4})
   \right] ,
\label {eq:seqint1}
\end {align}
which generalizes ref.\ \cite{dimreg} eq.\ (5.4).
The other type of integral we need, specifically for
the integral over $0 < \Delta t < a$, is the same as
ref.\ \cite{dimreg} eq.\ (5.5b):
\begin {equation}
   \Re \int_0^a d(\Delta t) \>
   \Delta t
   \left[\frac{d\Gamma}{dx\,d(\Delta t)}\right]_{E,x}
   =
   - \left( \frac{\mu}{E} \right)^{\eps}
   \frac{d\alphas P(x)}{2\pi E^{d-2}}
   \Re \left[
     \left( \frac{M}{2\pi i} \right)^{\!\!\frac{d}{2}-1}
     \int_0^a \frac{d(\Delta t)}{(\Delta t)^{d/2}}
   \right] ,
\label{eq:seqint2}
\end {equation}
where $a$ is a tiny cut-off used to isolate the UV-divergent pole terms.
Combining (\ref{eq:DxxyyetcDR}), (\ref{eq:seqint1}), and (\ref{eq:seqint2})
gives the following generalization of ref.\ \cite{dimreg} eq.\ (5.6):
\begin {align}
   \biggl[\Delta \frac{d\Gamma}{dx\,dy}\biggr]^{(\Delta t<a)}_{
        \begin{subarray}{} x\bar x y\bar y + x\bar x \bar y y \\
                        + \bar xx \bar yy + \bar x xy\bar y \end{subarray}
   }
   ={}&
   -
   \frac{\alphas^2 \mu^{2\eps} \, P(x) \, P(\yfrak)}{4\pi^2(1-x)}
   \left( \frac{d}{2} \right)^{\!2}
   \Beta(\tfrac12{+}\tfrac{d}{4},-\tfrac{d}{4})
   \left( \int_0^a \frac{d(\Delta t)}{(\Delta t)^{d/2}} \right)
\nonumber\\ & \quad \times
   \Biggl\{
     \Re\left[
       \left( \frac{M}{2\pi i} \right)^{\!\!\frac{d}{2}-1}
     \right]_{(1-x)E,\yfrak}
     \Re\left[
       \left( \frac{|M|\Omega}{2\pi} \right)^{\!\!\frac{d}{2}-1}
       i\Omega\sgn M
     \right]_{E,x}
\nonumber\\ & \qquad
     +
     \Re\left[
       \left( \frac{M}{2\pi i} \right)^{\!\!\frac{d}{2}-1}
     \right]_{E,x}
     \Re\left[
       \left( \frac{|M|\Omega}{2\pi} \right)^{\!\!\frac{d}{2}-1}
       i\Omega\sgn M
     \right]_{(1-x)E,\yfrak}
   \Biggr\}
   .
\label {eq:xxyyDR}
\end {align}

% --------------------------------------------------------------------------

\subsection{Total}

Adding together (\ref{eq:xyxyDR}) and (\ref{eq:xxyyDR}) gives
\begin {subequations}
\begin {align}
   2 \Re \left[\Delta \frac{d\Gamma}{dx\,dy}\right]^{(\Delta t<a)}_{
         x\bar xy\bar y + x\bar x\bar y y + x y \bar x\bar y
     }
   &
   \simeq \frac{\alphas^2 \mu^{2\eps} \, P(x) \, P(\yfrak)}{4\pi^2(1-x)}
   \,
   \left( \frac{d}{2} \right)^{\!2}
   \Beta(\tfrac12{+}\tfrac{d}{4},-\tfrac{d}{4})
   \int_0^a \frac{d(\Delta t)}{(\Delta t)^{d/2}}
\nonumber\\ & \hspace{6em} \times
   \Bigl[ {\cal Q} + \{ E,x \leftrightarrow (1{-}x)E,\yfrak \} \Bigr] ,
\label {eq:AseqQ}
\end {align}
where
\begin {multline}
  {\cal Q} \equiv
     \frac{2}{d}
     \Re\left[
       \left(
         \frac{\strut M_{(1-x)E,\yfrak}}{2\pi i} \,
         \frac{\bigl(|M|\Omega\bigr)_{E,x}}{2\pi}
       \right)^{\!\!\frac{d}{2}-1}
       i (\Omega\sgn M)_{E,x}
     \right]
\\
  -
     \Re\left[
       \left( \frac{M}{2\pi i} \right)^{\!\!\frac{d}{2}-1}
     \right]_{(1-x)E,\yfrak}
     \Re\left[
       \left( \frac{|M|\Omega}{2\pi} \right)^{\!\!\frac{d}{2}-1}
       i\Omega\sgn M
     \right]_{E,x} .
\label {eq:calQ}
\end {multline}
\end {subequations}
${\cal Q}$ vanishes for $d{=}2$, and so ${\cal Q} = O(\eps)$.
This means that we only need to keep the $O(1/\eps)$ piece of the
prefactors in (\ref{eq:AseqQ}):
\begin {equation}
   2 \Re \left[\Delta \frac{d\Gamma}{dx\,dy}\right]^{(\Delta t<a)}_{
         x\bar xy\bar y + x\bar x\bar y y + x y \bar x\bar y
     }
   =
   -\frac{\alphas^2 \, P(x) \, P(\yfrak)}{\pi^2(1-x) \eps}
   \Bigl[ {\cal Q} + \{ E,x \leftrightarrow (1{-}x)E,\yfrak \} \Bigr] .
\label {eq:AseqQ2}
\end {equation}
Expanding (\ref{eq:calQ}) to first order in $\eps$ gives
\begin {align}
  {\cal Q} &=
  \frac{\eps}{2}
  \left\{
    \Re[ i (\Omega\sgn M)_{E,x} ]
    -
    \Re\left[
      i (\Omega\sgn M)_{E,x}
      \ln\left( \frac{\sgn M_{(1-x)E,\yfrak}}{i} \right)
    \right]
  \right\}
\nonumber\\
  &=
  \frac{\eps}{2}
  \left\{
    \Re[ i (\Omega\sgn M)_{E,x} ]
    -
    \frac{\pi}{2}
    \Re[ (\Omega\sgn M)_{E,x} ]
    \sgn M_{(1-x)E,\yfrak}
  \right\}
  .
\label {eq:calQ2}
\end {align}
Using (\ref{eq:calQ2}) in (\ref{eq:AseqQ2}) gives our generalization
of ref.\ \cite{dimreg} eq.\ (5.20).
${\cal A_\seq^{\rm pole}}$ is half of the above.
Our generalization of ref.\ \cite{dimreg} eq.\ (7.4), and correspondingly
our correction to ref.\ \cite{qcd} eq.\ (A.37), is then
\begin {multline}
   {\cal A}_\seq^{\rm pole} =
   - \frac{\alphas^2 \, P(x) \, P(\yfrak)}{4\pi^2(1-x)}
   \Re\Bigl[
     i (\Omega\sgn M)_{E,x} (1+\tfrac{i\pi}{2} \sgn M_{(1-x)E,\yfrak})
\\
     +
     i (\Omega\sgn M)_{(1-x)E,\yfrak} (1+\tfrac{i\pi}{2} \sgn M_{E,x})
   \Bigr]
   .
\label {eq:seqDR}
\end {multline}
This correction is responsible for the $4\pi$ shift of fig.\ \ref{fig:cnew}
of this paper relative to fig.\ 20 of ref.\ \cite{qcd}.

We note in passing that,
because the complex phase of $\Omega$ is $e^{-(i\pi/4)\sgn M}$,
\begin {equation}
  \Re( \Omega ) = \Re( i \Omega\sgn M ) .
\label {eq:Rei}
\end {equation}
Using this, (\ref{eq:seqDR}) may also be written in the alternate form
\begin {multline}
   {\cal A}_\seq^{\rm pole} =
   - \frac{\alphas^2 \, P(x) \, P(\yfrak)}{4\pi^2(1-x)}
   \Re\bigl[
     i (\Omega\sgn M)_{E,x} + i (\Omega\sgn M)_{(1-x)E,\yfrak}
   \bigr]
\\ \times
   \bigl(1-\tfrac{\pi}{2} \sgn M_{E,x} \sgn M_{(1-x)E,\yfrak}\bigr)
   .
\label {eq:seqDR2}
\end {multline}

% ============================================================================
\section{Organization of IR divergences compared to ref.\ \cite{qcd}}
\label {app:TableComparison}

In this appendix, we will refer to table \ref{tab:cancel} of this paper,
which describes certain cancellations among IR single and double logarithms,
as the ``log table.''  We will refer to table 1 of ref.\ \cite{qcd},
which described cancellations among IR {\it power-law} divergences,
as the ``power-law table.''  The purpose of this appendix is to
describe the similarities and differences of the organization of
these two tables.

The power-law table gives, for example,
$y{\to}0$, $x{\to}0$ and $z{\to}0$ limits of the single diagram
$2\Re(yx\bar x\bar y)$.
The log table, in contrast, always refers to the softest gluon as
$y$ and so instead lists the $y{\to}0$ limit of the six diagrams
corresponding to the permutations of the daughters $(x,y,z)$
of $2\Re(yx\bar x\bar y)$.
The $y{\to}0$ limit of $2\Re(yx\bar x\bar y)$ in the power-law table
corresponds here to the average of the $(x,y)$ and $(z,y)$ columns in
the $2\Re(yx\bar x\bar y)$ row of our log table;
the $z{\to}0$ limit in the power-law table
corresponds to the average of the $(x,z)$ and $(z,x)$ columns;
and the $x{\to}0$ limit corresponds to the average of
the $(y,x)$ and $(y,z)$ columns.
We needed half as many columns in ref.\ \cite{qcd} because, in the
language of the current paper (where $y$ is always the softest gluon),
the power-law divergences of each individual
diagram have a symmetry under $x\to 1{-}x$ that the log divergences
do not.

For a slightly different example, consider the Class I virtual diagram
$2\Re(yx\bar x y)$.  Now, the $y{\to}0$ limit of the power-law table
corresponds to the average of the $(x,y)$ and $(1{-}x,y)$ columns of
the log table; and the $z{\to}0$ limit to the average of the
$(x,z)$ and $(1{-}x,x{-}y)$ columns.
For $2\Re(yx\bar x y)$, we do not list anything in
the log table corresponding to the $x{\to}0$ column of the
power-law table.  That's because even for power-law divergences,
the $x{\to}0$ behavior of virtual diagrams was not relevant to
checking the cancellation of IR divergences among diagrams,
for the reasons explained in ref.\ \cite{qcd}.%
\footnote{
  Specially, see section 3 and appendix E.2 of ref.\ \cite{qcd}.
}

As a final example, consider the Class II virtual diagram
$2\Re(x\bar y\bar y\bar x)$.  Since this is a virtual diagram,
the $x{\to}0$ column of the power-law table is irrelevant, as
mentioned above.  Because it is specifically a Class II diagram
(as defined in ref.\ \cite{qcd}), the $z{\to}0$ limit is also
irrelevant because there are no IR divergences in this limit:
no line of the diagram becomes soft in the limit $z{\to}0$.
So only the $y{\to}0$ column of the power-law diagram is relevant,
and this corresponds to the average of the $(x,y)$ and $(1{-}x,y)$ columns
of the log table.

We should clarify the meaning of the double-size boxes for
A1, B2, and C3 in our log table.  Each of these diagrams has two
large-$\Nc$ color orderings.  Consider A1, for example.
If one focuses on a particular large-$\Nc$
color ordering, then the $(x,y)$ and $(x,z)=(x,1{-}x{-}y)$ entries for A1
in table \ref{tab:cancel} represent different soft-gluon limits.
If one instead thinks of the full A1, summing both color
routings, then the two entries are the same and should not be double
counted.

% ============================================================================

\section{Small-\boldmath$y$ behavior of A3 diagram integrand}
\label {app:A3}

For the A3 diagram, the generic-$y$ formula for the $D$ of (\ref{eq:Cdef})
is \cite{2brem}
\begin {align}
   D(x_1,&x_2,x_3,x_4,\alpha,\beta,\gamma,\Delta t) =
\nonumber\\ &
   \frac{\CA^2 \alphas^2 M_\ix M_\fx}{32\pi^4 E^2} \, 
   ({-}x_1 x_2 x_3 x_4)
   \Omega_+\Omega_- \csc(\Omega_+\Delta t) \csc(\Omega_-\Delta t)
\nonumber\\ &\times
   \Bigl\{
     (\beta Y_\bx Y_\Ax + \alpha \Ybar_{\bx\Ax} Y_{\bx\Ax}) I_0
     + (\alpha+\beta+2\gamma) Z_{\bx\Ax} I_1
\nonumber\\ &\quad
     + \bigl[
         (\alpha+\gamma) Y_\bx Y_\Ax
         + (\beta+\gamma) \Ybar_{\bx\Ax} Y_{\bx\Ax}
        \bigr] I_2
     - (\alpha+\beta+\gamma)
       (\Ybar_{\bx\Ax} Y_\Ax I_3 + Y_\bx Y_{\bx\Ax} I_4)
   \Bigl\} .
\label {eq:summaryD}
\end {align}
A summary containing this formula and expressions for its
various elements may be found in
Appendices A.2.1--A.2.2 of ref.\ \cite{qcd}.
Of those, it will be useful to have at hand
\begin {subequations}
\label {eq:I}
\begin {equation}
    I_0 =
    \frac{4\pi^2}{(X_\bx X_\Ax - X_{\bx\Ax}^2)} \,,
  \qquad
    I_1 =
    - \frac{2\pi^2}{X_{\bx\Ax}}
    \ln\left( \frac{X_\bx X_\Ax - X_{\bx\Ax}^2}{X_\bx X_\Ax} \right) ,
\end {equation}
\begin {equation}
    I_2 = I_0 - \frac{I_1}{X_{\bx\Ax}} \,,
  \qquad
    I_3 = \frac{X_{\bx\Ax} I_0}{X_\Ax} \,,
  \qquad
    I_4 = \frac{X_{\bx\Ax} I_0}{X_\bx}
\end {equation}
\end {subequations}
and
\begin {subequations}
\label {eq:XYZdef}
\begin {align}
   \begin{pmatrix} X_\yx & Y_\yx \\ Y_\yx & Z_\yx \end{pmatrix} \,\,\,
   &\equiv
   \begin{pmatrix} |M_0|\Omega_0 & 0 \\ 0 & 0 \end{pmatrix}
     - i a_\yx^{-1\top}
     \begin{pmatrix} \Omega_+\cot_+ & \\ & \Omega_-\cot_- \end{pmatrix}
     a_\yx^{-1} ,
\label {eq:XYZy2}
\\
   \begin{pmatrix} X_\ybx & Y_\ybx \\ Y_\ybx & Z_\ybx \end{pmatrix} \,\,\,
   &\equiv
   \begin{pmatrix} |M_\fx|\Omega_\fx & 0 \\ 0 & 0 \end{pmatrix}
     - i a_\ybx^{-1\top}
     \begin{pmatrix} \Omega_+\cot_+ & \\ & \Omega_-\cot_- \end{pmatrix}
     a_\ybx^{-1} ,
\label {eq:XYZyb2}
\\
   \begin{pmatrix}
      X_{\yx\ybx} & Y_{\yx\ybx} \\
      \Ybar_{\yx\ybx} & Z_{\yx\ybx}
   \end{pmatrix}
   &\equiv
     - i a_\yx^{-1\top}
     \begin{pmatrix} \Omega_+\csc_+ & \\ & \Omega_-\csc_- \end{pmatrix}
     a_\ybx^{-1} ,
\label {eq:XYZyyb2}
\end {align}
\end {subequations}
where
\begin {equation}
   \csc_\pm \equiv \csc(\Omega_\pm\,\Delta t) ,
   \qquad
   \cot_\pm \equiv \cot(\Omega_\pm\,\Delta t) .
\end {equation}
For the A3 diagram of interest here,
\begin {equation}
   M_0 = x(1{-}x)E, \qquad M_\fx = x z (1{-}y) E .
\end {equation}
Both are positive, and so we will drop the absolute value signs
in (\ref{eq:XYZdef}).

From other formulas in Appendices A.2.1--A.2.2 of ref.\ \cite{qcd},
one may extract the small-$y$ limits for the complex frequencies
associated with 4-particle evolution:
\begin {subequations}
\begin {align}
  \Omega_+ &= \Omega_y [1 + O(y)] ,
\\
  \Omega_- &= (1-3s)^{1/2}\Omega_0 + O(y) ,
\end {align}
\end {subequations}
where $\Omega_0$ and $s$ were defined
previously in (\ref{eq:Om0}) and (\ref{eq:sxitau}).
One may similarly extract small-$y$ expansions of the
matrices $a_\yx$ and $a_\ybx$ of normal mode vectors.
We need their inverses in (\ref{eq:XYZdef}), whose
expansions we find to be
\begin {subequations}
\label {eq:ainv}
\begin {align}
  a_\ybx^{-1} &=
  \begin{pmatrix}
     -\tfrac12 x y^{1/2} & y^{1/2} \\[3pt]
     - x^{1/2} (1{-}x)^{1/2}(1{+}uy)~~~ & -\tfrac12 x^{1/2} (1{-x})^{-1/2} y
  \end{pmatrix}
  E^{1/2}
  + O(y^{3/2}) ,
\label {eq:aybinv}
\\
  a_\yx^{-1} &=
  \begin{pmatrix}
     \tfrac12 x y^{1/2} & (1{-}x) y^{1/2} \\[3pt]
     - x^{1/2} (1{-}x)^{1/2}(1{+}vy)~~~ & ~~~~\tfrac12 x^{1/2} (1{-}x)^{1/2} y~~
  \end{pmatrix}
  E^{1/2}
  + O(y^{3/2}) .
\label{eq:ayinv}
\end {align}
\end {subequations}
Above, $u{=}u(x)$ and $v{=}v(x)$ are some functions of $x$ which we did
not bother to determine. Though $u$ and $v$ will appear in some intermediate
formulas, each of their effects will cancel in final results.

% --------------------------------------------------------------------------

\subsection{$\Delta t \sim y$}
\label {app:A3y}

For $\Delta t \sim y$, we have $\Omega_\pm\,\Delta t \ll 1$ and so may
use small-argument expansions for all of
the trig functions in (\ref{eq:XYZdef}).
The small-$y$ expansions we will need, at the order that we will
need them, are
\begin {subequations}
\label{eq:XYZ}
\begin {align}
  X_\yx &= -\frac{i M_0}{\Delta t} + M_0 \Omega_0
             + \frac{i}{3} M_0 \Omega_0^2 \Delta t + O(y^2) ,
\label {eq:Xy}
\\
  X_\ybx &= -\frac{i M_\fx}{\Delta t} + M_\fx \Omega_\fx
             + \frac{i}{3} M_0 \Omega_0^2 \Delta t + O(y^2) .
\label {eq:Xyb}
\\
  X_{\yx\ybx} &=
    -\frac{i x z E}{\Delta t}
    - \frac{i}{6} M_0
      \Bigl( \Omega_0^2 + \frac{i\qhatA x^2}{4M_0} \Bigr)\Delta t
    + O(y^2) ,
\label {eq:Xyyb}
\\
  Y_\yx &= \tfrac{1}{12} x(1{-}x) \qhatA\,\Delta t + O(y),
\\
  Y_\ybx &= - \tfrac{1}{12} x \qhatA\,\Delta t + O(y),
\\
  Y_{\yx\ybx} &=
    -\frac{i x y E}{\Delta t}
    + O(y) ,
\\
  \Ybar_{\yx\ybx} &=
    +\frac{i x y z E}{\Delta t}
    + O(y) ,
\\
  Z_{\yx\ybx} &=
    -\frac{i y z E}{\Delta t} 
    - \frac{\qhatA(1{-}x)}{12} \Delta t
    + O(y^2) .
\end {align}
\end {subequations}
In general, we need the final expansion of (\ref{eq:summaryD}) to
next-to-leading order (NLO) in $y$, because of the (ultimately
canceling) spurious $y^{-2}$ divergence described in section \ref{sec:A3y}.
That requires NLO expressions for many of the $(X,Y,Z)$'s shown above.
However, for the $X$'s, we need next-to-next-to-leading (NNLO) order
in (\ref{eq:XYZ})
because the leading-order terms cancel in the combination
$X_\yx X_\ybx - X_{\yx\ybx}^2$.  We will need $X_\yx X_\ybx - X_{\yx\ybx}^2$
for the NLO evaluation of
the $2\gamma Z_{\yx\ybx} I_1 + \gamma \Ybar_{\yx\ybx} Y_{\yx\ybx} I_2$ terms
in (\ref{eq:summaryD}),
which are the terms that will contribute to the spurious $y^{-2}$ divergence.
The NLO expansion is
\begin {equation}
  X_\yx X_\ybx - X_{\yx\ybx}^2
  =
  - \frac{M_0 M_\fx}{(\Delta t)^2}
  \left\{
    \left[
      \frac{x y}{(1{-}x)(1{-}y)}
      +  i (\Omega_0{+}\Omega_\fx) \Delta t
    \right]
    - \left( 2 \Omega_0^2 + \frac{i \qhatA x^2}{12 M_0} \right) (\Delta t)^2
    + O(y^3)
  \right\}.
\end {equation}
This appears in $I_1$ in the combination
\begin {equation}
  \frac{X_\yx X_\ybx - X_{\yx\ybx}^2}{X_\yx X_\ybx}
  = \frac{x y}{(1{-}x)(1{-}y)}
    \bigl[
      (1 {+} \tau)
      - (1{+}s) \xi \tau^2
      + O(y^2)
    \bigr] ,
\label {eq:detXratio2}
\end {equation}
where $s$, $\xi$, and $\tau$ are defined as in (\ref{eq:sxitau})
and behave parametrically as
\begin {equation}
  s \sim y^0, \qquad \xi \sim y,
\end {equation}
and
\begin {equation}
  \tau \sim y^0 \quad \mbox{(for $\Delta t \sim y$)} .
\end {equation}

From the explicit formulas for $(\alpha,\beta,\gamma)$ give in
ref.\ \cite{qcd} eq.\ (A.23), parametrically
\begin {equation}
  \gamma \sim y^{-3}, \qquad
  \alpha \simeq -\beta \sim y^{-2}, \qquad
  \alpha{+}\beta \sim y^{-1} .
\label {eq:abcscale}
\end {equation}
Combined with the above formulas, the terms
\begin {equation}
  2\gamma Z_{\yx\ybx} I_1 \sim \gamma \Ybar_{\yx\ybx} Y_{\yx\ybx} I_2 \sim y^{-2}
  \qquad \mbox{(for $\Delta t \sim y$)}
\end {equation}
of (\ref{eq:summaryD}) contribute to $D$ starting at leading order
in $y$, and
\begin {equation}
  -\gamma \Ybar_{\yx\ybx} Y_\ybx I_3 \sim -\gamma Y_\yx Y_{\yx\ybx} I_4
      \sim y^{-1}
  \qquad \mbox{(for $\Delta t \sim y$)}
\end {equation}
contribute at NLO in $y$.  The rest of the terms in (\ref{eq:summaryD}) are
parametrically smaller and do
not contribute to IR logarithms.

Using the preceding expansions, we find contribution
\begin {multline}
  D_{(1)}
  \simeq
  - \frac{\CA^2\alphas^2}{8\pi^2(\Delta t)^2}
  (xyz)^2(1{-}x)(1{-}y)
  \gamma
\\ \times
  \biggl\{
     ( \red{1} + \xi - 2 s\xi\tau^2 )
       \ln\Bigl( 
         \frac{x y}{(1{-}x)(1{-}y)}
         (1 + \tau)
       \Bigr)
     + \red{ \frac{1}{1+\tau} }
\\
     + \xi
       \left[
          - \frac{2\tau}{1+\tau} 
          - \frac{(1+3s)\tau^2}{1+\tau}
          + \frac{(1+s)\tau^2}{(1+\tau)^2}
       \right]
  \biggr\}
\label {eq:D1}
\end {multline}
to $D$ from $2\gamma Z_{\yx\ybx} I_1 + \gamma \Ybar_{\yx\ybx} Y_{\yx\ybx} I_2$ and
\begin {equation}
  D_{(2)}
  \simeq
  - \frac{\CA^2\alphas^2}{4\pi^2(\Delta t)^2}
  x^2 y^2 (1{-}x)^3
  \gamma \,
  \frac{s \xi \tau^2}{(1+\tau)}
\label {eq:D2}
\end {equation}
from
$-\gamma(\bar Y_{\yx\ybx}Y_\ybx I_3\,{+}\,Y_\yx Y_{\yx\ybx} I_4)$.
The sum of (\ref{eq:D1}) and (\ref{eq:D2}) gives our
$\Delta t\sim y$ expansion of $D$ in (\ref{eq:DA3}).

% ---------------------------------------------------------------------------

\subsection{UV contribution}
\label {app:A3pole}

From eqs.\ (D.6--D.9) and (D.19) of ref.\ \cite{qcd}, the UV contribution
from the A3 diagram is
\begin {align}
   \left[ \frac{d\Gamma}{dx\,dy} \right]_{xy\bar y\bar x}^{(\Delta t<a)}
   =&
   \frac{\CA^2 \alphas^2}{16\pi^2}
   \left[ \frac{2}{\eps} + \ln\Bigl(\frac{\mu^4 a}{E^2}\Bigr)
          + 1 + \ln(2\pi^2) \right]
     \bigl[ (i\Omega_0)^{d/2} + (i\Omega_\fx)^{d/2} \bigr]
\nonumber\\ &\qquad\times
     x y z^2 (1{-}x)^2 (1{-}y)^2
       \left[
         (\alpha {+} \beta)
         + \frac{(\alpha {+} \gamma) x y}
                {(1{-}x)(1{-}y)}
       \right]
\nonumber\\
   &- \frac{i\CA^2 \alphas^2}{16\pi^2}
     ( \Omega_0 + \Omega_\fx )
     x y z^2 (1{-}x)^2 (1{-}y)^2
\nonumber\\ &\qquad\times
     \biggl\{
       \left[
         (\alpha + \beta)
         + \frac{(\alpha + \gamma) x y}
                {(1{-}x)(1{-}y)}
       \right]
       \ln(e^{-i\pi} x y z)
       - \red{2\gamma} 
%\nonumber\\ &\hspace{17em}
       + \frac{2 (\alpha+\gamma) x y}
                {(1{-}x)(1{-}y)}
    \biggr\}
  .
\label {eq:pole0}
\end {align}
We are only interested in terms that contribute $O(1/y)$ or larger
to (\ref{eq:pole0}), since those are the only terms that can generate
single or double log behavior.
Recalling the scaling (\ref{eq:abcscale})
of $(\alpha,\beta,\gamma)$ with $y$, we find that only the
terms proportional to $\gamma$ are large enough.  Most of those
terms are $O(\ln y/y)$ or $O(1/y)$, in which case one may make
small $y$ approximations in prefactors, such as $(1{-}y)^2 \simeq 1$.
The {\it only} term that's even bigger is the red $-2\gamma$ term in the last
line of (\ref{eq:pole0}), which generates spurious $1/y^2$ behavior,
for which NLO terms in $y$ are important.
So we can use the small-$y$ approximation
\begin {multline}
  \bigl[ (i\Omega_0)^{d/2} + (i\Omega_\fx)^{d/2} \bigr]
  \simeq
  2 (i\Omega_0)^{d/2}
  =
  2 i\Omega_0 \bigl[ 1 - \tfrac{\eps}{2} \ln(i\Omega_0) + O(\eps^2) \bigr]
\\
  \simeq
  (i\Omega_0 + i\Omega_\fx) 
    \bigl[ 1 - \tfrac{\eps}{2} \ln(i\Omega_0) + O(\eps^2) \bigr]
\label {eq:Omigapprox}
\end {multline}
in the {\it first}\/ line of (\ref{eq:pole0}).
(The last approximation in (\ref{eq:Omigapprox})
will be convenient for combining with other terms.)
Keeping all this in mind, we can rewrite (\ref{eq:pole0}) through
NLO in $y$ as
\begin {multline}
   \left[ \frac{d\Gamma}{dx\,dy} \right]_{xy\bar y\bar x}^{(\Delta t<a)}
   =
   \frac{\CA^2 \alphas^2}{16\pi^2}
   (x y z)^2 (1{-}x) (1{-}y) \gamma
   (i\Omega_0 + i\Omega_\fx)
\\ \times
   \biggl\{
     \left[ \frac{2}{\eps}
            + \ln\Bigl(\frac{\mu^4 a}{i\Omega_0 E^2}\Bigr)
            + 1 + \ln(2\pi^2) \right]
\\
     -
     \left[ \ln(e^{-i\pi} x y z)
            - \frac{2(1{-}x)(\red{1}{-}y)}{x y}
            + 2 \right] 
    \biggr\}
  .
\end {multline}
Taking $2\Re(\cdots)$, and using
\begin {equation}
   \frac{2(1{-}x)(1{-}y)}{x y}
   = \frac{2(z{+}xy)}{x y}
   = \frac{1}{\xi} + 2 ,
\end {equation}
gives (\ref{eq:A3poley}).

% ---------------------------------------------------------------------------

\subsection{$\Delta t \sim \sqrt{y}$}
\label {app:A3sqrty}

For $\Delta t \sim \sqrt{y}$, the power counting is a little different than
in section \ref{app:A3y}.  We still have $\Omega_-\,\Delta t \ll 1$,
but now $\Omega_+\,\Delta t \simeq \Omega_y \Delta t \sim 1$.
So not all trig functions can be expanded.
We will again need to evaluate $D$ to NLO in $y$, but this time NLO
will be suppressed compared to leading order by a factor of $\sqrt{y}$
rather than a factor of $y$.  That means that we will be able to
ignore corrections to $D$ that are suppressed by a full power of $y$.
However, similar to section \ref{app:A3y}, we will need to
expand $X$'s to NNLO because of cancellation of the
leading-order contributions in the combination $X_\yx X_\ybx - X_{\yx\ybx}^2$.
The expansions of (\ref{eq:XYZdef}) we will need are
\begin {subequations}
\label {eq:XYZ3}
\begin {align}
  X_\yx &=
   -i(1{+}2vy) M_0 (\Omega\cot)_-
   + (M\Omega)_0
   - \tfrac{i}{4} x^2 y E (\Omega\cot)_y 
   + O(y) ,
\label {eq:Xy2}
\\
  X_\ybx &=
   -i(1{+}2uy) M_0 (\Omega\cot)_-
   + (M\Omega)_0
   - \tfrac{i}{4} x^2 y E (\Omega\cot)_y 
   + O(y) ,
\label {eq:Xyb2}
\\
  X_{\yx\ybx} &=
   -i\bigl(1{+}(u{+}v)y\bigr) M_0 (\Omega\csc)_-
   + \tfrac{i}{4} x^2 y E (\Omega\csc)_y
   + O(y^{3/2}) ,
\label {eq:Xyyb2}
\\
  Y_\yx &= -(1{-}x) Y_\ybx [1 + O(y)] ,
\\
  Y_\ybx &= 
   \frac{i x y E}{2}
   \left[ (\Omega\cot)_y - \frac{1}{\Delta t} \right]
   + O(y^{3/2}) ,
\\
  Y_{\yx\ybx} &=
   -\frac{i x y E}{2} \left[ (\Omega \csc)_y + \frac{1}{\Delta t} \right]
   + O(y^{3/2}) ,
\\
  \Ybar_{\yx\ybx} &=
    -(1{-}x) Y_{\yx\ybx} [1 + O(y)] ,
\\
  Z_{\yx\ybx} &=
    -i y (1{-}x) E (\Omega\csc)_y
    + O(y^{3/2}) ,
\end {align}
\end {subequations}
where
\begin {equation}
   \csc_y \equiv \csc(\Omega_y \Delta t) ,
   \qquad
   \cot_y \equiv \cot(\Omega_y \Delta t) .
\end {equation}
Above, $\csc_-$ and $\csc_+$ should be expanded in the small argument
$\Omega_-\,\Delta t = O(y^{1/2})$, but it's algebraically convenient
to save that step until after one simplifies the combination $X_\yx
X_\ybx - X_{\yx\ybx}^2$.  That combination turns out to be
\begin {equation}
  X_\yx X_\ybx - X_{\yx\ybx}^2
  =
  M_0^2 \biggl[
    \red{-\frac{2 i \Omega_0}{\Delta t}}
    + \Omega_0^2 + \Omega_-^2
    - \frac{x y}{2(1{-}x)\Delta t}
        \Omega_y (\csc_y + \cot_y)
  \biggr]
  + O(y^{1/2}) ,
\end {equation}
and then
$S \equiv (X_\yx X_\ybx - X_{\yx\ybx}^2)/X_\yx X_\ybx$
is given through NLO by the $S$ shown in (\ref{eq:S}).
Note that the dependence on the $u$ and $v$ of (\ref{eq:ainv})
affected the NNLO contributions to the $X$'s in
(\ref{eq:XYZ3}), but their effects have canceled in the NLO
result for $X_\yx X_\ybx - X_{\yx\ybx}^2$ above, which is why they
do not affect our calculation.

Combining (\ref{eq:abcscale}) with the above formulas, the term
\begin {equation}
  2\gamma Z_{\yx\ybx} I_1 \sim y^{-2}
  \qquad \mbox{(for $\Delta t \sim \sqrt{y}$)}
\label {eq:ZI1sqrty}
\end {equation}
of (\ref{eq:summaryD}) contributes to $D$ starting at leading order
in $y$, and
\begin {equation}
  \gamma \Ybar_{\yx\ybx} Y_{\yx\ybx} I_2 \sim \gamma Y_\yx Y_\ybx I_2 \sim
  -\gamma \Ybar_{\yx\ybx} Y_\ybx I_3 \sim -\gamma Y_\yx Y_{\yx\ybx} I_4
      \sim y^{-3/2}
  \qquad \mbox{(for $\Delta t \sim y$)}
\label {eq:othersqrty}
\end {equation}
contribute at NLO in $y$.  The rest of the terms in (\ref{eq:summaryD}) are
parametrically smaller and do
not contribute to IR logarithms.%
\footnote{
  Remember that $D$ also contains prefactors in (\ref{eq:summaryD}) and
  is integrated over $\Delta t$ to get a rate $d\Gamma/dx\,dy$.
  Because of this, (\ref{eq:ZI1sqrty}) contributes $O(y^{-3/2})$ to the
  rate [which at leading order gives a power-law IR divergence],
  and (\ref{eq:othersqrty}) contributes $O(y^{-1})$ [which at leading
  order contributes IR logarithms].
}
Using the preceding expansions, we find contribution
\begin {equation}
  D_{\rm(A)}
  \simeq
  - \frac{\CA\alphas^2\,P(x)}{4\pi^2 y} \,
  (\Omega\csc)_y^2
  \ln S
\label {eq:DA}
\end {equation}
to $D$ from $2\gamma Z_{\yx\ybx} I_1$ and
\begin {equation}
  D_{\rm(B)}
  \simeq
  - \frac{\CA\alphas^2\,P(x)}{4\pi^2 y} \,
    \frac{\xi \Omega_y^3 \Delta t}{2 S} \csc_y (\csc_y+\cot_y)^2
\label {eq:DB}
\end {equation}
from
$\gamma (
  \Ybar_{\yx\ybx} Y_{\yx\ybx} I_2
  + Y_\yx Y_\ybx I_2
  - \bar Y_{\yx\ybx}Y_\ybx I_3 - Y_\yx Y_{\yx\ybx} I_4
 )$.
The sum of (\ref{eq:DA}) and (\ref{eq:DB}) gives our
$\Delta t\sim \sqrt{y}$ expansion of $D$ in (\ref{eq:DDA3}).

% ============================================================================

\section{Small-\boldmath$y$ behavior of A1 diagram integrand}

\subsection{$\Delta t \sim \sqrt{y}$}
\label {app:A1sqrty}

The extraction of the small-$y$ behavior for $\Delta t \sim \sqrt{y}$ for
the A1 diagram is very similar to that for the A3 diagram in
section \ref{app:A3sqrty}.  The function
\begin {align}
   \widetilde D_\new(x_1,x_2,x_3,x_4,\bar\alpha,\bar\beta,\bar\gamma,&\Delta t) =
\nonumber\\
   - \frac{\CA^2 \alphas^2 M_\ix^2}{32\pi^4 E^2} \, 
   ({-}x_1 x_2 x_3 x_4)
   &
     \Omega_+\Omega_- \csc(\Omega_+\Delta t) \csc(\Omega_-\Delta t)
\nonumber\\ &\quad \times
     \Bigl\{
       (\bar\beta Y_\yx^\new Y_\yx^\new
          + \bar\gamma \Ybar_{\yx\yx'}^{\,\new} Y_{\yx\yx'}^\new) I_0^\new
       + (2\bar\alpha{+}\bar\beta{+}\bar\gamma) Z_{\yx\yx'}^\new I_1^\new
\nonumber\\ &\quad\qquad
       + \bigl[
           (\bar\alpha{+}\bar\gamma) Y_\yx^\new Y_\yx^\new
           + (\bar\alpha{+}\bar\beta) \Ybar_{\yx\yx'}^{\,\new} Y_{\yx\yx'}^\new
          \bigr] I_2^\new
\nonumber\\ &\quad\qquad
       - (\bar\alpha{+}\bar\beta{+}\bar\gamma)
         (\Ybar_{\yx\yx'}^{\,\new} Y_\yx^\new I_3^\new
              + Y_\yx^\new Y_{\yx\yx'}^\new I_4^\new)
     \Bigl\}
\label {eq:Dnew}
\end {align}
is taken from eq.\ (A.64) of ref.\ \cite{qcd} {\it except}\/ that the
${\cal D}_2^{(\bbI)}$ term is excluded, as we have discussed in the
main text.  Above and below,
\begin {subequations}
\begin {equation}
   \begin{pmatrix} X_{\yx'}^\new & Y_{\yx'}^\new
                   \\ Y_{\yx'}^\new & Z_{\yx'}^\new \end{pmatrix}
   \equiv
   \begin{pmatrix} X_\yx^\new & Y_\yx^\new \\ Y_\yx^\new & Z_\yx^\new \end{pmatrix}
   \equiv
   \mbox{the} ~
   \begin{pmatrix} X_\yx & Y_\yx \\ Y_\yx & Z_\yx \end{pmatrix}
   ~\mbox{of eq.\ (\ref{eq:XYZy2}) ,}
\end {equation}
\begin {equation}
   \begin{pmatrix} X_{\yx\yx'}^\new & Y_{\yx\yx'}^\new \\[2pt]
                   \Ybar_{\yx\yx'}^\new & Z_{\yx\yx'}^\new \end{pmatrix}
   \equiv
     - i a_\yx^{-1\top}
     \begin{pmatrix} \Omega_+\csc_+ & \\ & \Omega_-\csc_- \end{pmatrix}
     a_\yx^{-1} .
\end {equation}
\end {subequations}
The relevant expansions are
\begin {subequations}
\label{eq:XYZ2}
\begin {align}
  X_{\yx'}^\new = X_\yx^\new &=
   -i(1{+}2vy) M_0 (\Omega\cot)_-
   + (M\Omega)_0
   - \tfrac{i}{4} x^2 y E (\Omega\cot)_y 
   + O(y) ,
\\
  X_{\yx\yx'}^\new &=
   -i(1{+}2vy) M_0 (\Omega\csc)_-
   - \tfrac{i}{4} x^2 y E (\Omega\csc)_+ 
   + O(y) ,
\\
  Y_{\yx'}^\new = Y_\yx^\new &=
   -\frac{i x y (1{-}x) E}{2}
   \left[ (\Omega\cot)_y - \frac{1}{\Delta t} \right]
   + O(y^{3/2}) ,
\\
  \Ybar_{\yx\yx'}^\new = Y_{\yx\yx'}^\new &=
   -\frac{i x y (1{-}x) E}{2}
       \left[ (\Omega \csc)_y - \frac{1}{\Delta t} \right]
   + O(y^{3/2}) ,
\\
  Z_{\yx\yx'}^\new &=
    -i y (1{-}x)^2 E (\Omega\csc)_y
    + O(y^{3/2}) ,
\end {align}
\end {subequations}
and thence
\begin {equation}
  \bigl( X_\yx X_{\yx'} - X_{\yx\yx'}^2 \bigr)^\new
  =
  M_0^2 \biggl[
    \red{-\frac{2 i \Omega_0}{\Delta t}}
    + \Omega_0^2 + \Omega_-^2
    - \frac{x y}{2(1{-}x)\Delta t}
        \Omega_y (-\csc_y + \cot_y)
  \biggr]
  + O(y^{1/2})
\end {equation}
and the
$S_\new \equiv \bigl[ (X_\yx X_{\yx'} - X_{\yx\yx'}^2)/X_\yx X_{\yx'} \bigr]^\new$
given by (\ref{eq:Snew}).

For small $y$, eq.\ (A.46) of ref.\ \cite{qcd} for
$(\bar\alpha,\bar\beta,\bar\gamma)$ has parametric behavior
\begin {equation}
  \bar\alpha \sim y^{-3}, \qquad
  \bar\beta \simeq -\bar\gamma \sim y^{-2}, \qquad
  \bar\beta{+}\bar\gamma \sim y^{-1} .
\label {eq:abcbarscale}
\end {equation}
The dominant contribution to (\ref{eq:Dnew}) is then
\begin {equation}
  \widetilde D_{\rm(A)}^\new
  \simeq
  \frac{\CA\alphas^2\,P(x)}{4\pi^2 y} \,
  (\Omega\csc)_y^2
  \ln S_\new.
\label {eq:D1A}
\end {equation}
from $2\bar\alpha (Z_{\yx\yx'} I_1)^\new$ and, contributing to NLO terms,
\begin {equation}
  \widetilde D_{\rm(B)}^\new
  \simeq
  -
  \frac{\CA\alphas^2\,P(x)}{4\pi^2 y} \,
    \frac{\xi \Omega_y^3 \Delta t}{2 S_\new} \csc_y (\csc_y-\cot_y)^2
\label {eq:D1B}
\end {equation}
from
$\bar\alpha (
  \Ybar_{\yx\yx'} Y_{\yx\yx'} I_2
  + Y_\yx Y_{\yx'} I_2
  - \bar Y_{\yx\yx'}Y_{\yx'} I_3 - Y_\yx Y_{\yx\yx'} I_4
 )^\new$.
The sum of (\ref{eq:D1A}) and (\ref{eq:D1B}) gives our
$\Delta t\sim \sqrt{y}$ expansion of $\widetilde D_\new$ in
(\ref{eq:DDA1}).

% ---------------------------------------------------------------------------

\subsection{UV contribution}
\label {app:A1pole}

Eq. (A.66) of ref.\ \cite{qcd} gives the pole piece of $A_\new(x,y)$ as%
\footnote{
  We've specialized here to the case $\sgn M > 0$.  That is, we are
  focused on the A1 diagram and not concerned here with front-end
  transformations.
}
\begin {equation}
   A^{\rm pole}_\new(x,y) =
    \frac{\alphas^2}{2\pi^2} \,
    \frac{P(x)\,P(\frac{y}{1-x})}{1-x} \,
    \Re \biggl\{
       i \Omega_0
       \Bigl[
         -  \Bigl(
               \tfrac{1}{\eps}
               + \ln\bigl(\tfrac{\pi\mu^2}{E\Omega_0} \bigr)
            \Bigr)
         + \tfrac12 \ln(x y z)
       \Bigr]
    \biggr\} .
\label {eq:Apolenew}
\end {equation}
But this%
\footnote{
  For (\ref{eq:Apoledef}),
  see the discussion of appendix D.4 of ref.\ \cite{qcd}.
  We've used the
  proportionality sign $\propto$ here just to
  avoid dwelling on the factor of 2 difference between $d\Gamma/dx\,dy$
  and $A_\new$ in (\ref{eq:dGfund}), having to do with which is defined
  to contain the amplitude loop symmetry factor of $\tfrac12$.
  This detail will not matter for the method we will use to
  remove the ${\cal D}_2$ contribution from (\ref{eq:Apolenew}).
}
\begin {equation}
  A^{\rm pole}_\new(x,y) \propto
  \lim_{\mbox{\small``$\scriptstyle{a\to 0}$''}} 2\Re\Biggl\{
    \left[ \frac{d\Gamma}{dx\,dy} \right]_{xyy\bar x_2}^{(\Delta t < a)}
    +
    \left[ \frac{d\Gamma}{dx,dy} \right]_{xyy\bar x_2}^{({\cal D}_2)}
  \Biggr\}
\label {eq:Apoledef}
\end {equation}
contains a ${\cal D}_2$ addition which we do not want here,
because we did not make the corresponding ${\cal D}_2$ subtraction in
(\ref{eq:Dnew}).
Instead, we just want the piece
\begin {equation}
  2\Re
    \left[ \frac{d\Gamma}{dx\,dy} \right]_{xyy\bar x_2}^{(\Delta t < a)} .
\end {equation}

To remove the ${\cal D}_2$ contribution from (\ref{eq:Apolenew}),
take note of eqs.\ (4.31) and (F.37) of \cite{QEDnf}:
\begin {equation}
  \bbI =
  2\pi^2 (i \Omega_0)^{d-1} \Bigl[
    - \Bigl( \frac{2}{\eps} - \gammaE + \ln(4\pi) \Bigr)
    - \frac{\ln(2 i \Omega_0 a)+1}{i \Omega_0 a}
    - \ln(i\Omega_0 a)
    + 3\ln(2\pi)
  \Bigr] ,
\label {eq:bbIresult2}
\end {equation}
\begin {equation}
   \int_a^\infty d(\Delta t) \> {\cal D}_2^{(\bbI)}(\Delta t)
   = 2\pi^2
     \Bigl(
        \frac{\ln(2 i\Omega_0 a) + 1}{a}
        + i\Omega_0 \bigl[ \ln(2 i\Omega_0 a) - 1 \bigr]
     \Bigr) ,
\label {eq:D2int}
\end {equation}
which combine to give
\begin {equation}
  \bbI + \int_a^\infty d(\Delta t) \> {\cal D}_2^{(\bbI)}(\Delta t)
  =
  2\pi^2 (i \Omega_0)^{d-1} \Bigl[
    - \Bigl( \frac{2}{\eps} - \gammaE + \ln(4\pi) \Bigr)
    + \ln2 - 1
    + 3\ln(2\pi)
  \Bigr] .
\label {eq:bbIwithD2}
\end {equation}
By comparing (\ref{eq:bbIresult2}) to (\ref{eq:bbIwithD2}), we
see that we can recover the result (\ref{eq:bbIresult2})
{\it without}\/ the ${\cal D}_2$ addition by taking
\begin {equation}
  \frac{1}{\eps}
  \longrightarrow
  \frac{1}{\eps}
    + \frac12 \left(
        \frac{\ln(2 i\Omega_0 a) + 1}{i\Omega_0 a}
        + \ln(i\Omega_0 a) + \ln 2 - 1
      \right)
\label {eq:epssub}
\end {equation}
in the result (\ref{eq:bbIwithD2}) {\it with}\/ the
${\cal D}_2$ addition.

Making the substitution (\ref{eq:epssub}) into the formula
(\ref{eq:Apolenew}) for $A_\new^{\rm pole}$ then gives us the
piece of $A_\new^{\rm pole}$ that does not involve ${\cal D}_2$:
\begin {multline}
    \frac{\alphas^2}{2\pi^2} \,
    \frac{P(x)\,P(\frac{y}{1-x})}{1-x} \,
    \Re \biggl\{
       i \Omega_0
       \biggl[
         -  \Bigl(
               \tfrac{1}{\eps}
               + \ln\bigl(\tfrac{\pi\mu^2}{E\Omega_0} \bigr)
            \Bigr)
         + \tfrac12 \ln(x y z)
\\
         - \frac12 \left(
             \frac{\ln(2 i\Omega_0 a) + 1}{i\Omega_0 a}
             + \ln(i\Omega_0 a) + \ln 2 - 1
           \right)
       \biggr]
    \biggr\} .
\label {eq:ApolenewNoD2}
\end {multline}
Taking the small-$y$ limit and reorganizing gives
(\ref{eq:ApolenewNoD2LO}).

% ============================================================================

\section{Small-\boldmath$y$ behavior of ${\cal A}_\seq(y,x)$}
\label {app:Aseq}

In this appendix, we investigate the small-$y$ expansion of
${\cal A}_\seq(y,x)$ and its front-end transformation, which appear
as two of the four entries in the dark pink ($\beta$) section of
table \ref{tab:cancel}.  We want analytic formulas for
the small-$y$ expansion of ${\cal A}_\seq(y,x)$ that will work for
both positive and negative values of $y$.

% ---------------------------------------------------------------------------

\subsection{$A_\seq(y,x)$ vs.\ $A_\seq(x,y)$}

Most formulas concerning ${\cal A}_\seq$ in earlier papers
\cite{seq,dimreg,qcd} are written for ${\cal A}_\seq(x,y)$ rather than
${\cal A}_\seq(y,x)$.  Furthermore, ${\cal A}_\seq(x,y)$ has the same
longitudinal momentum fractions $(x_1,x_2,x_3,x_4)=(-1,y,z,x)$
for 4-particle
evolution as both the A1 and A3 diagrams discussed earlier in this
paper.  Because it is quicker and easier to re-use previous
formulas, it will be more convenient to study the small-$y$ behavior
of ${\cal A}_\seq(x,y)$ instead of ${\cal A}_\seq(y,x)$.  Fortunately, we
have checked with numerics that the small-$y$ expansions of the two
are equal to each other up to and including $O(y^{-1})$ --- that is,
they produce the same IR power-law divergences and IR logarithms.  So,
for the sake of simplicity, in this appendix we will study the
small-$y$ expansion analytically for ${\cal A}_\seq(x,y)$
(for both positive and negative values of $y$) instead of directly
for the ${\cal A}_\seq(y,x)$ appearing in table \ref{tab:cancel}.

\subsection{Setup for $A_\seq(x,y)$}

Unlike our analysis of the A1 and A3 diagrams in this paper,
we will carefully
keep track of the signs needed
in our analysis of ${\cal A}_\seq(x,y)$ to handle
either sign of $y$.

The basic formulas for ${\cal A}_\seq(x,y)$ are summarized in
eqs.\ (A.32--A.36) of ref.\ \cite{qcd}, which are
\begin {equation}
   {\cal A}_\seq(x,y)
   =
   {\cal A}^{\rm pole}_\seq(x,y)
   + \int_0^{\infty} d(\Delta t) \>
     \Bigl[
        2 \Re \bigl( B_{\rm seq}(x,y,\Delta t) \bigr)
        + F_{\rm seq}(x,y,\Delta t)
     \bigr] ,
\label {eq:Aseq}
\end {equation}
\begin {equation}
   B_\seq(x,y,\Delta t) =
       C_\seq({-}1,y,z,x,\bar\alpha,\bar\beta,\bar\gamma,\Delta t) ,
\label {eq:Bseq}
\end {equation}
\begin {equation}
   C_\seq = D_\seq - \lim_{\hat q\to 0} D_\seq ,
\end {equation}
\begin {align}
   D_\seq(x_1,&x_2,x_3,x_4,\bar\alpha,\bar\beta,\bar\gamma,\Delta t) =
\nonumber\\ &
   \frac{\CA^2 \alphas^2 M_\ix M_\fx^\seq}{32\pi^4 E^2} \, 
   ({-}x_1 x_2 x_3 x_4)
   \Omega_+\Omega_- \csc(\Omega_+\Delta t) \csc(\Omega_-\Delta t)
\nonumber\\ &\times
   \Bigl\{
     (\bar\beta Y_\yx^\seq Y_\xbx^\seq
        + \bar\alpha \Ybar_{\yx\xbx}^{\,\seq} Y_{\yx\xbx}^\seq) I_0^\seq
     + (\bar\alpha+\bar\beta+2\bar\gamma) Z_{\yx\xbx}^\seq I_1^\seq
\nonumber\\ &\quad
     + \bigl[
         (\bar\alpha+\bar\gamma) Y_\yx^\seq Y_\xbx^\seq
         + (\bar\beta+\bar\gamma) \Ybar_{\yx\xbx}^{\,\seq} Y_{\yx\xbx}^\seq
        \bigr] I_2^\seq
\nonumber\\ &\quad
     - (\bar\alpha+\bar\beta+\bar\gamma)
       (\Ybar_{\yx\xbx}^{\,\seq} Y_\xbx^\seq I_3^\seq
            + Y_\yx^\seq Y_{\yx\xbx}^\seq I_4^\seq)
   \Bigl\} ,
\label {eq:Dseq}
\end {align}
\begin {align}
   F_\seq(x,y,\Delta t) =
   \frac{\alphas^2 P(x) P(\yfrak)}{4\pi^2(1-x)}
   \Bigl[ &
      \Re\bigl(i(\Omega\sgn M)_{E,x}\bigr) \,
      \Re\bigl( \Delta t \, \Omega_{(1-x)E,\yfrak}^2
                \csc^2(\Omega_{(1-x)E,\yfrak} \, \Delta t) \bigr)
\nonumber\\
      + &
      \Re\bigl(i(\Omega\sgn M)_{(1-x)E,\yfrak}\bigr) \,
      \Re\bigl( \Delta t \, \Omega_{E,x}^2
                \csc^2(\Omega_{E,x} \, \Delta t) \bigr)
   \Bigr] ,
\label {eq:Fseq}
\end {align}
where
\begin {equation}
  M_\fx^\seq = y z (1{-}x) E.
\end {equation}
For ${\cal A}^{\rm pole}$, we want to use the corrected formula given by
eq.\ (\ref{eq:seqDR2}) of this paper instead of the formula in
ref.\ \cite{qcd}.
Other formulas of interest from appendix A.2.3 of ref.\ \cite{qcd} are
\begin {subequations}
\label {eq:XYZseqdef}
\begin {align}
   \begin{pmatrix} X_\yx^\seq & Y_\yx^\seq \\ Y_\yx^\seq & Z_\yx^\seq \end{pmatrix}
   &\equiv
   \mbox{the} ~
   \begin{pmatrix} X_\yx & Y_\yx \\ Y_\yx & Z_\yx \end{pmatrix}
   ~\mbox{of eq.\ (\ref{eq:XYZy2}) ,}
\\
   \begin{pmatrix}
        X_\xbx^\seq & Y_\xbx^\seq \\ Y_\xbx^\seq & Z_\xbx^\seq
   \end{pmatrix}
   &\equiv
   \begin{pmatrix} |M_\fx^\seq|\Omega_\fx^\seq & 0 \\ 0 & 0 \end{pmatrix}
     - i (a_\xbx^\seq)^{-1\top}
     \begin{pmatrix} \Omega_+\cot_+ & \\ & \Omega_-\cot_- \end{pmatrix}
     (a_\xbx^\seq)^{-1} ,
\\
   \begin{pmatrix} X_{\yx\xbx}^\seq & Y_{\yx\xbx}^\seq \\[2pt]
                   \Ybar_{\yx\xbx}^\seq & Z_{\yx\xbx}^\seq \end{pmatrix}
   &\equiv
     - i a_\yx^{-1\top}
     \begin{pmatrix} \Omega_+\csc_+ & \\ & \Omega_-\csc_- \end{pmatrix}
     (a_\xbx^\seq)^{-1} ,
\end {align}
\end {subequations}
where
\begin {equation}
   a_\xbx^\seq \equiv 
   \begin{pmatrix} 0 & 1 \\ 1 & 0 \end{pmatrix} a_\yx .
\label {eq:axb}
\end {equation}
Above, $a_\ybx$ is the same matrix that appeared in our
analysis of the A3 diagram in appendix \ref{app:A3}.

For small $y$, it turns out that
$\Delta t \times 2\Re B_\seq(x,y,\Delta t)$
transitions at $\Delta t \sim \sqrt{y}$ between
a constant value for $\Delta t \ll \sqrt{y}$ and zero
for $\Delta t \gg \sqrt{y}$.  For our purposes, we may
therefore adequately approximate
the behavior of $\Delta t \times 2\Re B_\seq$ for all $\Delta t$
by finding its small-$y$ approximation for $\Delta t \sim \sqrt{y}$.
It turns out that
$\Delta t \times F_\seq(x,y,\Delta t)$ does something similar
at $\Delta t \sim y^0$,
and so we may adequately approximate its behavior for all $\Delta t$
by finding its small-$y$ approximation for $\Delta t \sim y^0$.
Once we have those approximations, we will
integrate the sum $2\Re B_\seq(x,y,\Delta t) + F_\seq(x,y,\Delta t)$
over $\Delta t$ [which is equivalent to integrating
$\Delta t \times 2\Re B_\seq(x,y,\Delta t)
  + \Delta t \times F_\seq(x,y,\Delta t)$
over $\ln(\Delta t)$]
as in (\ref{eq:Aseq}).

% ---------------------------------------------------------------------------

\subsection{$B_{\rm seq}$ for $\Delta t \sim \sqrt{y}$}

Using (\ref{eq:ainv}) and (\ref{eq:axb}), we find that the expansions we
need of
(\ref{eq:XYZseqdef}) for $\Delta t \sim \sqrt{y}$ are
\begin {subequations}
\begin {align}
  X_\yx^\seq &=
   -i \frac{M_0}{\Delta t}
   + \bigl(M\Omega\bigr)_0
   + O(y^{1/2}) ,
\\
  X_\xbx^\seq = Z_{\yx\xbx}^\seq &=
   (M\Omega)_\fx^\seq (\sgn y - i\cot_y) + O(y^{3/2})
,
\\
  Y_{\yx\xbx}^\seq &=
   - \frac{iM_0}{\Delta t} + O(y^{1/2}) ,
\\
  \Ybar_{\yx\ybx}^\seq &=
    -i M_\fx^\seq (\Omega\csc)_y + O(y^{3/2}) ,
\end {align}
and the parametric result that
\begin {equation}
   X_{\yx\xbx}^\seq \sim Y_\yx^\seq \sim Y_\xbx^\seq \sim y^{1/2} .
\end {equation}
\end {subequations}
Above, we have taken $M_0 = x(1{-}x)E$ to be positive, but
$y$ and so $M_\fx^\seq = y (1{-}x) (1{-}x{-}y)$ may have either sign.
Note in particular that $|M_\fx^\seq| = M_\fx^\seq \sgn y$.
Unlike the analysis of the A1 and A3 diagrams, we do {\it not} need
expansions of the $X$'s to NNLO in order to get a NLO result for
$(X_\yx X_\xbx - X_{\yx\xbx}^2)^\seq$ because here the leading-order
behavior of the $X$'s does not cancel in the combination.

Using the scaling (\ref{eq:abcbarscale}) of $(\bar\alpha,\bar\beta,\bar\gamma)$,
the only term of (\ref{eq:Dseq}) that contributes up to
NLO in $\sqrt{y}$ is the $\bar\alpha \Ybar_{\yx\xbx} Y_{\yx\xbx} I_0$ term.
Because we want to allow for front-end transformations, we need to use
the version of $\bar\alpha$ in eq.\ (A.46) of
ref.\ \cite{qcd} that is consistent with
front-end transformations, which involves appropriate absolute value signs.
The small-$y$ limit is
\begin {equation}
  \bar\alpha = \frac{2\,P(x)}{\CA x^2(1{-}x)^4 |y|^3} \, \bigl[ 1 + O(y) \bigr],
\end {equation}
instead of the less-general (\ref{eq:baralphaapprox}) that we used for
the A1 diagram (where $y$ was positive).
We then find
\begin {equation}
  D_\seq \simeq 
  \frac{\CA \alphas^2 \, P(x)}{4\pi^2|y|\,\Delta t} \,
     (\red{1} - i\Omega_0\,\Delta t) \,\Omega_y (\cot_y - i\sgn y)
  .
\end {equation}
Subtracting the vacuum limit $\hat q \to 0$ then gives
\begin {equation}
  B_\seq = C_\seq \simeq 
  \frac{\CA \alphas^2 \, P(x)}{4\pi^2|y|\,\Delta t}
  \left[
     (\red{1} - i\Omega_0\,\Delta t) \, \Omega_y (\cot_y - i\sgn y)
     - \red{\frac{1}{\Delta t}}
   \right] .
\label {eq:Bseqy}
\end {equation}

% ---------------------------------------------------------------------------

\subsection{Assembling ${\cal A}_\seq(x,y)$}

As mentioned earlier, the behavior of $F_\seq(x,y,\Delta t)$ is non-trivial
only at $\Delta t \sim y^0$.
So expand (\ref{eq:Fseq}) for small $y$ at $\Delta t \sim y^0$.
The gluon
splitting function $P(\xi)$ also has an absolute value sign in its
definition in eq.\ (A.5) of ref.\ \cite{qcd}, in order to facilitate front-end
transformations.  Accounting for this, the expansion of $F_\seq$ is
\begin {equation}
  F_{\rm seq} \simeq
  \frac{\CA \alphas^2 \, P(x)}{2\pi^2|y|}
  \biggl[
     \red{\Re(i\Omega_y\sgn y) \Re\bigl( \Delta t \, (\Omega\csc)_0^2 \bigr)}
     +
     \Re(i\Omega_0) \Re\bigl( \Delta t \, (\Omega\csc)_y^2 \bigr)
  \biggr]
\label {eq:Fseqy}
\end {equation}
through NLO.

The small-$y$ expansion of the pole term (\ref{eq:seqDR2}) is
\begin {equation}
   {\cal A}_\seq^{\rm pole} \simeq
   - \frac{\CA\alphas^2 \, P(x)}{2\pi^2|y|}
   \Re\bigl[
     \red{i \Omega_y \sgn y} + i \Omega_0
   \bigr]
   \bigl(1-\tfrac{\pi}{2} \sgn y \bigr)
   .
\label {eq:seqDR2y}
\end {equation}
Combining (\ref{eq:Bseqy}), (\ref{eq:Fseqy}) and
(\ref{eq:seqDR2y}), the small-$y$ expansion of (\ref{eq:Aseq}) is
\begin {align}
   {\cal A}_\seq(x,y)
   \simeq &
   \frac{\CA \alphas^2 \, P(x)}{2\pi^2|y|}
   \biggl\{
      -\Re(\red{i\Omega_y\sgn y} + i\Omega_0)
           \bigl( 1 - \tfrac{\pi}{2} \sgn y \bigr)
\nonumber\\ & \quad
      + \int_0^{\infty} d(\Delta t) \>
      \biggl[
        \red{
          \Re\left(
             \frac{\Omega_y(\cot_y-i\sgn y)}{\Delta t}
          \right)
          - \frac{1}{(\Delta t)^2}
}\nonumber\\ &\hspace{9em} {} \red{
          + \Re(i\Omega_y\sgn y)
            \Re\left(
               \Delta t \, (\Omega\csc)_0^2
            \right)
        }
\nonumber\\ &\hspace{9em} {}
          - \Re\bigl( i\Omega_0 \Omega_y (\cot_y - i\sgn_y) \bigr)
          + \Re(i\Omega_0) \Re\bigl( \Delta t \, (\Omega\csc)_y^2 \bigr)
     \biggr]
  \biggr\}.
\label {eq:Aseq2}
\end {align}
The $\Delta t$ integral of individual terms in the integrand above
would be
divergent due to their $\Delta t{\to}0$ or $\Delta t{\to}\infty$
behavior.
But the combination of terms in the integrand conspires so that
the total integral is convergent.

% --------------------------------------------------------------------------

\subsection{Integration}

The integral may be performed using the same techniques as appendix B
of ref.\ \cite{qcd}.  First, add an unnecessary regulator:
multiply the integrand by $(\Delta t)^\eps$, with the understanding that
we will take $\eps\to 0$ at the end of the calculation.
If we treat this regulator with the same logic as dimensional regularization
(i.e.\ calculate integrals for values of $\eps$ where they converge and then
analytically continue to $\eps\to 0$), we may the split up the integral
into regulated integrals of the individual terms in the integrand and
tackle those integrals one at a time.

% ...........................................................................

\subsubsection{Leading-order terms}

For instance, defining%
\footnote{
  Recall that the complex phase of $\Omega_y$ is $e^{-i(\pi/4)\sgn y}$.
  The $\sgn y$ is needed in the definition of $\tilde\tau$ in
  (\ref{eq:tildetausgn}) so that, in the first step of (\ref{eq:intA}),
  one does not deform the contour through a region where the integrand blows
  up as $|\Delta t| \to \infty$.
}
\begin {equation}
  \tilde\tau \equiv i (\Omega_y \sgn y) \Delta t ,
\label {eq:tildetausgn}
\end {equation}
we may write
\begin{align}
  \int_0^\infty d(\Delta t) & \, (\Delta t)^\eps \,
  \frac{\Omega_y(\cot_y-i\sgn y)}{\Delta t}
  =
  (i\Omega_y\sgn y)^{1-\eps}
  \int_0^\infty d\tilde\tau \> \tilde\tau^{-1+\eps} \,
     (\coth\tilde\tau - 1)
\nonumber\\ &
  = (i\Omega_y\sgn y)^{1-\eps}
     \int_0^\infty d\tilde\tau \> \tilde\tau^{-1+\eps} \,
     2 \sum_{n=1}^\infty e^{-2n\tilde\tau}
  = (2 i \Omega_y\sgn y)^{1-\eps} \, \Gamma(\eps) \, \zeta(\eps)
\nonumber\\ &
  = i\Omega_y\sgn y
    \left[
       - \frac{1}{\eps}
       + \ln\Bigl( \frac{i\Omega_y\sgn y}{\pi} \Bigr)
       + \gammaE
       + O(\eps)
    \right] .
\label {eq:intA}
\end {align}
By logic similar to dimensional regularization, integrals of powers
like $1/(\Delta t)^2$ give zero.  For the next integral, switch to
$\sigma \equiv i\Omega_0\,\Delta t$ and use the integral
from eq.\ (B.13) of ref.\ \cite{qcd}:
\begin {equation}
  \int_0^\infty d(\Delta t) \> (\Delta t)^\eps \,
     \Delta t \, (\Omega\csc)_0^2
  = (i\Omega_0)^{-\eps} \int_0^\infty d\sigma
     \> \frac{\sigma^{1+\eps}}{\sh^2\sigma}
  = \frac{1}{\eps} - \ln(2i\Omega_0) + 1 + O(\eps) .
\end {equation}
Combining these integration results gives the convergent integral of the
leading (red)
terms in (\ref{eq:Aseq2}):
\begin {multline}
   \int_0^{\infty} d(\Delta t) \>
      \biggl[
%        \red{
          \Re\left(
             \frac{\Omega_y(\cot_y-i\sgn y)}{\Delta t}
          \right)
          - \frac{1}{(\Delta t)^2}
          + \Re(i\Omega_y\sgn y)
            \Re\left(
               \Delta t \, (\Omega\csc)_0^2
            \right)
%        }
      \biggr]
\\
   =
   \Re\bigl( i\Omega_y\sgn y \ln(i\Omega_y\sgn y) \bigr)
   + \Re( i\Omega_y\sgn y )
     \Re \bigl[ - \ln(i\Omega_0) - \ln(2\pi) + \gammaE + 1 \bigr]
   .
\label {eq:AseqintLO}
\end {multline}

% ...........................................................................

\subsubsection{NLO terms}

Similarly, the integrals we need for the NLO (black) terms in (\ref{eq:Aseq2})
are
\begin {align}
  \int_0^\infty d(\Delta t) \> (\Delta t)^\eps \,
     \Omega_y (\cot_y - i\sgn_y)
  &= (2 i \Omega_y\sgn y)^{-\eps} \, \Gamma(1{+}\eps) \, \zeta(1{+}\eps)
\nonumber\\ & \qquad\qquad
  = \frac{1}{\eps} - \ln(2i\Omega_y\sgn y)
       + O(\eps) ,
\\
  \int_0^\infty d(\Delta t) \> (\Delta t)^\eps \,
     \Delta t \, (\Omega\csc)_y^2
  &= \frac{1}{\eps} - \ln(2i\Omega_y\sgn y) + 1 + O(\eps) ,
\end {align}
which gives
\begin {multline}
   \int_0^{\infty} d(\Delta t) \>
   \Bigl[
          - \Re\bigl( i\Omega_0 \Omega_y (\cot_y - i\sgn_y) \bigr)
          + \Re(i\Omega_0) \Re\bigl( \Delta t \, (\Omega\csc)_y^2 \bigr)
   \Bigr]
\\
   =
   \Re\bigl( i\Omega_0 \ln(i\Omega_y\sgn y) \bigr)
   + \Re( i\Omega_0 )
     \Re\bigl( -\ln(i\Omega_y\sgn y) + 1 \bigr)
   .
\label {eq:AseqintNLO}
\end {multline}

% ---------------------------------------------------------------------------

\subsection{Final expansion and implications}

Now use (\ref{eq:AseqintLO}) and (\ref{eq:AseqintNLO}) in (\ref{eq:Aseq2})
to get the small-$y$ expansion of ${\cal A}_\seq(x,y)$.  We can isolate
the relative phases of the $y>0$ and $y<0$ case by using the fact
that the complex phase of $\Omega_y$ is $e^{-i(\pi/4)\sgn y}$ to write
the final result in the form
\begin {equation}
   {\cal A}_\seq(x,y)
   \simeq
   \frac{\CA \alphas^2 \, P(x)}{2\pi^2|y|}
   \biggl\{
     \red{
       \Re( i\Omega_{|y|} )
     }
       \left[
         \ln\Bigl( \frac{|\Omega_y|}{2\pi|\Omega_0|} \Bigr) + \gammaE
         - \frac{\pi}{4} + \frac{\pi}{2} \sgn y
       \right]
     +
       \blue{\Re( i\Omega_0 )}
       \frac{\pi}{4} \sgn y
  \biggr\} .
\label {eq:Aseq3NEW}
\end {equation}

The discussion surrounding
eqs.\ (\ref{eq:frontendy}--\ref{eq:leftover}) in the main text qualitatively
explained the imperfect cancellation, in the small-$y$ limit,
of the ${\cal A}_\seq(y,x)$
diagrams with the virtual diagrams given
by the front-end transformation of ${\cal A}_\seq(y,x)$.
In the small-$y$ limit (up to and including the order relevant for
IR logarithms), that front-end
transformation involves an overall minus sign for the diagrams and the
negation of $y$ as in (\ref{eq:frontendsmally}).  So the sum of the
original and front-end transformed diagrams gives
\begin {equation}
   {\cal A}_\seq(y,x) + \operatorname{frEnd}[{\cal A}_\seq(y,x)]
   \simeq
   {\cal A}_\seq(y,x) - {\cal A}_\seq(-y,x) ,
\label {eq:FEcancel}
\end {equation}
where now $y>0$.
As discussed at the start of this appendix, ${\cal A}_\seq(y,x)$ has the
same small-$y$ behavior as ${\cal A}_\seq(x,y)$.
The explicit formula (\ref{eq:Aseq3NEW}) for the latter
then gives that the sum
of diagrams (\ref{eq:FEcancel}) is equivalent to
\begin {equation}
   {\cal A}_\seq(x,y) - {\cal A}_\seq(x,-y)
   \simeq
   \frac{\CA \alphas^2 \, P(x)}{2\pi^2 y}
   \biggl\{
     \red{
       \Re( i\Omega_{y} )
     }
     \pi
     +
       \blue{\Re( i\Omega_0 )}
       \frac{\pi}{2}
  \biggr\} .
\label {eq:Aseqshift}
\end {equation}
As discussed in the main text but now seen explicitly,
everything has canceled {\it except}\/ for $\pi$ terms that originated
from logarithms of complex phases that were changed in some
way by $y \to -y$.
From numerics, we know that these $\pi$ terms cancel only when the
larger set of diagrams depicted by the dark pink ($\beta$) regions of
table \ref{tab:cancel} are added together.

% ----------------------------------------------------------------------------

\subsection {Implication for $\bar s(x)$ and $c(x)$ in fig.\ \ref{fig:cnew}}
\label {app:cshift}

In arriving at (\ref{eq:Aseqshift}), we used the corrected version
(\ref{eq:seqDR2}) of ${\cal A}_\seq^{\rm pole}$.
If we had instead used the original, uncorrected version for eq.\ (A.37)
of ref.\ \cite{qcd}, we would have obtained
\begin {equation}
   \left[ {\cal A}_\seq(x,y) - {\cal A}_\seq(x,-y) \right]^{\rm wrong}
   \simeq
   \frac{\CA \alphas^2 \, P(x)}{2\pi^2 y}
   \biggl\{
     \red{
       \Re( i\Omega_{y} )
     }
     \pi
     -
       \blue{\Re( i\Omega_0 )}
       \frac{3\pi}{2}
  \biggr\} .
\label {eq:AseqshiftWrong}
\end {equation}
Taking the difference of (\ref{eq:Aseqshift}) and (\ref{eq:AseqshiftWrong}),
the effect of the change on the total small-$y$ rate (including all diagrams
and not just ABC diagrams) is
\begin {equation}
  \left[ \frac{d\Gamma}{dx\,dy} \right]
  -
  \left[ \frac{d\Gamma}{dx\,dy} \right]^{\rm wrong}
  \simeq
   \frac{\CA \alphas^2 \, P(x)}{\pi y}
       \Re( i\Omega_0 )
  =
   \frac{\CA \alphas}{y}
       \left[ \frac{d\Gamma}{dx} \right]_{\rm LO}
  .
\end {equation}
Comparison to (\ref{eq:total2}), ignoring the subscript ABC there,
yields
\begin {equation}
  \bar s(x)
  = [\bar s(x)]^{\rm wrong} - 4\pi .
\end {equation}
This is the origin of the $4\pi$ downward shift of our fig.\ \ref{fig:cnew}
compared to the earlier fig.\ 20 of ref.\ \cite{qcd}.

%%%%%%%%%%%%%%%%%%%%%%%%%%%%%%%%%%%%%%%%%%%%%%%%%%%%%%%%%%%%%%%%%%%%%%%%%%%%%%


\begin{thebibliography}{99}

\bibitem{LP1}
  L.~D.~Landau and I.~Pomeranchuk,
  ``Limits of applicability of the theory of bremsstrahlung electrons and
  pair production at high-energies,''
  Dokl.\ Akad.\ Nauk Ser.\ Fiz.\  {\bf 92} (1953) 535.

\bibitem{LP2}
  L.~D.~Landau and I.~Pomeranchuk,
  ``Electron cascade process at very high energies,''
  Dokl.\ Akad.\ Nauk Ser.\ Fiz.\  {\bf 92} (1953) 735.

\bibitem{Migdal}
  A.~B.~Migdal,
  ``Bremsstrahlung and pair production in condensed media at high-energies,''
   Phys.\ Rev.\  {\bf 103}, 1811 (1956);

\bibitem{LPenglish}
  L. Landau,
  {\sl The Collected Papers of L.D. Landau}\/
  (Pergamon Press, New York, 1965).

\bibitem{BDMPS1}
  R.~Baier, Y.~L.~Dokshitzer, A.~H.~Mueller, S.~Peigne and D.~Schiff,
  ``The Landau-Pomeranchuk-Migdal effect in QED,''
  Nucl.\ Phys.\  B {\bf 478}, 577 (1996)
  [arXiv:hep-ph/9604327];

\bibitem{BDMPS2}
  R.~Baier, Y.~L.~Dokshitzer, A.~H.~Mueller, S.~Peigne and D.~Schiff,
  ``Radiative energy loss of high-energy quarks and gluons in a
    finite volume quark - gluon plasma,''
  Nucl.\ Phys.\  B {\bf 483}, 291 (1997) [arXiv:hep-ph/9607355].
  %%CITATION = NUPHA,B483,291;%%

\bibitem{BDMPS3}
  R.~Baier, Y.~L.~Dokshitzer, A.~H.~Mueller, S.~Peigne and D.~Schiff,
  ``Radiative energy loss and $p_\perp$-broadening of high energy partons in
    nuclei,''
  {\it ibid.}\ {\bf 484} (1997)
  [arXiv:hep-ph/9608322].
  %%CITATION = NUPHA,B484,265;%%

\bibitem{Zakharov1}
 B.~G.~Zakharov,
 ``Fully quantum treatment of the Landau-Pomeranchuk-Migdal effect in
   QED and QCD,''
 JETP Lett.\  {\bf 63}, 952 (1996)
 [arXiv:hep-ph/9607440].

\bibitem{Zakharov2}
 B.~G.~Zakharov,
 ``Radiative energy loss of high-energy quarks in finite size nuclear matter
   and quark-gluon plasma,''
 JETP Lett.\  {\bf 65}, 615 (1997)
 [Pisma Zh.\ Eksp.\ Teor.\ Fiz.\  {\bf 63}, 952 (1996)]
 [arXiv:hep-ph/9607440].
 %%CITATION = JTPLA,63,952.%%

\bibitem{Blaizot}
  J.~P.~Blaizot and Y.~Mehtar-Tani,
  ``Renormalization of the jet-quenching parameter,''
  Nucl.\ Phys.\ A {\bf 929}, 202 (2014)
  [arXiv:1403.2323 [hep-ph]].
  %%CITATION = ARXIV:1403.2323;%%

\bibitem{Iancu}
  E.~Iancu,
  ``The non-linear evolution of jet quenching,''
  JHEP \textbf{10}, 95 (2014)
  [arXiv:1403.1996 [hep-ph]].
  %%CITATION = ARXIV:1403.1996;%%

\bibitem{Wu}
  B.~Wu,
  ``Radiative energy loss and radiative $p_{\bot}$-broadening of
    high-energy partons in QCD matter,''
  JHEP \textbf{12}, 081 (2014)
  [arXiv:1408.5459 [hep-ph]].
  %%CITATION = ARXIV:1408.5459;%%

\bibitem{Wu0}
  T.~Liou, A.~H.~Mueller and B.~Wu,
  ``Radiative $p_\bot$-broadening of high-energy quarks and gluons in
    QCD matter,''
  Nucl.\ Phys.\ A {\bf 916}, 102 (2013)
  [arXiv:1304.7677 [hep-ph]].
  %%CITATION = ARXIV:1304.7677;%%
  %15 citations counted in INSPIRE as of 27 Oct 2014

\bibitem{2brem}
  P.~Arnold and S.~Iqbal,
  ``The LPM effect in sequential bremsstrahlung,''
  JHEP \textbf{04}, 070 (2015)
  [{\it erratum} JHEP \textbf{09}, 072 (2016)]
  %doi:10.1007/JHEP09(2016)072, 10.1007/JHEP04(2015)070
  [arXiv:1501.04964 [hep-ph]].
  %%CITATION = doi:10.1007/JHEP09(2016)072, 10.1007/JHEP04(2015)070;%%
  %9 citations counted in INSPIRE as of 20 Sep 2016

\bibitem{seq}
  P.~Arnold, H.~C.~Chang and S.~Iqbal,
  ``The LPM effect in sequential bremsstrahlung 2: factorization,''
  JHEP \textbf{09}, 078 (2016)
  [arXiv:1605.07624 [hep-ph]].
  %%CITATION = ARXIV:1605.07624;%%

\bibitem{dimreg}
  P.~Arnold, H.~C.~Chang and S.~Iqbal,
  ``The LPM effect in sequential bremsstrahlung: dimensional regularization,''
  JHEP \textbf{10}, 100 (2016)
  %doi:10.1007/JHEP10(2016)100
  [arXiv:1606.08853 [hep-ph]].
  %%CITATION = ARXIV:1606.08853;%%
  %10 citations counted in INSPIRE as of 25 May 2020

\bibitem{4point}
  P.~Arnold, H.~C.~Chang and S.~Iqbal,
  ``The LPM effect in sequential bremsstrahlung: 4-gluon vertices,''
  JHEP \textbf{10}, 124 (2016)
  %doi:10.1007/JHEP10(2016)124
  [arXiv:1608.05718 [hep-ph]].
  %%CITATION = doi:10.1007/JHEP10(2016)124;%%
  %4 citations counted in INSPIRE as of 10 Jan 2018

\bibitem{QEDnf}
  P.~Arnold and S.~Iqbal,
  ``In-medium loop corrections and longitudinally polarized gauge bosons
    in high-energy showers,''
  JHEP \textbf{12}, 120 (2018)
  %doi:10.1007/JHEP12(2018)120
  [arXiv:1806.08796 [hep-ph]].
  %%CITATION = doi:10.1007/JHEP12(2018)120;%%

\bibitem{qedNfstop}
  P.~Arnold, S.~Iqbal and T.~Rase,
  ``Strong- vs. weak-coupling pictures of jet quenching: a dry run using QED,''
  JHEP \textbf{05}, 004 (2019)
  %doi:10.1007/JHEP05(2019)004
  [arXiv:1810.06578 [hep-ph]].

\bibitem{qcd}
P.~Arnold, T.~Gorda and S.~Iqbal,
``The LPM effect in sequential bremsstrahlung:
  nearly complete results for QCD,''
JHEP \textbf{11}, 053 (2020)
%doi:10.1007/JHEP11(2020)053
[arXiv:2007.15018 [hep-ph]].

\bibitem{logs2}
  P. Arnold,
  ``Universality (beyond leading log) of soft radiative corrections
    to $\hat q$ in $p_\perp$ broadening and energy loss,''
  [arXiv:2111.05348 [hep-ph]].

%\bibitem{1overN}
%P.~Arnold and O.~Elgedawy, in preparation.

\bibitem{qcdI}
P.~Arnold, T.~Gorda and S.~Iqbal, in preparation.

\end{thebibliography}
\end{document}